\documentclass[aps,amsmath,amssymb,showpacs,showkeys]{revtex4}
\usepackage[dvips]{graphicx,color}
\usepackage{times}
\usepackage{multirow}
\usepackage{xcolor}
\usepackage[%
  colorlinks=true,
  urlcolor=blue,
  linkcolor=red,
  citecolor=blue
]{hyperref}

\newcommand{\orcid}[1]{\href{https://orcid.org/#1}{\includegraphics[width=8pt]{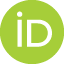}}}

\begin{document}
\title{Tideless Traversable Wormholes surrounded by cloud of strings in f(R) gravity}

\author{Dhruba Jyoti Gogoi \orcid{0000-0002-4776-8506}}
\email[Email: ]{moloydhruba@yahoo.in}

\affiliation{Department of Physics, Dibrugarh University,
Dibrugarh 786004, Assam, India}

\author{Umananda Dev Goswami \orcid{0000-0003-0012-7549}}
\email[Email: ]{umananda2@gmail.com}
\affiliation{Department of Physics, Dibrugarh University,
Dibrugarh 786004, Assam, India}

\begin{abstract}
We study the tideless traversable wormholes in the $f(R)$ gravity metric 
formalism. First we consider three shape functions of wormholes and study 
their viabilities and structures. The connection between the $f(R)$ gravity 
model and wormhole shape function has been studied and the dependency of the 
$f(R)$ gravity model with the shape function is shown. We also obtain a 
wormhole solution in the $f(R)$ gravity Starobinsky model surrounded by a 
cloud of strings. In this case, the wormhole shape function depends on both 
the Starobinsky model parameter and the cloud of strings parameter. The 
structure and height of the wormhole is highly affected by the cloud of 
strings parameter, while it is less sensitive to the Starobinsky model 
parameter. The energy conditions have been studied and we found the ranges of 
the null energy condition violation for all wormhole structures. The 
quasinormal modes from these wormhole structures for the scalar and Dirac 
perturbations are studied using higher order WKB approximation methods. The 
quasinormal modes for the toy shape functions depend highly on the model 
parameters. In case of the Starobinsky model's  wormhole the quasinormal 
frequencies and the damping rate increase with an increase in the Starobinsky 
model parameter in scalar perturbation. Whereas in Dirac perturbation, with 
an increase in the Starobinsky model parameter the quasinormal frequencies 
decrease and the damping rate increases. The cloud of strings parameter also 
impacts prominently and differently the quasinormal modes from the wormhole 
in the Starobinsky model.
\end{abstract}

\keywords{Modified Gravity; Quasinormal modes; Wormholes}

\maketitle

\section{Introduction}\label{section01}
Recent experimental observations suggest that the universe is undergoing a 
phase of accelerated expansion \cite{Riess_2001, Perlmutter_1999, Bennett_2003}.
Since General Relativity (GR) can't explain this current phase of expansion of
the universe, several modifications to GR have been introduced including 
dark energy models and Modified Gravity Theories (MGTs) \cite{Sami}. In MGTs, 
the curvature part of Lagrangian is modified to explain the observational 
results. One of the attractive and simplest forms of such extensions is the 
$f(R)$ theories of gravity, where the Ricci scalar $R$ in the Lagrangian or 
action of GR is replaced by an arbitrary function of $R$. Some of the 
promising $f(R)$ gravity models are: Starobinsky model \cite{Starobinsky1980}, 
Hu-Sawicki model \cite{hu_sawicki}, Tsujikawa model \cite{Tsujikawa}, 
Gogoi-Goswami model \cite{gogoi2} etc. Of late, there are many works in 
which the viabilities and different aspects of these models have been 
explicitly studied \cite{gogoicosmology,jbora,BHu,Martinelli2009, Asaka2016,Cen2019,Baruah2021, Parbin2022}.  Apart from these, various other $f(R)$ gravity models are studied in different perspectives \cite{Sotiriou2006, Huang2014, Lu2016, Nojiri2007, Cognola2008, Felice2010, Bamba2011, Bamba2010, Bamba2012, Sebastiani2014, Thakur2013, Bamba2014, Sebastiani2018, Motohashi2017, Astashenok2017, Bahamonde2017, Faraoni2017, Abbas2017, Sussman2017,Mongwane2017, Mansour2018, Papagiannopoulos2018, Oikonomou2018, Castellanos2018, Samanta2020, Kalita2021, Sarmah2022, Kalita2022}.

Like GR, MGTs also show the possibilities of black holes and wormholes. Recent 
studies show that the MGTs play a very important role in the 
study of black holes and wormholes. It is seen that the black hole solutions, 
black hole thermodynamics and quasinormal modes are affected by the type of 
modifications introduced in the extended forms of gravity 
\cite{gogoi3, Gogoi_bumblebee}. Lately, it was found that the polarization 
modes of Gravitational Waves (GWs) increase from two to three in $f(R)$ 
theory metric formalism due to the existence of extra degrees of freedom. The 
extra polarization mode is a scalar massive polarization mode 
which is a mixture of the massless transverse but not traceless breathing 
scalar mode and the massive longitudinal scalar mode \cite{gogoi2, gogoi1}. 
In recent times there are plenty of works in literature that are related to the
study of different aspects of wormholes in the $f(R)$ theory of gravity. For
example, in Ref.~\cite{Lobo2009}, the traversable wormholes have been studied 
in the framework of $f(R)$ gravity. In this work, the authors have explicitly 
studied the factors responsible for the null energy condition violation 
supporting the 
existence of wormholes for different shape functions. The cosmological model 
with a traversable wormhole has been studied in Ref.~\cite{Kim1996}. In 
Ref.~\cite{Bronnikov:2010tt}, existence of wormholes in scalar tensor theory 
and $f(R)$ gravity has been studied. In another study, the cosmological 
evolution of wormhole solutions in $f(R)$ gravity has been explored 
\cite{Bahamonde2016}. This study deals with the construction of dynamical
wormhole that asymptotically approaches a Friedmann-Lema\^itre-Robertson-Walker
universe. Traversable wormholes in $f(R)$ gravity with non-commutative geometry
 has been studied in Ref.~\cite{Kuhfittig2018}. In this work, the author has 
studied the relation of wormhole shape function with the $f(R)$ gravity model 
explicitly and also studied the associated energy conditions at the throat of 
the wormhole. Exactly traversable wormholes in bumblebee 
gravity have been obtained recently in Ref.~\cite{Jusufi19}. The thin-shell 
wormholes in quadratic $f(R)$ gravity have been studied in 
Ref.~\cite{Eiroa2016}, where the authors present a class of spherically 
symmetric Lorentzian wormholes in inflationary Starobinsky type model with and 
without charge. The authors constructed the wormholes by cutting and pasting 
two manifolds with different constant curvatures into a hypersurface 
representing the throat of the wormhole which is not symmetric across the 
throat. Wormholes in $f(R)$ gravity sourced by a phantom scalar field have been 
recently studied in Ref.~\cite{Karakasis2022}, where the authors obtained exact 
wormhole solutions and studied the energy conditions explicitly by considering 
a scalar field with negative kinetic energy, a phantom scalar field which 
has a self interacting potential. In their study, they have obtained the 
wormhole solution without specifying the actual form of the $f(R)$ function. 
They show that the presence of such a scalar field has impacts on the scalar 
curvature and the size of the wormhole throat. An increase in the strength of 
the scalar field results in wormholes with larger throat radii, which 
eventually decreases the curvature near the wormhole. The traversable wormhole 
geometries using Karmarkar condition have been studied in 
Ref.~\cite{Shamir2020}. In Ref.~\cite{Chervon2021}, the authors have studied 
the wormholes and black holes in $f(R)$ gravity with a kinetic curvature scalar.
Traversable wormholes in the $R+\alpha R^n$ model were studied by N.~Godani and 
G.~C.~Samanta in Ref.~\cite{Godani2020}. Deflection angle of the 
Brane-Dicke 
wormhole in the weak field limit has been studied in Ref.~\cite{Javed2019}. In 
another work, the weak deflection angle by black holes and wormholes has been 
extensively studied with different examples and wormhole structures 
\cite{Ovgun202101}. Apart from these, there are several studies dealing with 
deflection angle in the domain of black holes and wormholes 
\cite{Javed2022,Ovgun2020d,Pantig202202,Javed202201,Javed202202}. Traversable 
wormholes in extended teleparallel gravity with matter coupling is studied in 
Ref.~\cite{Mustafa2021}, where the authors explored mainly non-commutative 
Lorentzian and Gaussian distributed wormholes explicitly. An evolving wormhole 
hole configuration is studied in the dark matter halo in 
Ref.~\cite{Ovgun202102}.

It is worth to be mentioned that the wormhole solutions are primarily useful 
as ``gedanken-experiments'' and as a theoretical probe of the possible results 
of GR. In basic GR, wormholes are supported by exotic matters. These exotic 
matters involve a stress-energy tensor which can violate the Null Energy 
Condition (NEC) \cite{Morris:1988cz,Visser}. The NEC is defined by 
$T_{\mu\nu}k^\mu k^\nu \geq 0$, where $k^\mu$ is any null vector. Hence, in 
wormhole physics it is an important challenge to find any realistic matter 
source which can violate the NEC.

The quasinormal modes of black holes have been studied extensively in different
 MGTs \cite{gogoi3, qnm_bumblebee, Bhatta2018, Liu2022, Ovgun201802, Okyay2022, Karmakar2022, Gonzal2010, Pantig202201, Yang2022}. In such a study, it was seen 
that the quasinormal modes from black holes may play a very important part in 
distinguishing GR from $f(R)$ gravity \cite{Bhatta2018}. Along with the black 
holes, the quasinormal modes of wormholes also have received significant 
attention from the researchers. The quasinormal modes of a natural anti-de 
Sitter wormhole in Einstein-Born-Infeld gravity have been extensively studied 
in Ref.~\cite{Kim2018}. Authors in this paper studied the dependencies of the 
quasinormal modes from the wormholes on the mass of the scalar field as well 
as on other wormhole parameters. Recently, quasinormal modes from a wormhole 
in bumblebee gravity have been studied in Ref.~\cite{Oliveira2019}. It has 
been shown recently that similar to black holes, the arbitrarily long lived modes or quasi-resonances can also exist in case of a wormhole if it does not have a constant red-shift function \cite{Churilova2020}. The quasinormal modes, echoes and shadows of wormholes without exotic matter were studied in Ref.~\cite{Churilova2021}. 
Apart from these studies, there are many recent studies which involves different properties of wormholes as well as quasinormal modes and echoes \cite{Kim2008, Jusufi2021, Dutta2020, Gonz2022, Aneesh2018, Blaz2018, Bronnikov2021, Volkel2018, Bueno2018, Bronnikov2020, Ovgun2018, Jusufi2018, Halilsoy2014, Jusufi2017, Ovgun2016, Ovgun2016_2, Liu2021, Junior2020}.
Being motivated from these studies, we study wormholes in the $f(R)$ gravity 
Starobinsky model and in the models defined by three shape functions in this 
work. At first, we use three different types of wormhole shape functions and 
study their viability conditions. After that we show the dependency of the shape functions with $f(R)$ gravity model in presence of cloud of strings. In the next stage, we consider the $f(R)$ 
gravity Starobinsky model and obtain the wormhole solution surrounded by a 
cloud of strings. The idea of cloud of strings in GR has been implemented for 
the first time in Ref.~\cite{Letelier1979}. Since then the cloud of strings 
has been used by many researchers in different perspectives \cite{Ganguly2014, Bronnikov2016, Herscovich2010, Ghosh2014, Younesizadeh2022, Badawi2022, Belhaj2022, Sood2022, Ali2021, Singh2020}. The impact of the cloud of strings on 
wormholes in GR has been studied previously in Ref.~\cite{Richarte2008}. 
However, in our work, we shall 
consider this relic in $f(R)$ gravity to obtain the wormhole solutions. Such 
a study will help us to see the impact of clouds of strings on the shape of 
the wormholes. 
Apart from this, we shall study the quasinormal modes of such 
wormholes using the WKB approximation method. A comparative study with the 
previously used wormhole shape functions will be done to get a clear view on 
the quasinormal mode dependencies on the shape functions. This study will also 
provide some important insights on the possibilities of wormholes and chances 
of detecting them using the quasinormal modes for the considered $f(R)$ 
gravity models. It will also provide the possibilities of differentiating a 
wormhole and a black hole in terms of quasinormal modes. In our work, we shall 
consider two types of perturbations. The general scalar perturbation and Dirac 
perturbation. The scalar field perturbation is commonly used in different 
studies to check the behaviour of the quasinormal modes and a study of scalar 
quasinormal modes in wormhole spacetime will help us to compare the results 
easily with quasinormal modes from black holes. On the other hand, interaction 
of the Dirac field with gravity has been studied in Ref.s \cite{Finster1999, Finster1999_2}. It was found that the time-periodic solutions in various black 
hole spacetimes do not exist \cite{Finster2000, Finster2000_2}. This implies 
that any Dirac particles, such as electrons, neutrinos etc.~cannot remain on a 
periodic orbit around a black hole. So, if such particles around a black hole 
or wormhole collapses gravitationally, they should vanish inside the event 
horizon of a black hole or they can escape to infinity. Hence a study of the 
Dirac fields in curved background spacetimes can provide interesting results. 
In this study, we shall study such Dirac field perturbations on the wormhole 
spacetimes considered in the work.

We have already discussed that quasinormal modes from wormholes have been 
studied widely in different literatures. Although wormholes are still considered
 to be hypothetical, one can see that GR exhibits both black holes and 
wormholes as promising solutions of the field equation. Since GR has been one 
of the most successful theories of gravity till now in predicting physically
realizable things, the possibility of existence of wormholes in the Universe 
can not be nullified. If that is so, tools and methods to probe wormholes and 
also differentiating them from black holes is an essential need for the 
scientific community. As previous studies show that wormholes also can emit 
quasinormal modes when a perturbation is introduced to its background, we 
believe that studies dealing with quasinormal modes from wormholes can play a 
significant role in differentiating black holes from wormholes in the near 
future when suitable observational data of quasinormal modes will be obtained.

The primary motivation of this work is to see how the structure of a wormhole 
behaves in presence of a cloud of strings in $f(R)$ gravity and possibility of 
experimental detection of the impacts of a cloud of strings in the wormhole 
background using quasinormal modes as a probe. The study will also focus on 
the variation of the quasinormal modes with the cloud of string parameter and 
the $f(R)$ gravity model parameters in order to understand how they can affect 
the ringdown GWs. The justification for using the $f(R)$ gravity Starobinsky 
model is its simplicity and versatility. Since the Starobinsky model has been 
very successful in explaining the inflationary epoch of the Universe and 
several other observational aspects, it has been the choice of many 
researchers to consider it in different directions of study. This 
investigation dealing with the Starobinsky model will put some light on the 
possible configuration of wormholes in this model as well as its behaviour and 
properties of quasinormal modes.

The rest of the structure of this paper is as follows. We have given a brief 
introduction of wormholes in $f(R)$ gravity in section \ref{section02}. Here 
we have studied the field equations and connection between shape function and $f(R)$ gravity model. We have obtained a wormhole solution in 
$f(R)$ gravity Starobinsky model in this section and then studied the energy conditions in brief for a general idea. In section \ref{section03}, we have studied 
quasinormal modes of the wormhole solution obtained in the Starobinsky model 
along with three other toy shape functions for the scalar perturbations and 
Dirac perturbations. The time domain analysis part has been included in section 
\ref{section04}. We have summarized the results of our work with a brief 
conclusion in section \ref{section05}.

\section{Wormholes in $f(R)$ gravity} \label{section02}
In this work, we shall use the metric formalism in which the action of the 
theory is varied with respect to the metric $g^{\mu\nu}$. The $f(R)$ gravity in 
metric formalism has been widely studied in the field of black holes and 
wormholes previously. Other formalisms frequently used in literature are the 
Palatini formalism \cite{Sotiriou2005} and the metric-affine formalism 
\cite{Sotiriou:2006qn}. In the Palatini formalism, the metric and the 
connections are considered as independent or separate variables and the 
matter action is independent of the connection, while in the metric-affine 
formalism the matter part of the action is also varied with respect to the 
connection. In our work, we shall restrict to the study of wormholes in 
$f(R)$ gravity metric formalism only.

The action in $f(R)$ gravity is given by
\begin{equation}\label{action}
S=\frac{1}{2\kappa}\int d^4x\sqrt{-g}\;f(R)+\int d^4x\sqrt{-g}\;{\cal
L}_m(g_{\mu\nu},\psi)\,,
\end{equation}
here $\kappa =8\pi G$ and from hereafter we shall consider $\kappa=1$. 
${\cal L}_m$ is the matter Lagrangian density, in which matter is minimally 
coupled to the metric $g_{\mu\nu}$ and $\psi$ denotes the matter fields. Now 
varying the action \eqref{action} with respect to the metric $g^{\mu\nu}$, we 
obtain the field equations in $f(R)$ gravity metric formalism as given by
\begin{equation}
\mathcal{F} R_{\mu\nu}-\frac{1}{2}f\,g_{\mu\nu}-\nabla_\mu \nabla_\nu
\mathcal{F}+g_{\mu\nu}\Box \mathcal{F}=\,T^{(m)}_{\mu\nu} \,,
    \label{field:eq}
\end{equation}
where $\mathcal{F}\equiv df(R)/dR$ and $T^{(m)}_{\mu\nu} = - \frac{2}{\sqrt{-g}}\frac{\delta\left(\sqrt{-g}\,{\cal L}_m\right)}{\delta g^{\mu\nu}}$ is the
stress-energy tensor of the matter. Taking trace of this Eq.~(\ref{field:eq}), 
we have
\begin{equation}
\mathcal{F} R-2f+3\,\Box \mathcal{F}=\,T .
 \label{trace}
\end{equation}
Using this trace equation we can rewrite the field Eq.~\eqref{field:eq} in the 
following possible form:
\begin{equation}
G_{\mu\nu}\equiv R_{\mu\nu}-\frac{1}{2}R\,g_{\mu\nu}= T^{{\rm
eff}}_{\mu\nu}\,.
    \label{field:eq2}
\end{equation}
Here the effective stress-energy tensor is $T^{{\rm
eff}}_{\mu\nu}= T^{(c)}_{\mu\nu}+\tilde{T}^{(m)}_{\mu\nu}$, where
$\tilde{T}^{(m)}_{\mu\nu}=T^{(m)}_{\mu\nu}/\mathcal{F}$ and the curvature
stress-energy tensor,
\begin{equation}
T^{(c)}_{\mu\nu}=\frac{1}{\mathcal{F}}\left[\nabla_\mu \nabla_\nu \mathcal{F}
-\frac{1}{4}g_{\mu\nu}\left(R\mathcal{F}+\Box \mathcal{F}+T\right) \right].
    \label{gravfluid}
\end{equation}
The associated conservation law with the modified field equation can be given
by
\begin{equation}\label{conserv-law}
\nabla^\mu T^{(c)}_{\mu\nu}=\frac{1}{\mathcal{F}^2}
T^{(m)}_{\mu\nu}\nabla^\mu \mathcal{F}.
\end{equation}

As mentioned earlier, since the possibility that wormholes are supported by 
$f(R)$ theories of gravity, here our intention is to explore the distinguishing 
characteristics of a few specific forms of wormholes in this area of MGTs. One 
should note that it is the effective stress energy of $f(R)$ gravity, which 
may be interpreted as a gravitational fluid, is responsible for the null 
energy condition violation. This leads to the non-standard wormhole geometries, 
fundamentally different from their counterparts in GR \cite{Lobo2009}. However, we 
demand that the matter threading the wormhole satisfies the energy 
conditions. For the purpose of our study we consider the following ansatz,
\begin{equation}
ds^2=-e^{2\Phi(r)}dt^2+\frac{dr^2}{1-b(r)/r}+r^2\,(d\theta^2 +\sin
^2{\theta} \, d\phi ^2) \,,
    \label{metric}
\end{equation}
where $\Phi(r)$ and $b(r)$ are arbitrary functions of the radial coordinate 
$r$, referred to as the lapse function and the shape function respectively 
\cite{Morris:1988cz}. This ansatz represents a static and spherically 
symmetric wormhole in spacetime. The lapse function determines the 
red-shift effect and tidal force associated with the wormhole spacetime. If 
the lapse function $\Phi(r) = 0$ or $e^{2\Phi(r)} = 1$, the wormhole is said 
to be tideless \cite{Konoplya2018}.

It is to be noted here is that the radial coordinate $r$ is non-monotonic 
which decreases from infinity to minimum value $r=r_0$ at the throat of the 
wormhole, defined by $b(r_0)=r_0$, and it then again increases to infinity. 
Hence, the shape function $b(r)$ has the minimum value $r_0$ at the throat of 
the wormhole. In general to have a wormhole solution certain conditions should 
be satisfied including the flaring out condition of the throat, given by
$(b-b^{\prime}r)/b^{2}>0$ \cite{Morris:1988cz} at the throat $b(r_0)=r_0$ and 
the condition $b^{\prime}(r_{0})<1$. These conditions impose the NEC violation 
in the classical GR. Another condition for a stable wormhole is $1-b(r)/r>0$. 
Most importantly, for the wormhole to be traversable, there should be no 
horizons present, which are defined on the spacetime by 
$e^{2\Phi}\rightarrow0$, so that $\Phi(r)$ must be finite everywhere. In view 
of this, we consider a constant redshift function i.e.\ $\Phi'=0$. This 
simplifies the calculations associated with the field equations and provide 
interesting wormhole solutions.

The unusual structure of wormhole suggests that the distribution of matter
threading the wormhole is anisotropic and the stress-energy tensor for such 
distribution of matter is given by
\begin{equation}
T^{(m)}_{\mu\nu}=(\rho+p_t)U_\mu \, U_\nu+p_t\,
g_{\mu\nu}+(p_r-p_t)\chi_\mu \chi_\nu \,,
\end{equation}
where $U^\mu$ represents the four-velocity of matter field, 
$\chi^\mu=\sqrt{1-b(r)/r}\,\delta^\mu{}_r$ represents the unit spacelike 
vector along the radial direction, $\rho(r)$ represents the
energy density, $p_r(r)$ represents the radial pressure along the direction
of $\chi^\mu$, and $p_t(r)$ represents the transverse pressure along the
direction orthogonal to $\chi^\mu$. Using this stress-energy tensor of matter, 
we can write the field equations \eqref{field:eq2} in the following forms: 

\begin{align}
\frac{b'}{r^2}&=\frac{\rho}{\mathcal{F}}+\frac{\left(\mathcal{F}R+\Box \mathcal{F} +T\right)}{4\mathcal{F}}
  \,,    \label{fieldtt}
     \\
-\frac{b}{r^3}&=\frac{p_r}{\mathcal{F}}+\frac{1}{\mathcal{F}}\Bigg\{\left(1-\frac{b}{r}\right)\left[\mathcal{F}'' -\mathcal{F}'\frac{b'r-b}{2r^2(1-b/r)}\right] -\frac{1}{4}\left(\mathcal{F}R+\Box \mathcal{F} +T\right)\Bigg\}
  \,,  \label{fieldrr} \\
-\frac{b'r-b}{2r^3}
     &=\frac{p_t}{\mathcal{F}}+\frac{1}{\mathcal{F}}\left[\left(1-\frac{b}{r}\right)
     \frac{\mathcal{F}'}{r}
     -\frac{1}{4}\left(\mathcal{F}R+\Box \mathcal{F} +T\right)\right]
     \label{fieldthetatheta},
\end{align}
where a prime denotes a derivative with respect to the radial coordinate $r$. 
In the above equations, $\Box \mathcal{F}$ is given by
\begin{equation}
\Box \mathcal{F}=\left(1-\frac{b}{r}\right)\left[\mathcal{F}''
-\frac{b'r-b}{2r^2(1-b/r)}\,\mathcal{F}'+\frac{2\mathcal{F}'}{r}\right]
\end{equation}
and the Ricci curvature scalar $R$ is given by
\begin{eqnarray}
R&=& \frac{2b'}{r^2}
    \,.
    \label{Ricciscalar}
\end{eqnarray}
Rearranging the field equations \eqref{fieldtt}, \eqref{fieldrr} and 
\eqref{fieldthetatheta} we may obtain the following expressions for $\rho$, 
$p_r$ and $p_t$ \cite{Lobo2009} as
\begin{align}
\label{fieldeq01} \rho&=\frac{\mathcal{F}b'}{r^2}\,,
       \\
\label{generic2}
p_r&=-\frac{b\mathcal{F}}{r^3}+\frac{\mathcal{F}'}{2r^2}(b'r-b)-\mathcal{F}''\left(1-\frac{b}{r}\right),   \\
\label{generic3}
p_t&=-\frac{\mathcal{F}'}{r}\left(1-\frac{b}{r}\right)+\frac{\mathcal{F}}{2r^3}(b-b'r).
\end{align}
These are the generic forms of expressions of the energy density and pressures
of the matter threading the wormhole in $f(R)$ gravity metric formalism as a 
function of the wormhole shape function and $\mathcal{F}(r)$.

\subsection{Toy Models of Wormhole Shape Function}
\begin{figure}[t!]
\centerline{
   \includegraphics[scale = 0.32]{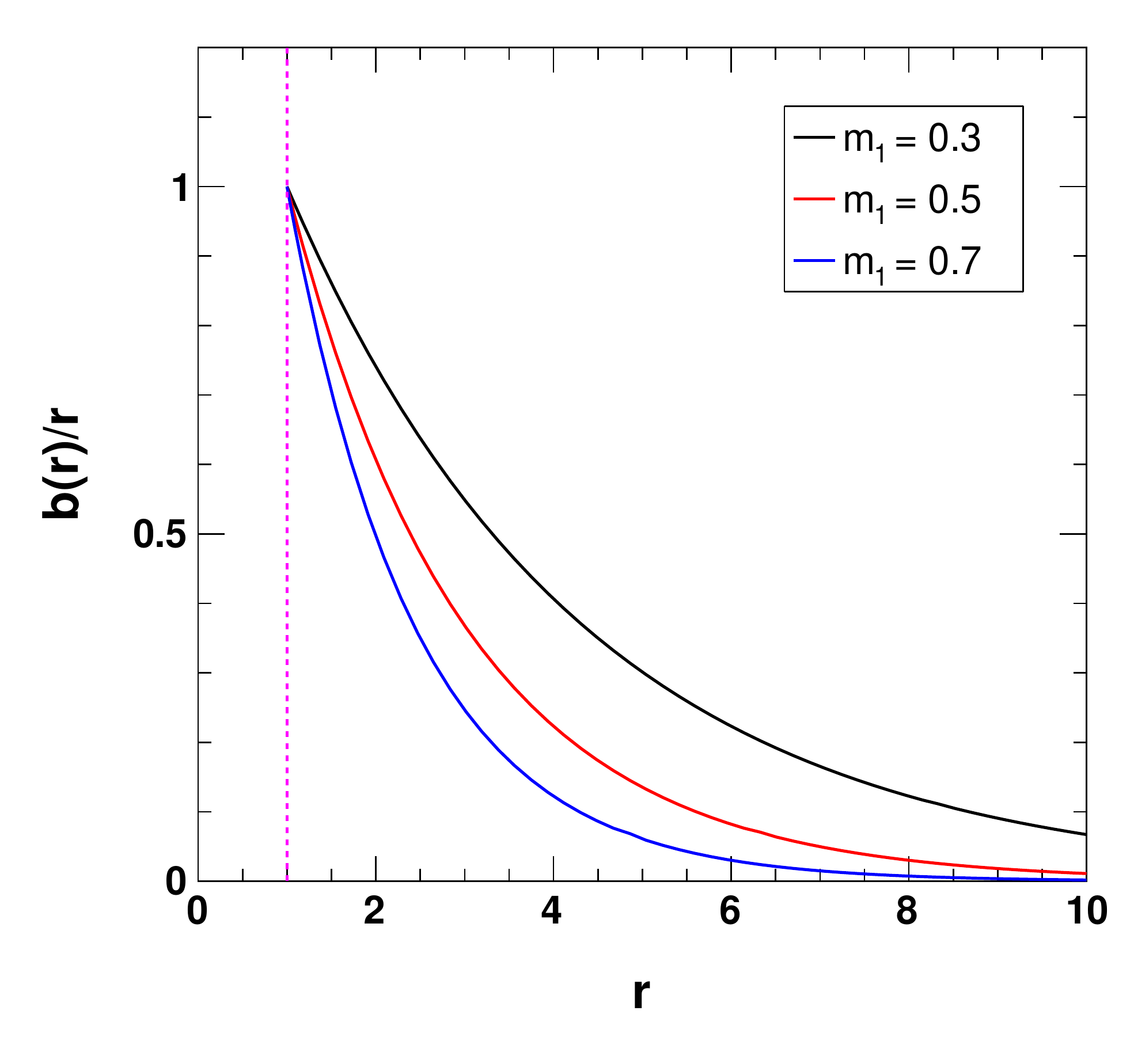}\hspace{1cm}
   \includegraphics[scale = 0.32]{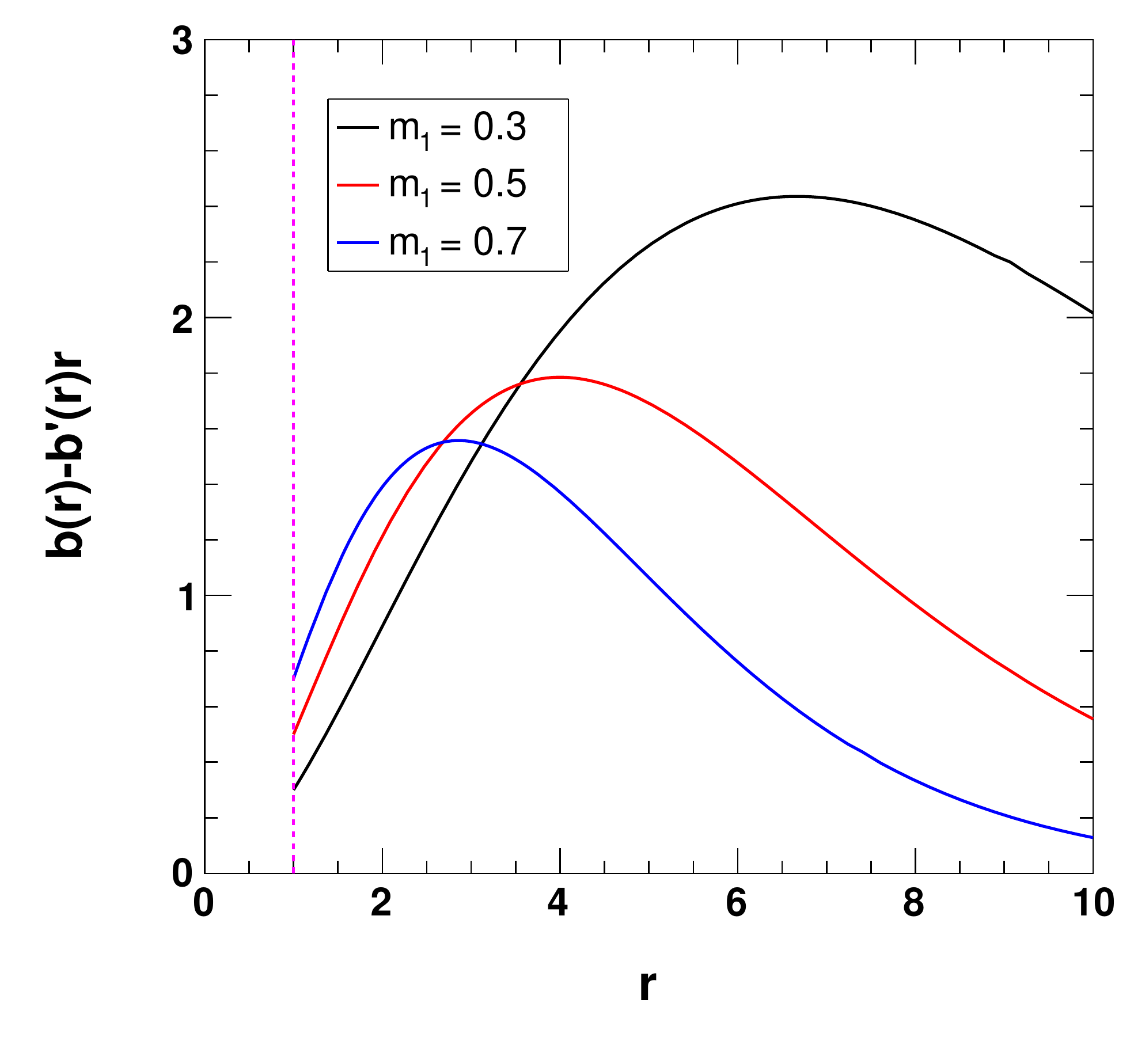}}
\vspace{-0.2cm}
\caption{Plots of $b(r)/r$ vs.~$r$ (on the left panel) and $b(r) - b'(r) r$
vs.~$r$ (on the right panel) for the wormhole shape function
\eqref{toy_shape01} with different values of the parameter $m_1$ and throat
radius $r_0=1$. The vertical dotted line in each plot represents the position
of the throat of the wormhole.}
\label{shape01}
\end{figure}

\begin{figure}[t!]
\centerline{
   \includegraphics[scale = 0.32]{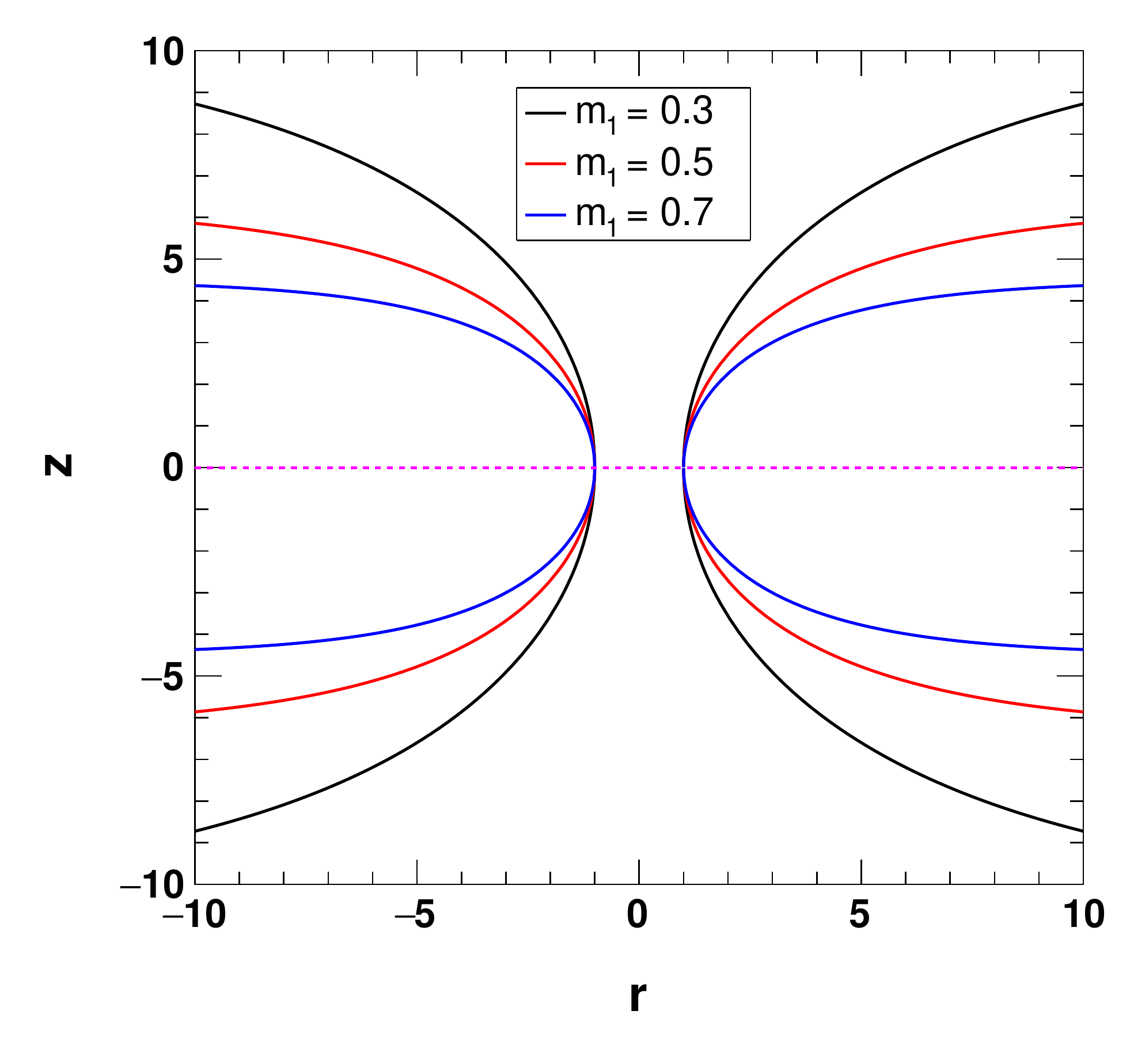}\hspace{1cm}
   \includegraphics[scale = 0.35]{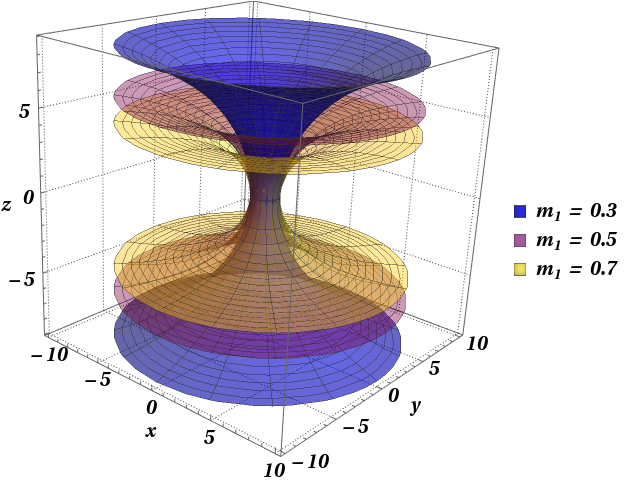}}
\vspace{-0.2cm}
\caption{Embedded 2-D and 3-D plots of the wormhole defined by the shape
function \eqref{toy_shape01} with different values of the parameter $m_1$ and
throat radius $r_0=1$.}
\label{shape02}
\end{figure}

\begin{figure}[h!]
\centerline{
   \includegraphics[scale = 0.32]{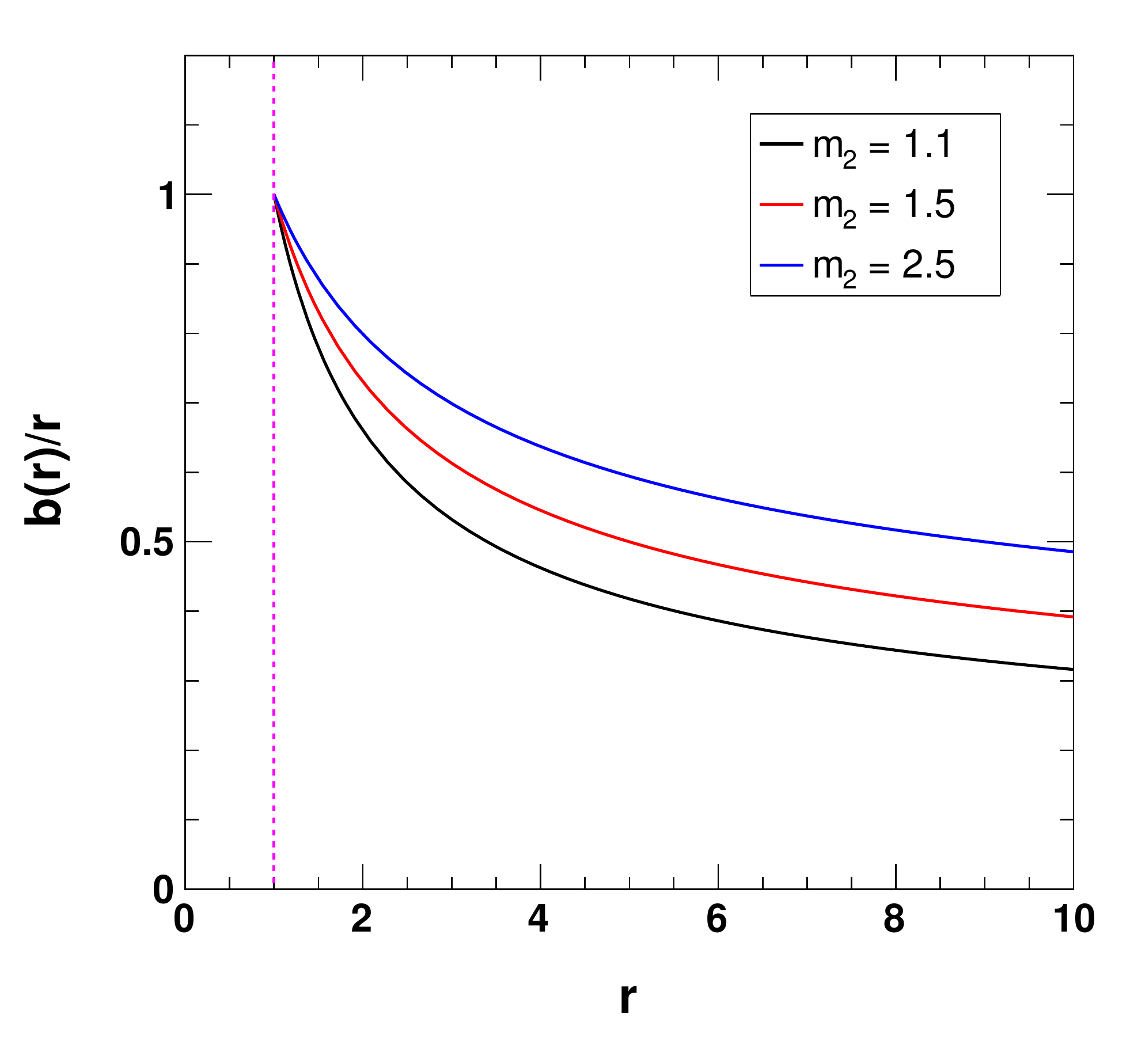}\hspace{1cm}
   \includegraphics[scale = 0.32]{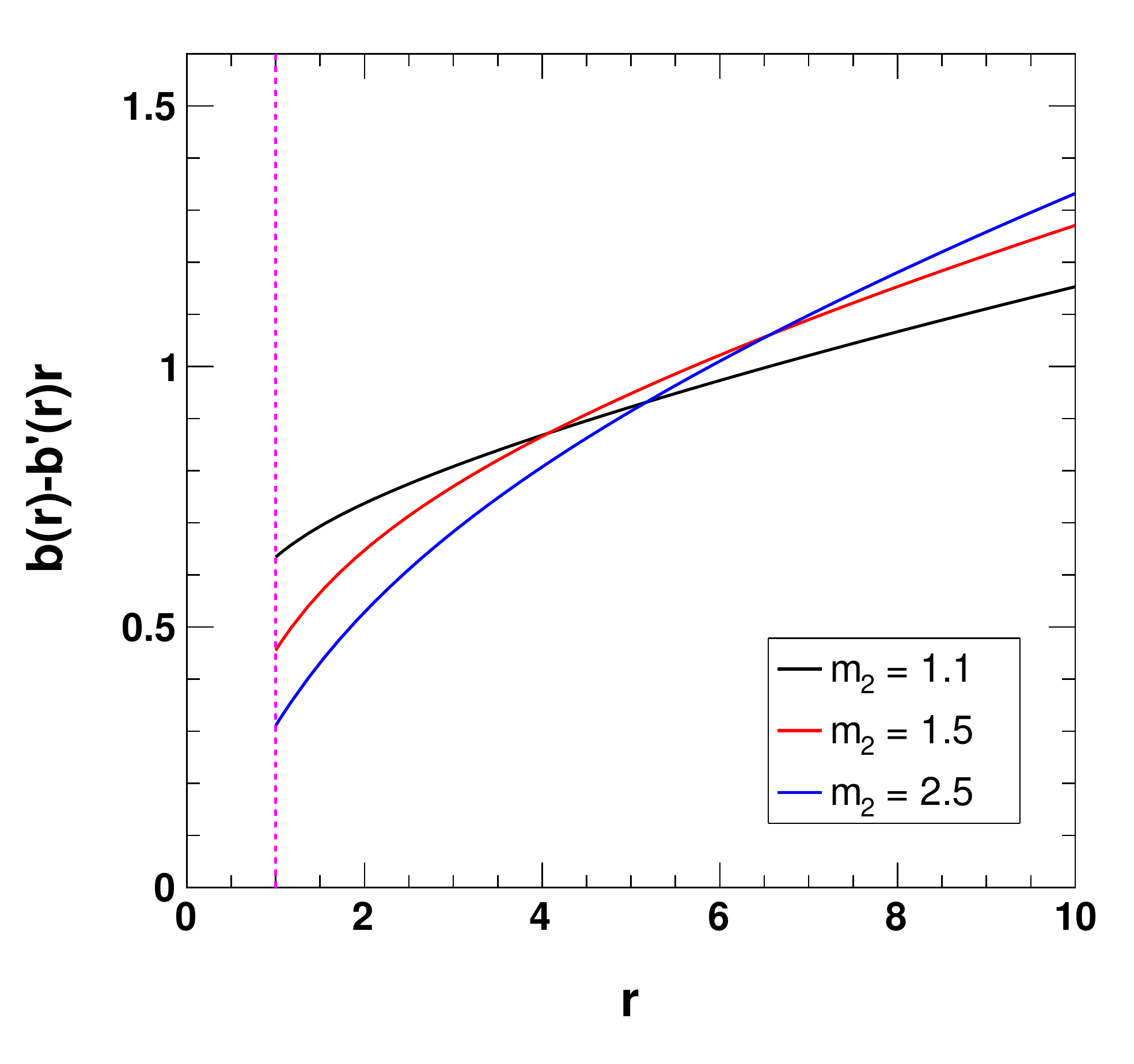}}
\vspace{-0.2cm}
\caption{Plots of $b(r)/r$ vs.~$r$ (on the left panel) and $b(r) - b'(r) r$
vs.~$r$ (on the right panel) for the wormhole shape function
\eqref{toy_shape02} with different values of the parameter $m_2$ and throat
radius $r_0=1$. The vertical dotted line in each plot represents the position
of the throat of the wormhole.}
\label{shape03}
\end{figure}

\begin{figure}[h!]
\centerline{
   \includegraphics[scale = 0.32]{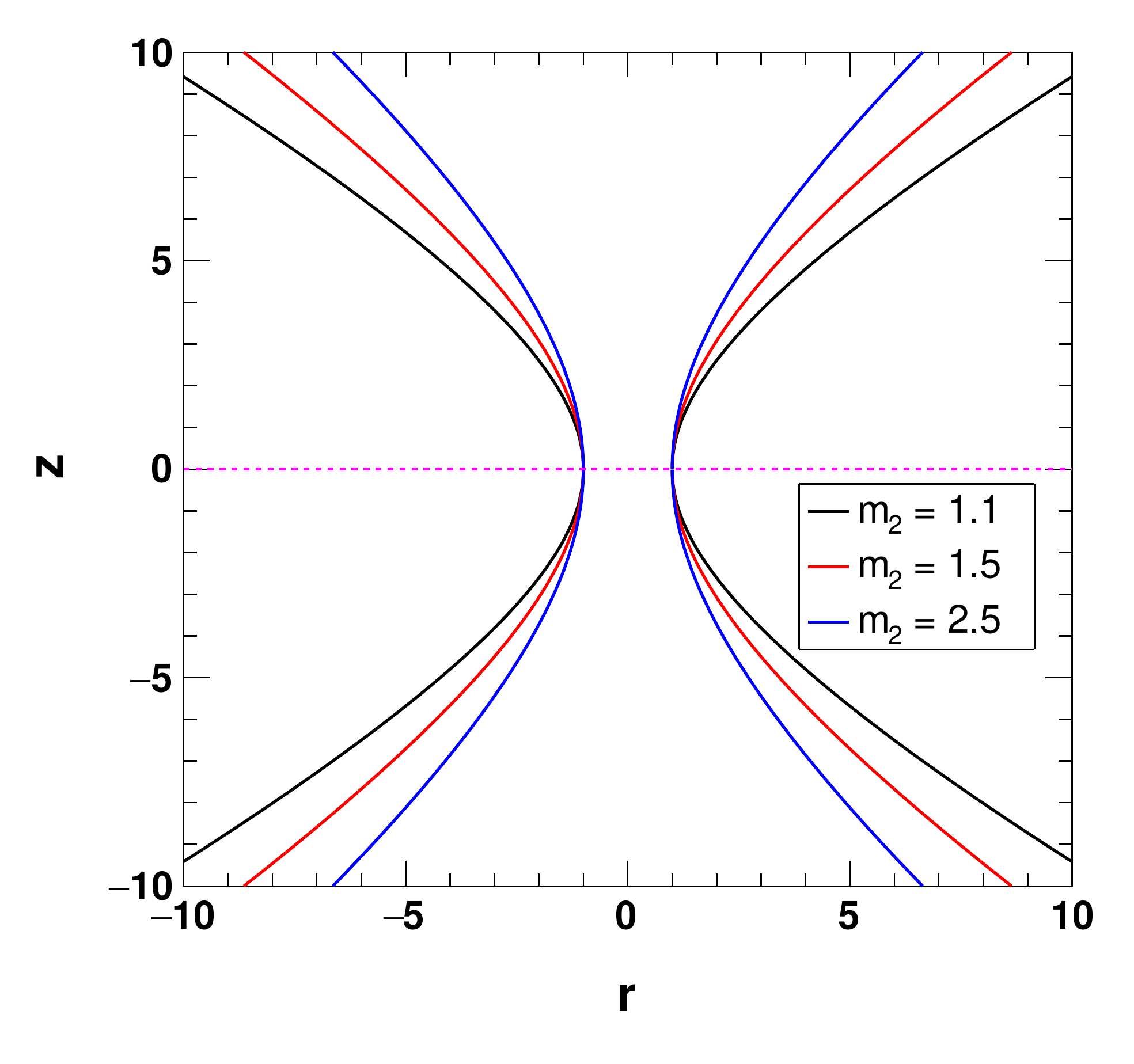}\hspace{1cm}
   \includegraphics[scale = 0.35]{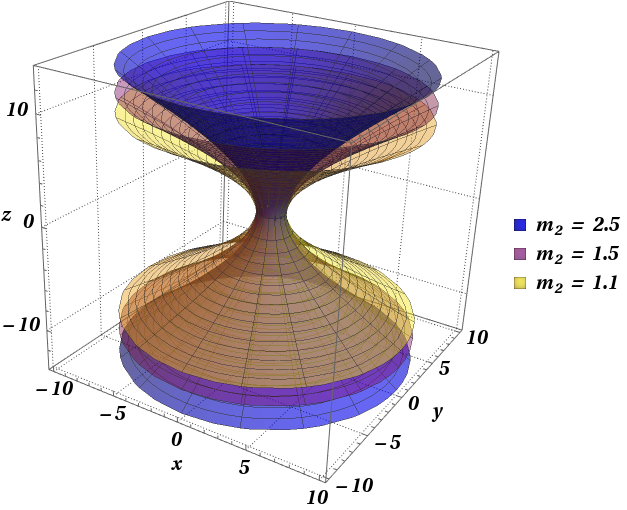}}
\vspace{-0.2cm}
\caption{Embedded 2-D and 3-D plots of the wormhole defined by
\eqref{toy_shape02} with different values of the parameter $m_2$ and throat
radius $r_0=1$.}
\label{shape04}
\end{figure}

\begin{figure}[h!]
\centerline{
   \includegraphics[scale = 0.32]{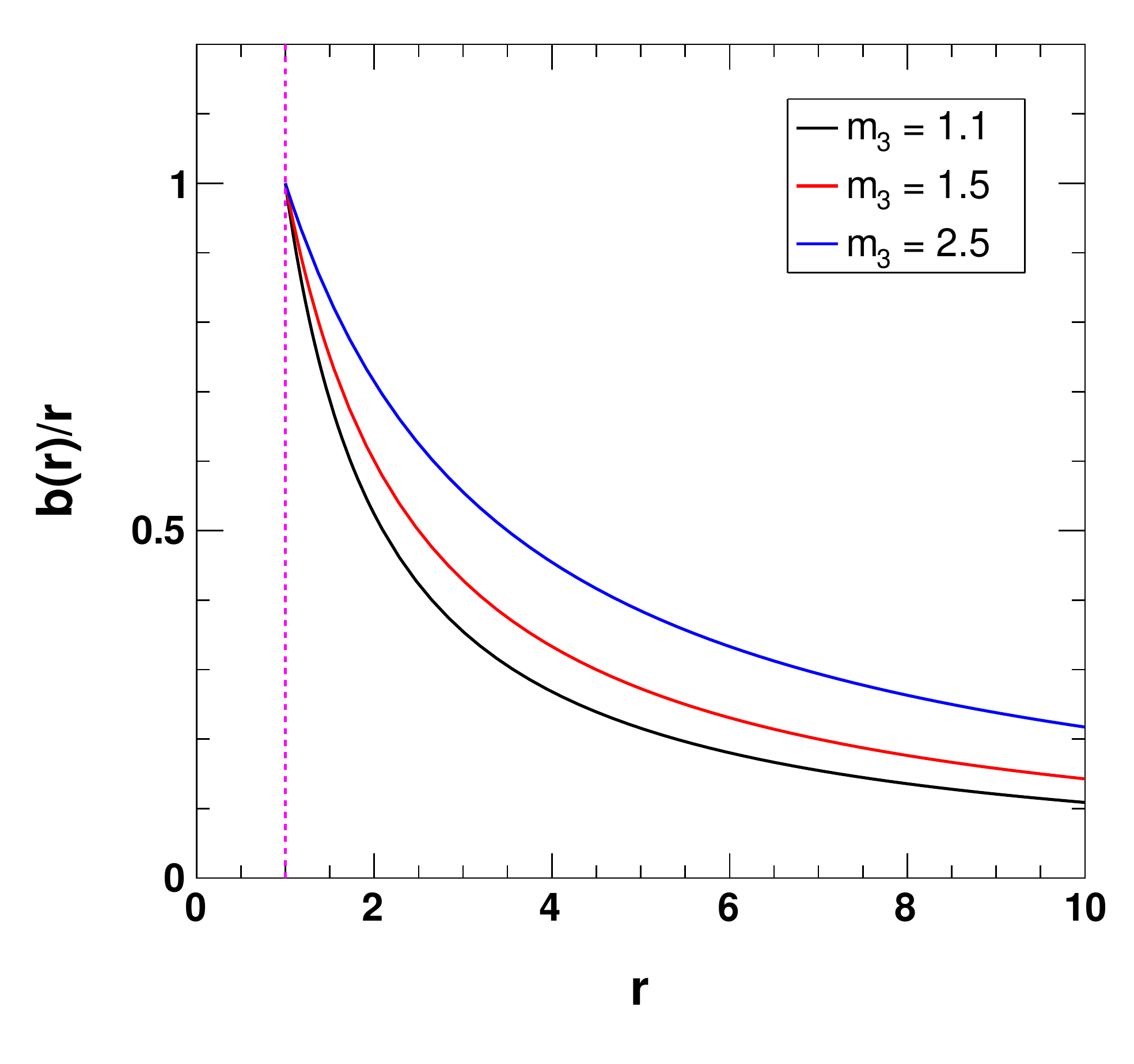}\hspace{1cm}
   \includegraphics[scale = 0.32]{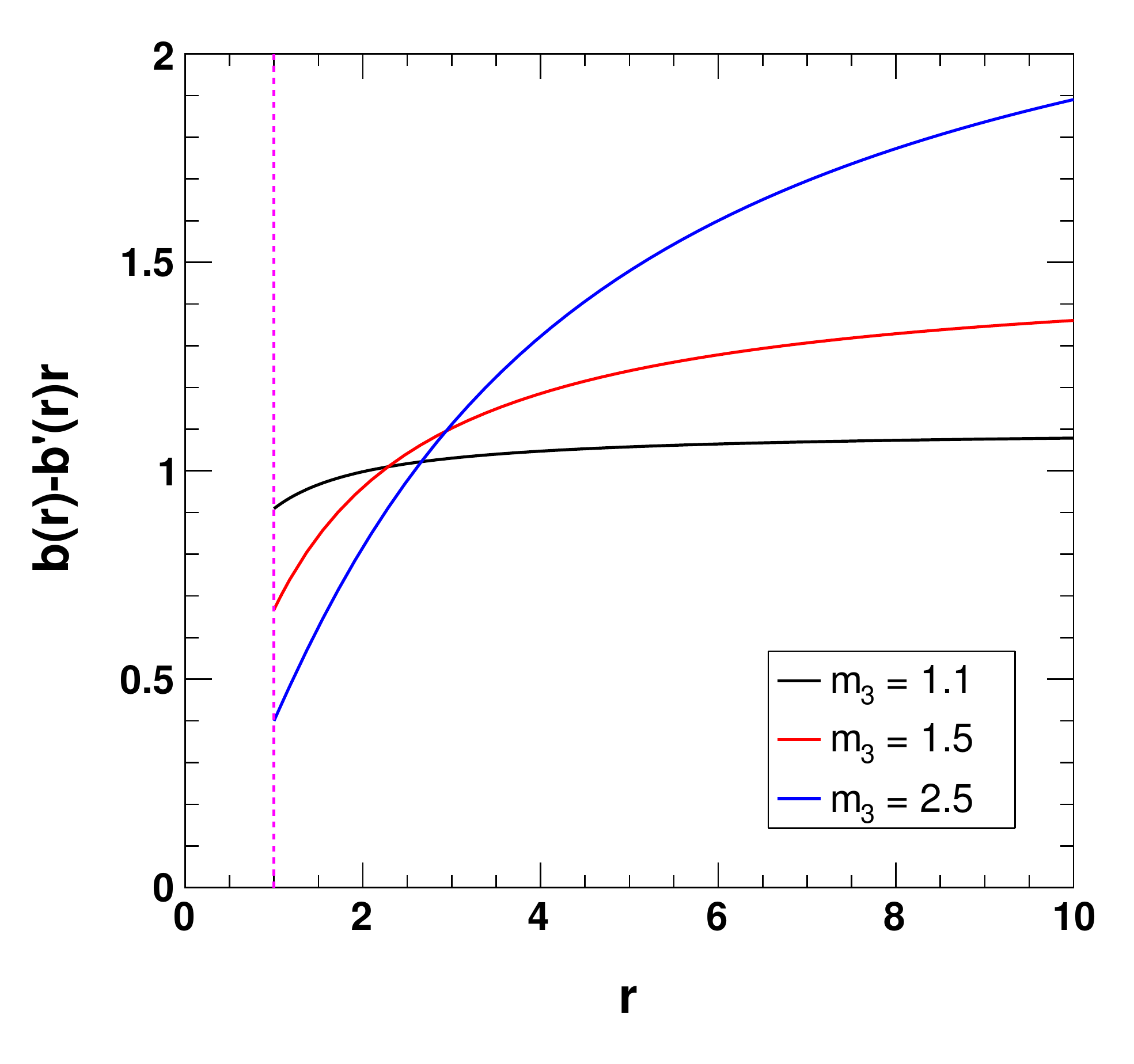}}
\vspace{-0.2cm}
\caption{Plots of $b(r)/r$ vs.~$r$ (on the left panel) and $b(r) - b'(r) r$
vs.~$r$ (on the right panel) for the wormhole shape function
\eqref{toy_shape03} with different values of the parameter $m_3$ and throat
radius $r_0=1$. The vertical dotted line in each plot represents the position
of the throat of the wormhole.}
\label{shape05}
\end{figure}

\begin{figure}[h!]
\centerline{
   \includegraphics[scale = 0.32]{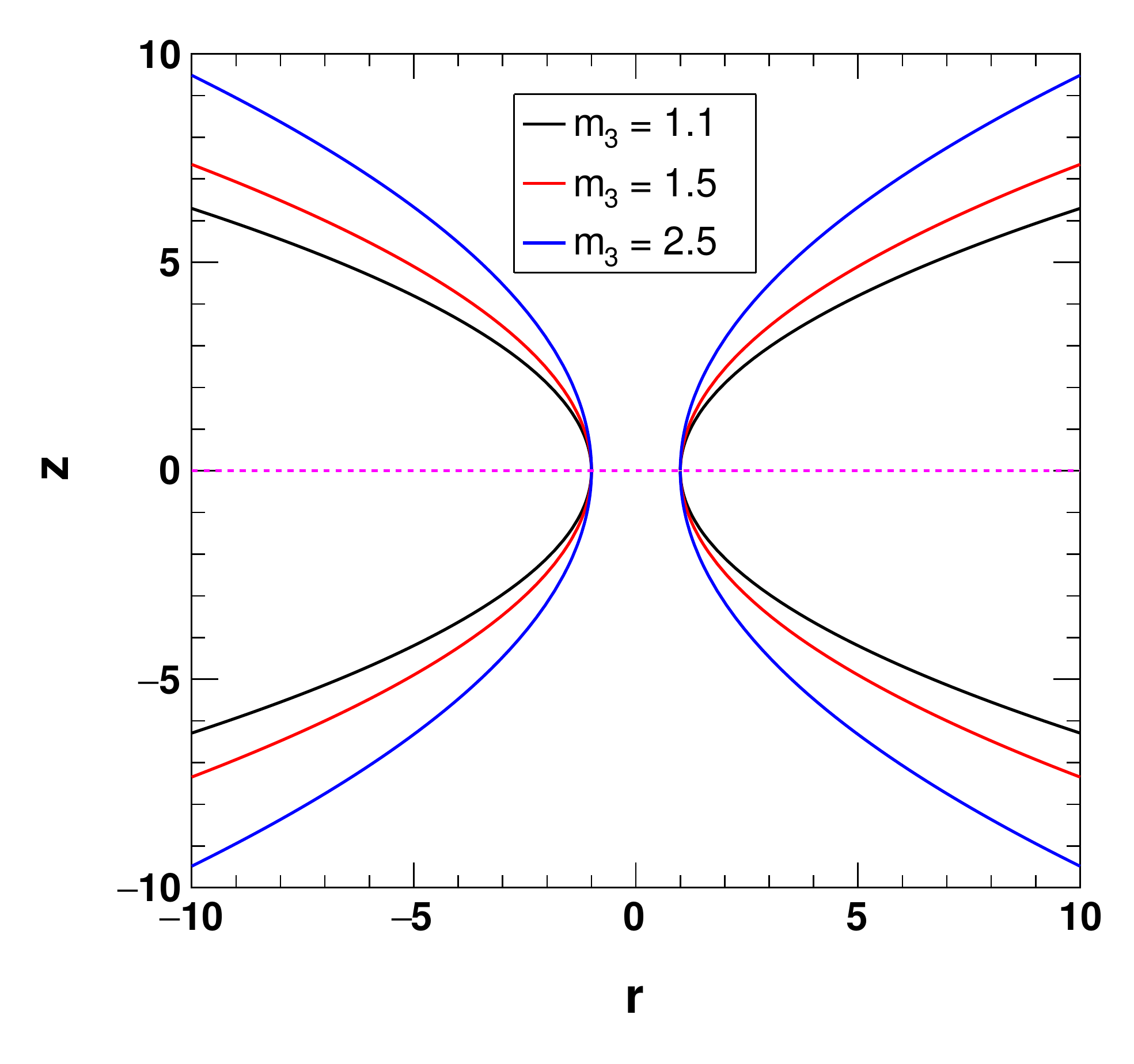}\hspace{1cm}
   \includegraphics[scale = 0.38]{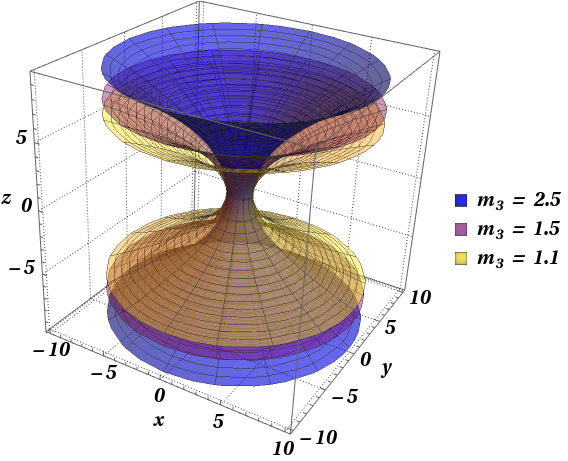}}
\vspace{-0.2cm}
\caption{Embedded 2-D and 3-D plots of the wormhole defined by
\eqref{toy_shape03} with different values of the parameter $m_3$ and throat
radius $r_0=1$.}
\label{shape06}
\end{figure}

As the shape function of a wormhole is the deciding factor for its particular
construction or structure, it is necessary to study the behaviours of this
function of a wormhole under certain required conditions. It is to be noted 
that in order to have a consistent wormhole structure the shape function 
should satisfy the following conditions or properties: $(i)\; b(r)/r < 1$ for 
$r>r_0,$ $(ii) \; b(r) = r_0$ at $r=r_0,$ $(iii) \; b(r)/r \rightarrow 0 $ as 
$r \rightarrow \infty,$ $(iv) \; b(r) - b'(r) r > 0$ and $(v) \; b'(r) < 1$ 
at $r=r_0.$ In this study, we consider three toy wormhole shape functions as 
given by
\begin{equation}\label{toy_shape01}
b_1(r) = r \exp\left[{m_1 \left(r_0-r\right)}\right],
\end{equation}
\begin{equation}\label{toy_shape02}
b_2(r) = \frac{r \log \left(2 m_2 \ r_0\right)}{\log \left[m_2 \left(r+r_0\right)\right]},
\end{equation}
and
\begin{equation}\label{toy_shape03}
b_3(r) = \frac{m_3\ r}{m_3+r-r_0}.
\end{equation}
Here $m_1, m_2$ and $m_3$ are the model parameters and $r_0$ denotes the 
throat radius of the wormholes. The first toy model has been widely used in 
different works \cite{exp_shape01, exp_shape02, exp_shape03}. Whereas other 
two toy functions have been introduced by us as two possible shapes or 
structures of wormhole. First, to check the viabilities of these toy models, 
we check the viability 
conditions mentioned above for them. For this purpose, we have plotted the 
functions $b(r)/r $ and $b(r) - b'(r) r$ with respect to $r$ for these 
functions and the corresponding embedded diagrams of the wormholes in 
Fig.s \ref{shape01}, \ref{shape02}, \ref{shape03}, \ref{shape04}, 
\ref{shape05}  and \ref{shape06} respectively. From Fig.~\ref{shape01}, we can 
see that the first toy model can effectively satisfy the conditions for a 
viable wormhole. The right panel of this figure shows that the function has a 
peak which moves toward the throat of the wormhole with an increase in the 
value of the model parameter $m_1$. Although the function decreases gradually 
for higher values of $r$, it remains positive satisfying the condition 
$b(r) - b'(r) r>0$. However, the behaviours of this test function 
$b(r) - b'(r) r$ are totally different 
for the second and third toy shape functions respectively. In Fig.s 
\ref{shape03} and \ref{shape05}, we have plotted the test functions for the 
second and the third toy models or shape functions respectively. Here, one can 
see that the test functions show suitable behaviours for $m_2>1$ and $m_3> 1$ 
respectively. In both cases, the parameters $m_2$ and $m_3$ impose similar 
signatures. On the left panel of Fig.~\ref{shape03}, we can see that with an 
increase in the model parameter $m_2$, the function $b(r)/r$ increases at 
$r> r_0$. On the right panel, near the throat, with an increase in $m_2$, the 
test function $b(r) - b'(r) r$ decreases initially, but at a significantly far 
distance away from the throat, the opposite trend comes into picture. For the 
third shape function also, we observe a similar behaviour.

Next, in order to visualise the embedded diagrams of the wormholes represented
by toy shape functions \eqref{toy_shape01}, \eqref{toy_shape02} and 
\eqref{toy_shape03}, we use an equatorial slice $\theta = \pi/2$ at a 
fixed time or at $t=$ constant. This gives us the privilege to reduce the 
metric \eqref{metric} into the following form:
\begin{equation}\label{reduced_metric}
ds^2 = \frac{dr^2}{1-b(r)/r} + r^2 d\phi^2.
\end{equation}
In cylindrical coordinates, we can write the above equation as
\begin{equation}\label{cylindrical_metric}
ds^2 = dz^2 + dr^2 + r^2 d\phi^2.
\end{equation}
As the embedded surface in three dimensional Euclidean space is expressed by 
$z = z(r)$, we can rewrite Eq.~\eqref{cylindrical_metric} as
\begin{equation}\label{recasted_metric}
ds^2 = \left[1 + \left(dz/dr \right)^2 \right] dr^2 + r^2 d\phi^2.
\end{equation}
Finally, comparing the recasted metric \eqref{recasted_metric} with the reduced metric \eqref{reduced_metric}, we get,
\begin{equation}\label{embedded_eq}
\dfrac{dz}{dr} = \pm \left[ \dfrac{r}{b(r)} - 1 \right]^{-1/2}.
\end{equation}
This relation gives the embedded surface of the wormhole. From the flare-out 
condition, we can see that the inverse of the embedding function $r(z)$ 
satisfies $d^2r/dz^2 > 0$ near or at the throat of the wormhole. More 
explicitly, differentiating inverse of the Eq.~\eqref{embedded_eq} with 
respect to $z$, we get,
\begin{equation}
\dfrac{d^2r}{dz^2} = \dfrac{b(r) - r b'(r)}{2 b(r)^2} > 0.
\end{equation} 
Apart from this, from Eq.~\eqref{embedded_eq} one can see that 
$dz/dr \rightarrow \infty $ at the wormhole throat and the wormhole space is 
asymptotically flat as $r \rightarrow \infty$. Using Eq.~\eqref{embedded_eq} 
we have plotted the embedded diagrams of the wormholes in Fig.s \ref{shape02}, 
\ref{shape04} and \ref{shape06} for the toy shape functions 
\eqref{toy_shape01}, \eqref{toy_shape02} and \eqref{toy_shape03} respectively. 
These embedded diagrams show the impact of the model parameters on the shape 
of the wormhole. For the shape function \eqref{toy_shape01} it is seen that 
with an increase in the model parameter $m_1$, height of the wormhole gets 
reduced. On the other hand, for the shape functions \eqref{toy_shape02} and 
\eqref{toy_shape03}, it is clear from the diagrams \ref{shape04} and 
\ref{shape06} respectively that the impact of the model parameters $m_2$ and 
$m_3$ on the respective wormholes is similar in nature. In both cases, with 
an increase in the model parameters, the height of the wormholes increases 
gradually.

\subsection{$f(R)$ gravity model from toy shape function of wormhole 
surrounded by a cloud of strings}
Under some suitable conditions, it is possible to obtain the $f(R)$ gravity 
model for a wormhole shape function satisfying the field equations. Here, we 
shall consider the third toy shape function \eqref{toy_shape03} as an example 
to obtain the corresponding $f(R)$ gravity model in presence of a cloud of 
strings. A cloud of strings is a kind of fluid model in which one-dimensional 
strings are distributed in a given direction, which may exist in different 
geometrical shapes, such as spherical, axisymmetric and planar 
\cite{Graca2018}. In our work we shall consider spherically symmetric cloud of 
strings distribution. The solution for a spherically symmetric strings cloud 
configuration was obtained in Ref.~\cite{Graca2018}. The only non-null 
components of the energy-momentum tensor of a cloud of strings can be shown as
\begin{equation}
T^t_t = T^r_r = -\, \dfrac{\eta^2}{r^2},
\end{equation}
where $\eta$ is a constant, which is related to the energy of the cloud of 
strings. For the shape function \eqref{toy_shape03}, the field 
Eq.~\eqref{fieldeq01} can be written as
\begin{equation}
\frac{(m_3^2 - m_3\, r_0 ) \mathcal{F}(r)}{r^2 \left(m_3+r-r_0\right){}^2}+\frac{\eta ^2}{r^2}=0.
\end{equation}
Now, from this equation one can obtain an expression for $\mathcal{F}(r)$ 
given by 
\begin{equation}\label{eq27}
\mathcal{F}(r)=-\frac{\eta ^2 \left(m_3+r-r_0\right)^2}{m_3 \left(m_3-r_0\right)}.
\end{equation}
In this expression, $r$ can be replaced by the Ricci curvature using the 
definition \eqref{Ricciscalar} of Ricci curvature as
\begin{equation}
R=\frac{2 m_3 \left(m_3-r_0\right)}{r^2 \left(m_3+r-r_0\right){}^2}.
\end{equation}
Solving this equation for $r$ one can find,
\begin{equation}
r=\frac{1}{2} \left(\mp\, \sqrt{\left(m_3-r_0\right)^2 \mp 4 \sqrt{2\,m_3 \left(m_3-r_0\right)/R}}-m_3+r_0\right).
\end{equation}
For the mathematical feasibility we choose the solution given by
\begin{equation}\label{eq30}
r=\frac{1}{2} \left(\sqrt{4\sqrt{2\,m_3 \left(m_3-r_0\right)/R}+\left(m_3-r_0\right){}^2}-m_3+r_0\right).
\end{equation}
One can obtain the throat radius of the wormhole in terms of background Ricci 
curvature using the limit $r\rightarrow r_0$ and $R \rightarrow R_0$ in the 
above Eq.~\eqref{eq30}. Using that relation for the throat radius, we have the 
final relation for $r$ as
\begin{equation}
r = \frac{A}{2\,m_3R_0}\left[\left\{4\sqrt{2}\,m_3^2R^{-1/2}\left(R_0/A\right)^{3/2}+ 1 \right\}^{1/2}-1\right],
\end{equation}
where $A = m_3^2 R_0-\sqrt{2\, m_3^2 R_0+1}+1$. Hence, Eq.~\eqref{eq27} 
can be written as
\begin{equation}
\mathcal{F}(R) = -\,\frac{\eta^2A\left[1 + \sqrt{1+4\sqrt{2}\,m_3^2R^{-1/2}\left(R_0/A\right)^{3/2}}\right]^2}{4\,m_3^2R_0}. 
\end{equation}
Integrating above expression with respect to $R$, we obtain the $f(R)$ 
function for the wormhole with the shape function \eqref{toy_shape03} as
\begin{align}\label{reverse_model}
f(R) &= - (2 A^2 m_3^2 R_0)^{-1} \eta ^2 \Bigg[\;\; \sqrt{A^5 R \left(4 \sqrt{2} m_3^2 \sqrt{\frac{R R_0^3}{A}}+A R\right)}+4 \sqrt{2} m_3^2 \sqrt{A^3 R R_0^3} \\ \notag &-8 m_3^4 R_0^3 \log \left(\sqrt{A^3 R R_0}+A \sqrt{R_0 \left(4 \sqrt{2} m_3^2 R_0 \sqrt{\frac{R R_0}{A}}+A R\right)}+2 \sqrt{2} m_3^2 R_0^2\right)+A^3 R \\ \notag &+2 A m_3^3 R_0^2 \sqrt{\frac{2 A R}{m_3^2 R_0}+8 \sqrt{2} \sqrt{\frac{R R_0}{A}}}\;\;\Bigg] + C_1.
\end{align}
Here $C_1$ is an integration constant. In this function the background 
curvature is connected with the throat of the wormhole by the following 
relation:
\begin{equation}
R_0=\frac{2 \left(m_3-r_0\right)}{m_3 r_0^2}.
\end{equation}
Obviously, the $f(R)$ function or model in Eq.~\eqref{reverse_model} depends 
on the shape function of the wormhole. One may note that the cloud of strings 
parameter $\eta$ appears in the $f(R)$ model explicitly because of the 
presence of a cloud of strings in the wormhole spacetime as clear from our 
consideration above. Whereas the fact is that the shape function of the 
wormhole is considered to be independent of the cloud of strings parameter. In 
such a situation, the cloud of strings parameter contributes to the geometry 
modification only as shown in Eq.~\eqref{reverse_model} and the basic 
properties of the wormhole remain independent of it. This type of situation 
arises when we define the shape function at first. So, for the toy shape 
functions of the wormholes, we shall not see any cloud of strings dependency 
with the quasinormal modes in general. Hence, such type of ad-hoc wormhole 
definitions may not be very feasible to study the impacts of any surrounding 
relics on the quasinormal modes from the wormhole spacetime. Any additional 
relic such as cloud of strings etc.~will appear as a spacetime modification 
imprinted in the $f(R)$ gravity model due to fixing the shape function 
initially.

\subsection{Wormhole solution in $f(R)$ gravity Starobinsky model surrounded by a cloud of strings}
In this section, we shall obtain the wormhole solutions in $f(R)$ gravity 
Starobinsky model surrounded by a cloud of strings. Here the approach will be opposite to the previous case where we fixed the shape function at first and then calculated the corresponding $f(R)$ gravity model. This approach will help us obtain a wormhole shape function which has a dependency with the surrounding relic if any. In this work we shall use the Starobinsky's inflationary model 
\cite{Starobinsky1980}, which is given by
\begin{equation}\label{model}
 f(R) = R+ \alpha R^2, 
\end{equation}
where $\alpha$ is the constant model parameter. Using this model \eqref{model} 
in Eq.~\eqref{fieldeq01}, we obtain,
\begin{equation}
4 \alpha  b'(r)^2+r^2 b'(r)+r^2\eta ^2=0.
\end{equation}
This equation can be solved for $b'(r)$ as
\begin{equation}
b'(r) = \frac{-\, r^2 \pm r\, \sqrt{r^2-16 \alpha  \eta ^2}}{8 \alpha }.
\end{equation}
For a feasible situation, we pick only the solution
\begin{equation}
b'(r) = \frac{-\, r^2 + r\, \sqrt{r^2-16 \alpha  \eta ^2}}{8 \alpha }
\end{equation}
and the solution for $b(r)$ with the boundary condition $b(r_0)=r_0$ yields,
\begin{equation}\label{staro_shape}
b(r) = \frac{-\,r^3+\left(r^2-16 \alpha  \eta ^2\right)^{3/2}-\left(r_0^2-16 \alpha  \eta ^2\right){}^{3/2}+24 \alpha  r_0+r_0^3}{24 \alpha }.
\end{equation}
One may note that in this shape function of the wormhole, the cloud of strings parameter and $f(R)$ gravity model parameter appear explicitly and any modification in these parameters will affect the geometry of the wormhole and the corresponding quasinormal modes.
\begin{figure}[htbp]
\centerline{
   \includegraphics[scale = 0.3]{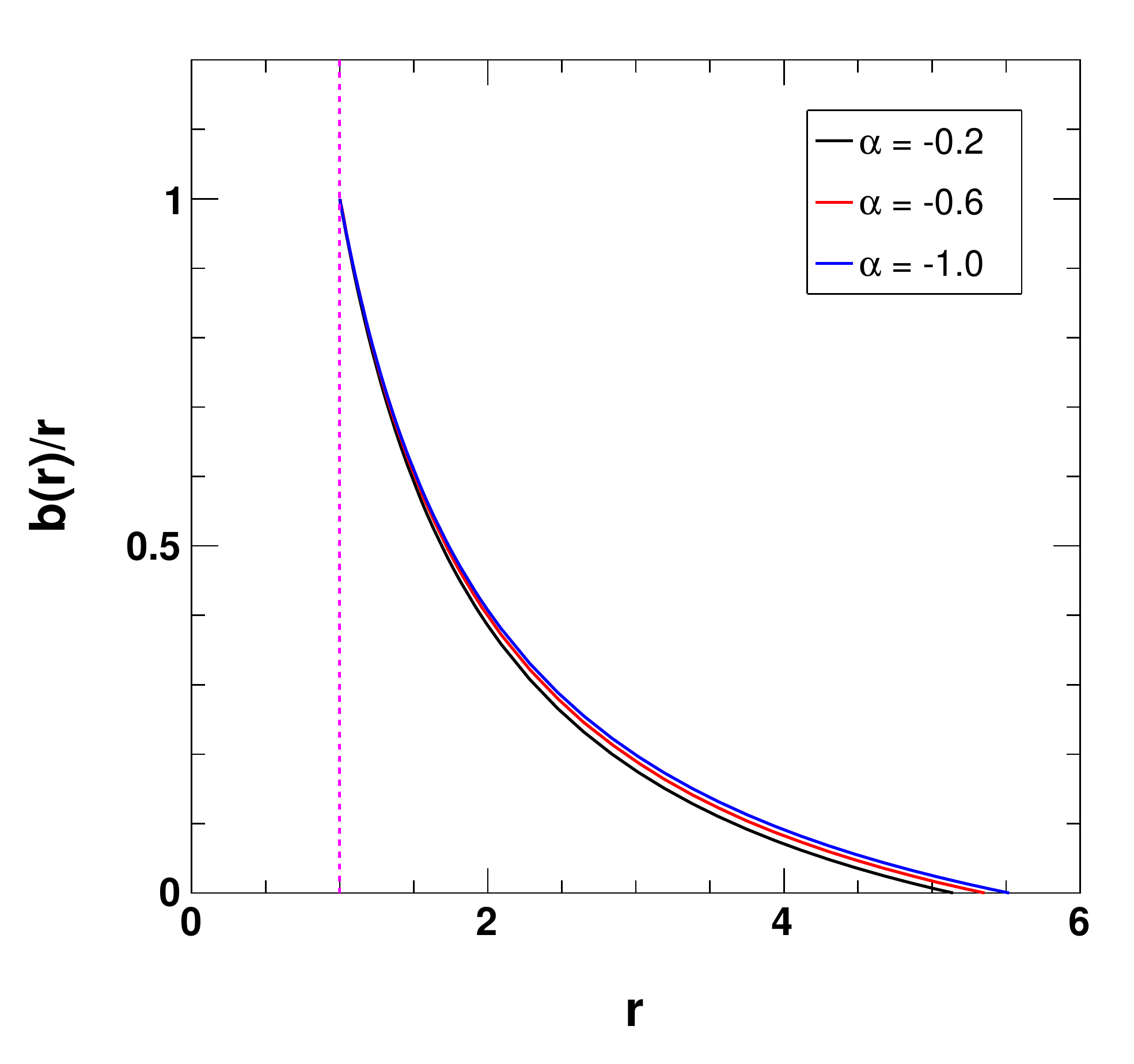}\hspace{1cm}
   \includegraphics[scale = 0.3]{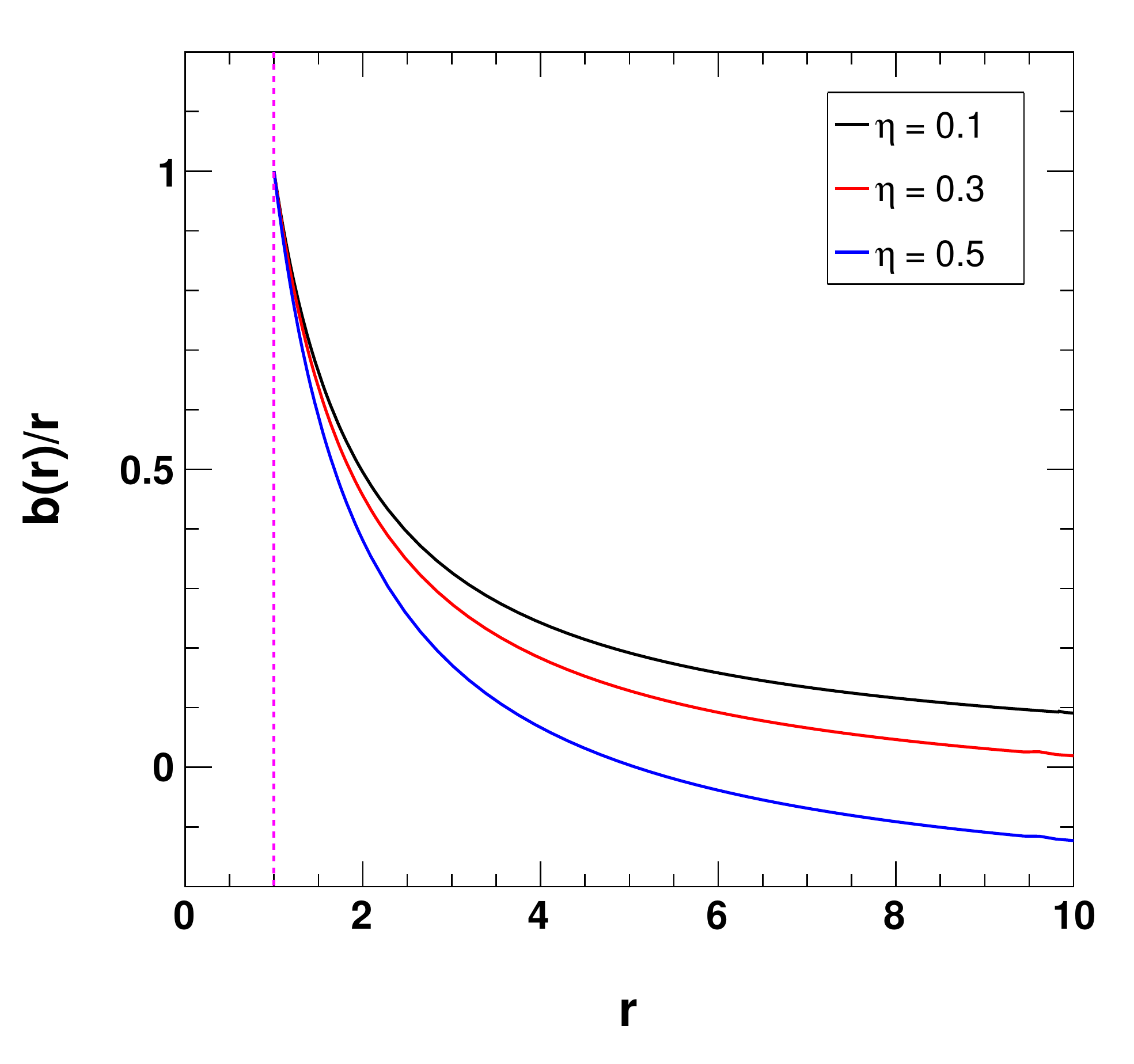}}
\vspace{-0.2cm}
\caption{Plots of $b(r)/r$ vs.~$r$ with $\eta=0.5$ (on the left panel) and 
with $\alpha= -0.10$ (on the right panel) for the wormhole shape function 
\eqref{staro_shape} with throat radius $r_0=1$. The vertical dotted line in 
each plot represents the position of the throat of the wormhole.}
\label{staro_shape_c1}
\end{figure}

\begin{figure}[htbp]
\centerline{
   \includegraphics[scale = 0.3]{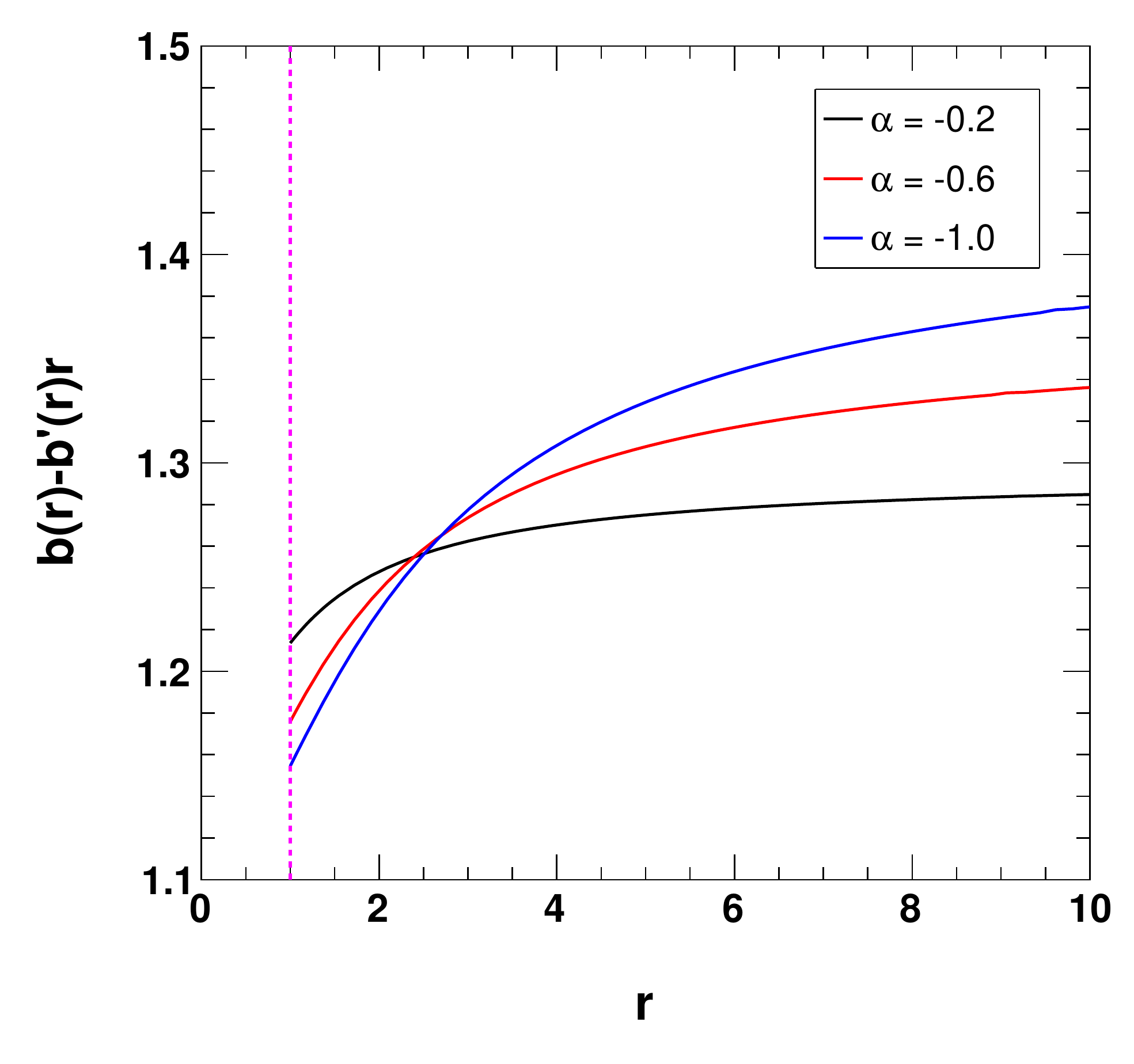}\hspace{1cm}
   \includegraphics[scale = 0.3]{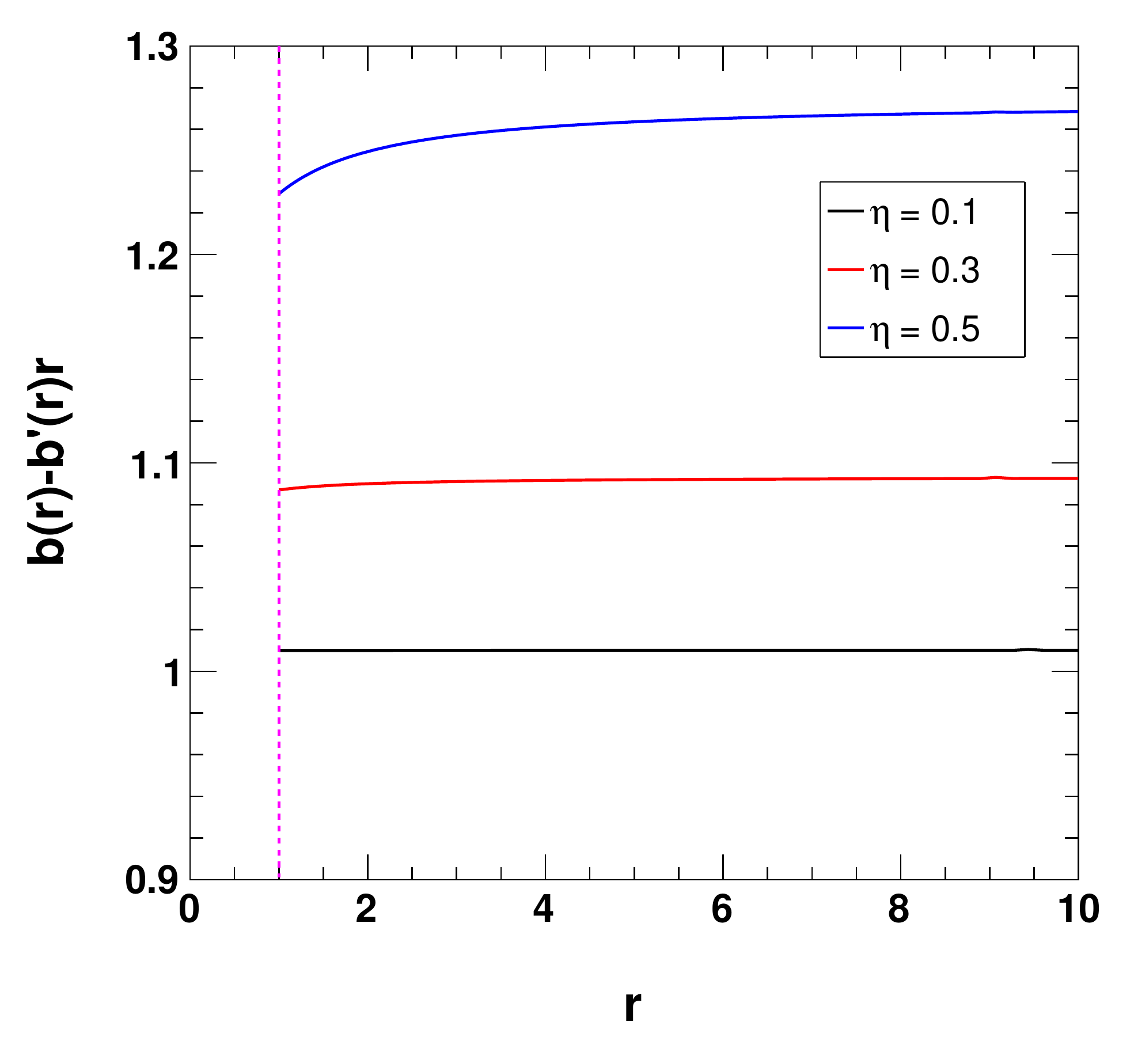}}
\vspace{-0.2cm}
\caption{Plots of $b(r) - b'(r) r$ vs.~$r$ with $\eta=0.5$ (on the left panel) 
and with $\alpha= -0.10$ (on the right panel) for the wormhole shape function 
\eqref{staro_shape} with throat radius $r_0=1$. The vertical dotted line in 
each plot represents the position of the throat of the wormhole.}
\label{staro_shape_c2}
\end{figure}

\begin{figure}[htbp]
\centerline{
   \includegraphics[scale = 0.3]{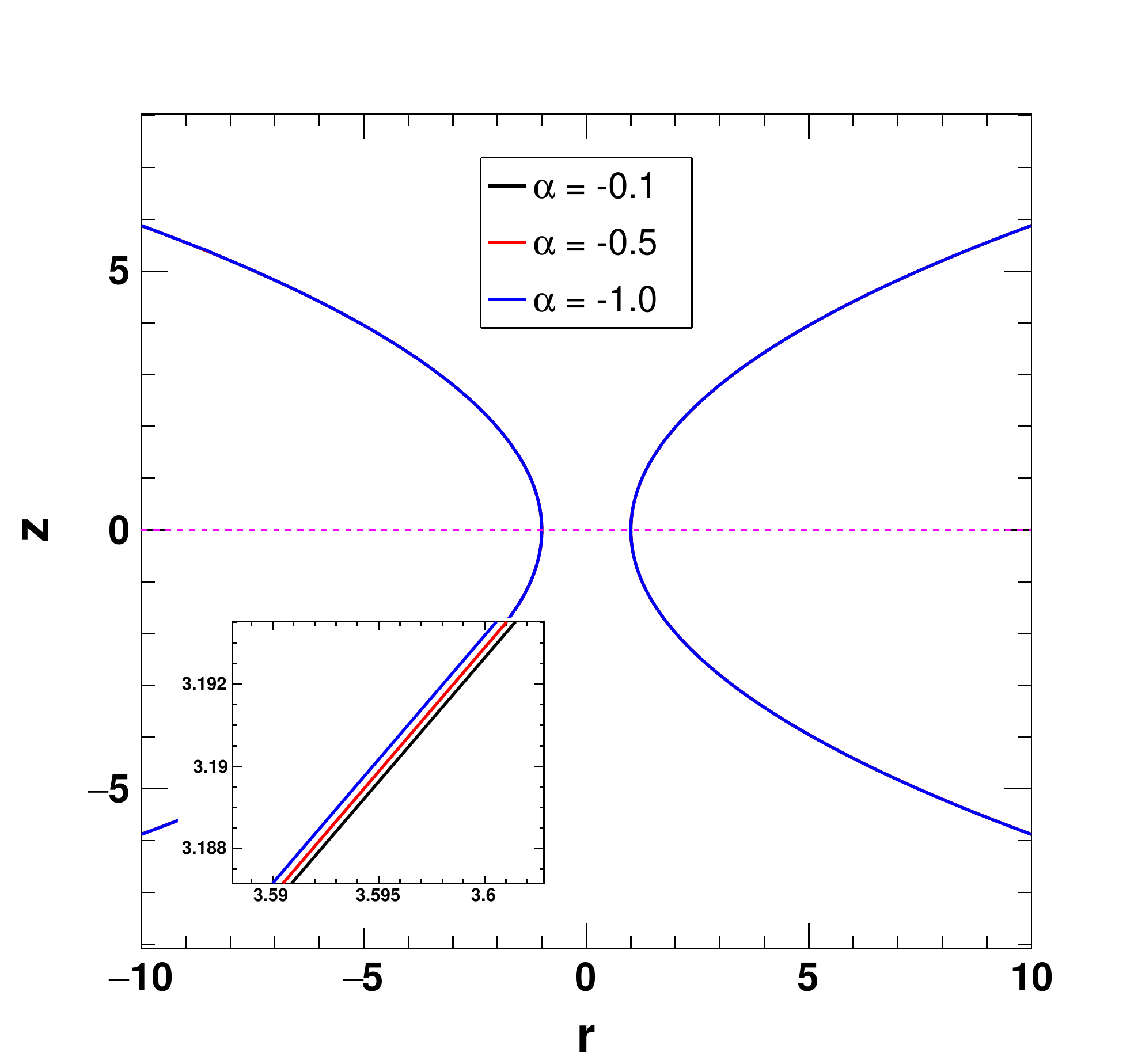}\hspace{1cm}
   \includegraphics[scale = 0.3]{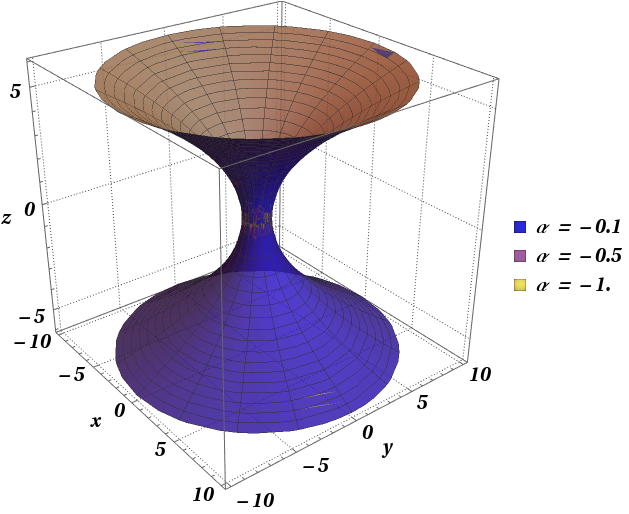}}
\vspace{-0.2cm}
\caption{Embedded 2-D and 3-D plots of the wormhole defined by the shape 
function \eqref{staro_shape} with throat radius $r_0=1$ and $\eta=0.1$ for 
different values of $\alpha$.}
\label{staro_shape_a1}
\end{figure}

\begin{figure}[h!]
\centerline{
   \includegraphics[scale = 0.3]{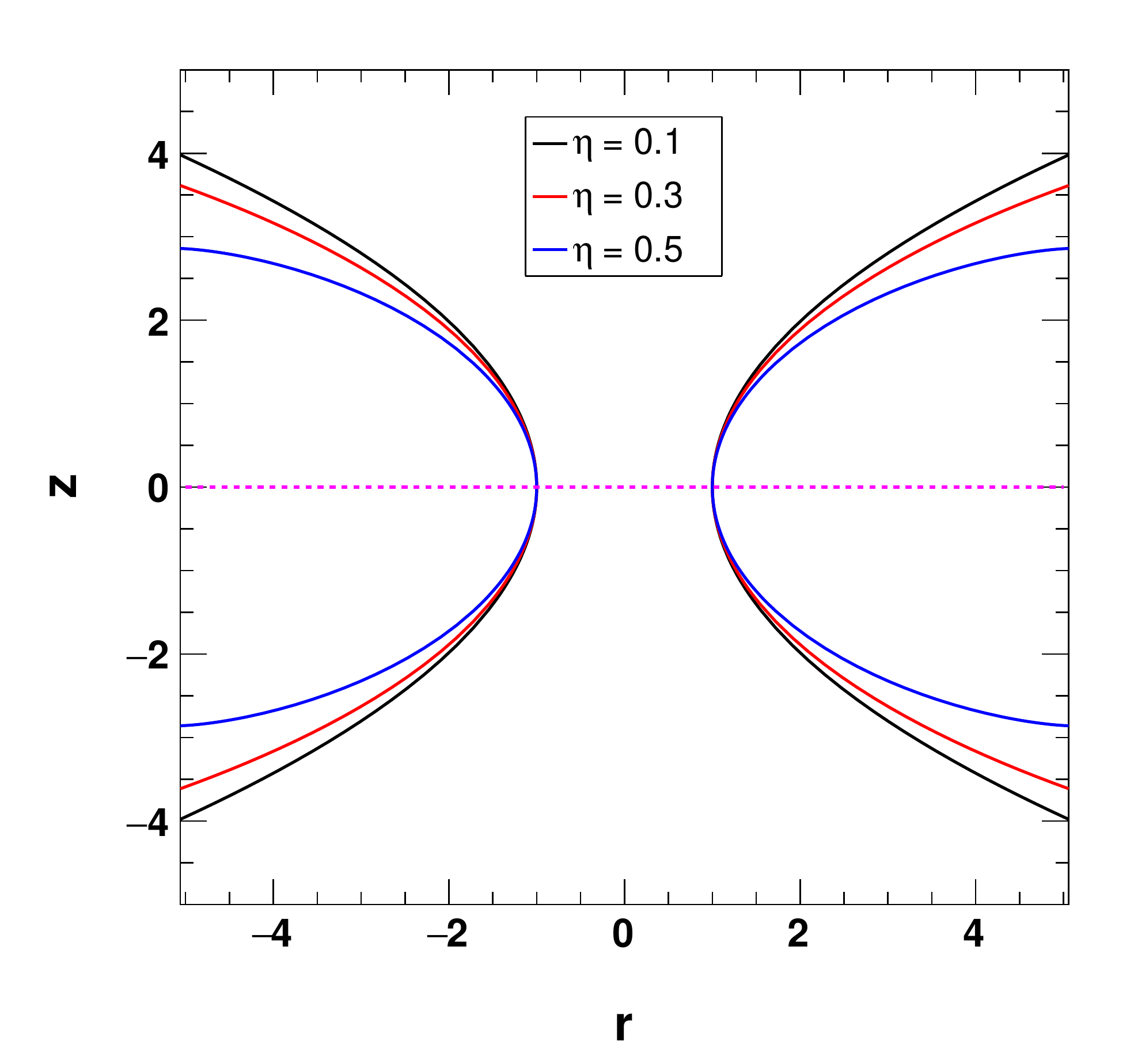}\hspace{1cm}
   \includegraphics[scale = 0.3]{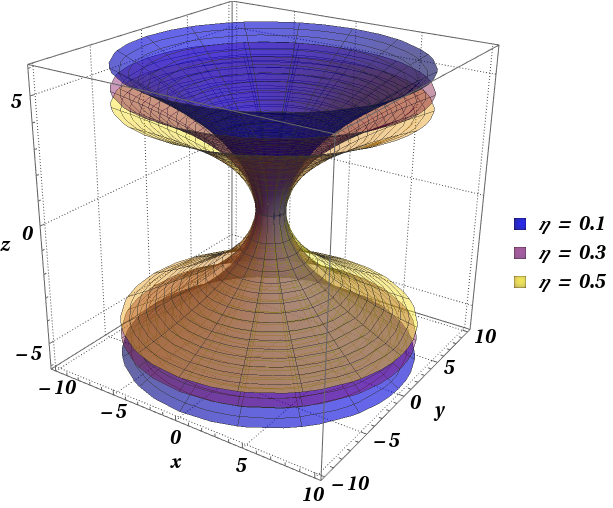}}
\vspace{-0.2cm}
\caption{Embedded 2-D and 3-D plots of the wormhole defined by the shape 
function \eqref{staro_shape} with throat radius $r_0=1$ and $\alpha= -0.10$ 
for different values of $\eta$.}
\label{staro_shape_eta}
\end{figure}

To check the viability of the new shape function \eqref{staro_shape}, we have 
checked the necessary conditions mentioned earlier and two of the important 
conditions are shown in the Fig.s \ref{staro_shape_c1} and \ref{staro_shape_c2} for different values of shape function parameters
$\alpha$ and $\eta$. In Fig.~\ref{staro_shape_c1}, we have plotted the 
function $b(r)/r$ vs.~$r$ for different values of wormhole shape function 
parameters $\alpha$ and $\eta$. We see that the function shows a maximum value 
near the throat of the wormhole and it decreases gradually with an increase in 
the value of $r$ in both cases. Thus, clearly one can realise that $b(r)/r < 1$ for $r>r_0$, which is a necessary condition for the formation of a wormhole. 
It is to be noted that here the impact of the Starobinsky model parameter 
$\alpha$ is very small in comparison to the cloud of strings 
parameter.  In Fig.~\ref{staro_shape_c2}, we have plotted $b(r) - b'(r) r$ 
vs.~$r$ for the shape function for different values of $\alpha$ and $\eta$. 
The Starobinsky model parameter $\alpha$ shows a similar impact on the function 
$b(r) - b'(r) r$ with the impacts of model parameters $m_2$ and $m_3$ of the 
toy shape functions defined in Eq.~\eqref{toy_shape02} and \eqref{toy_shape03}
respectively. With an increase in the values of $\alpha$, the function 
increases slowly and the value of the function is always greater than $0$. One 
may note that as previously mentioned for a viable wormhole, one needs 
$b(r) - b'(r) r >0$. Moreover, the test function $b(r) - b'(r) r$ and hence 
the shape function has a higher dependency on the other model parameter 
i.e.~cloud of strings parameter $\eta$ as seen from the 
Fig.~\ref{staro_shape_c2}. Here also, the function is always greater than $0$ 
for any $r>r_0$ justifying the viability of the shape function 
\eqref{staro_shape}. Now, to have a better visualisation of the wormhole shape 
function \eqref{staro_shape}, we plot the embedded diagrams of the wormhole 
using the Eq.~\eqref{embedded_eq} after solving it numerically. The embedded 
diagrams are shown in Fig.~\ref{staro_shape_a1} and in 
Fig.~\ref{staro_shape_eta} for different values of the Starobinsky model 
parameter $\alpha$ and cloud of strings parameter $\eta$ respectively. From 
Fig.~\ref{staro_shape_a1}, one can see that the Starobinsky model parameter 
$\alpha$ has a very small impact on the structure of the wormhole as already 
clear from above. However, with decrease in $\alpha$, the height of the 
wormhole increases very slowly. Moreover, we have seen that for a viable 
wormhole one must have, $\alpha < 0$. From these observations, we can infer 
that the Starobinsky model parameter has a very minimal impact over the 
structure of the wormhole. On the other hand, the cloud of strings parameter 
has a significant impact over the structure of the wormhole. As seen from the 
Fig.~\ref{staro_shape_eta}, with an increase in the cloud of strings parameter, 
the height of the wormhole decreases gradually. Hence, presence of a 
surrounding field e.g.~a cloud of strings may have significant influences over 
different properties of the wormhole. The impact of these two model parameters 
on the quasinormal modes of wormhole will be elaborately studied in the next 
section.

\subsection{Energy conditions}
Here we shall study the energy conditions of the wormholes in $f(R)$ gravity 
metric formalism briefly. In GR, a fundamental requirement in wormholes is the 
violation of energy conditions \cite{Morris1988, Bolokhov2021}. Whereas in 
$f(R)$ theories of gravity, the required condition $R_{\mu\nu} k^\mu k^\nu \geq 0 $ results null energy condition of the form $T^{eff}_{\mu\nu} k^\mu k^\nu \geq 0$ with the modified field equations. This expression can be used to study 
the NEC. Now, using a radial null vector, violation of NEC can be expressed as
\begin{equation}\label{nec}
\rho^{eff} + p_r^{eff} = \dfrac{\rho + p_r}{\mathcal{F}} + \dfrac{1}{\mathcal{F}} \left\lbrace \mathcal{F}'' \,\left(1-b(r)/r\right) - \mathcal{F}' \, \left(b'(r) r - b(r)\right)/(2 r^2) \right\rbrace < 0.
\end{equation}
At the throat of the wormhole, this relation reduces to the following form:  
\begin{equation}\label{throat_nec}
\rho^{eff} + p_r^{eff}|_{r\, =\, r_0} = \dfrac{\rho + p_r}{\mathcal{F}}|_{r\, =\, r_0} + \dfrac{1 - b'(r_0)}{2 r_0} \dfrac{\mathcal{F}'}{\mathcal{F}}|_{r\, =\, r_0} <0.
\end{equation}
One may note that for a Morris-Throne wormhole in GR, the flare out condition 
is sufficient to yield a violation of the NEC at the throat. It can be easily 
checked by using $\mathcal{F} = 1$ in Eq.s \eqref{fieldeq01} and 
\eqref{generic2} that
\begin{equation}
\rho(r_0) + p_r(r_0) = \dfrac{r_0 b'(r_0) - b(r_0)}{b(r_0)^2\,r_0}<0,
\end{equation}
as we have $b(r_0) = r_0$ at the throat of the wormhole. But from 
Eq.~\eqref{nec} one can see that in $f(R)$ gravity, the violation of NEC 
depends on the form of the $f(R)$ gravity model also apart from the shape 
function.

\begin{figure}[htbp]
\centerline{
   \includegraphics[scale = 0.3]{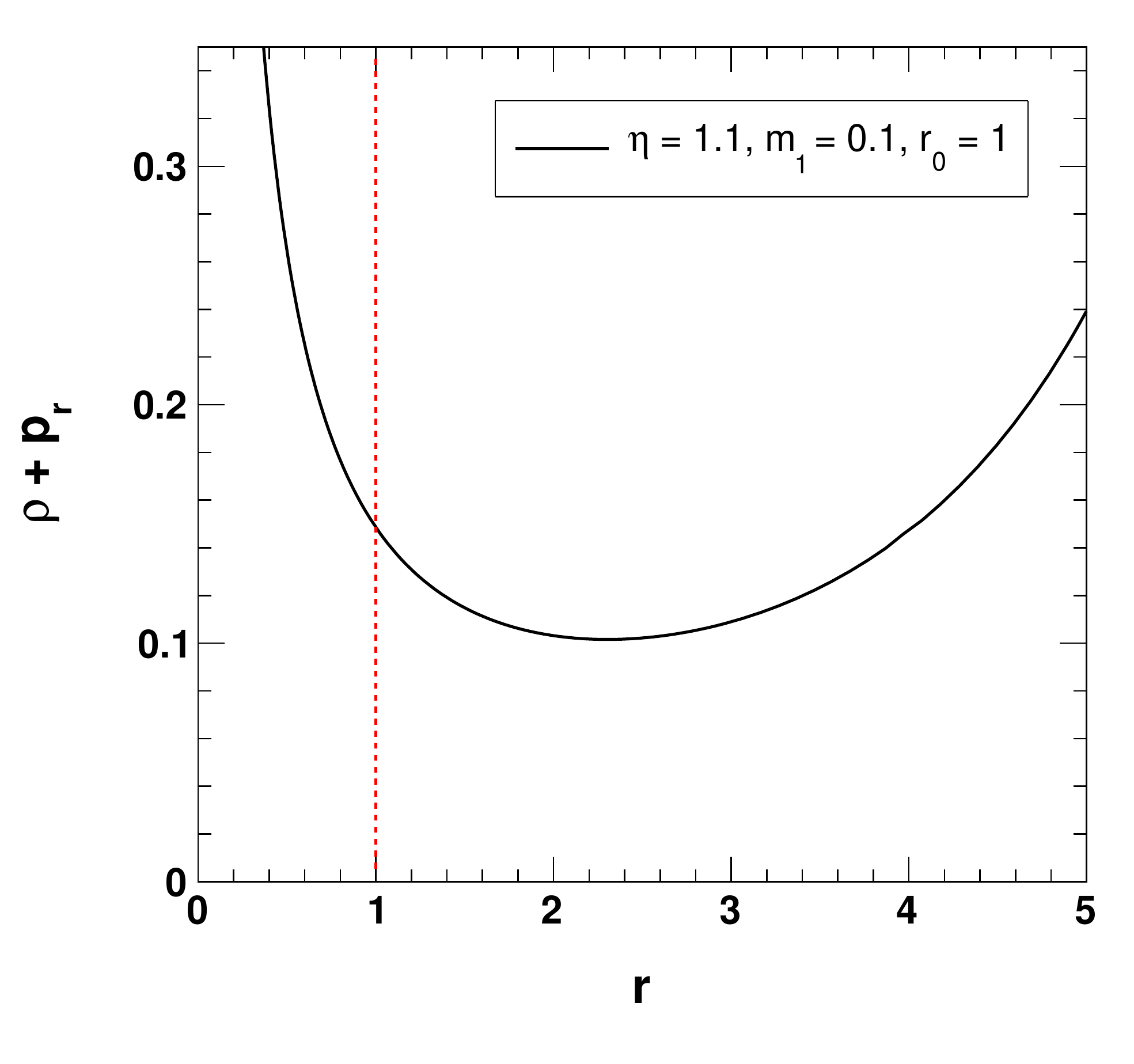}\hspace{1cm}
   \includegraphics[scale = 0.3]{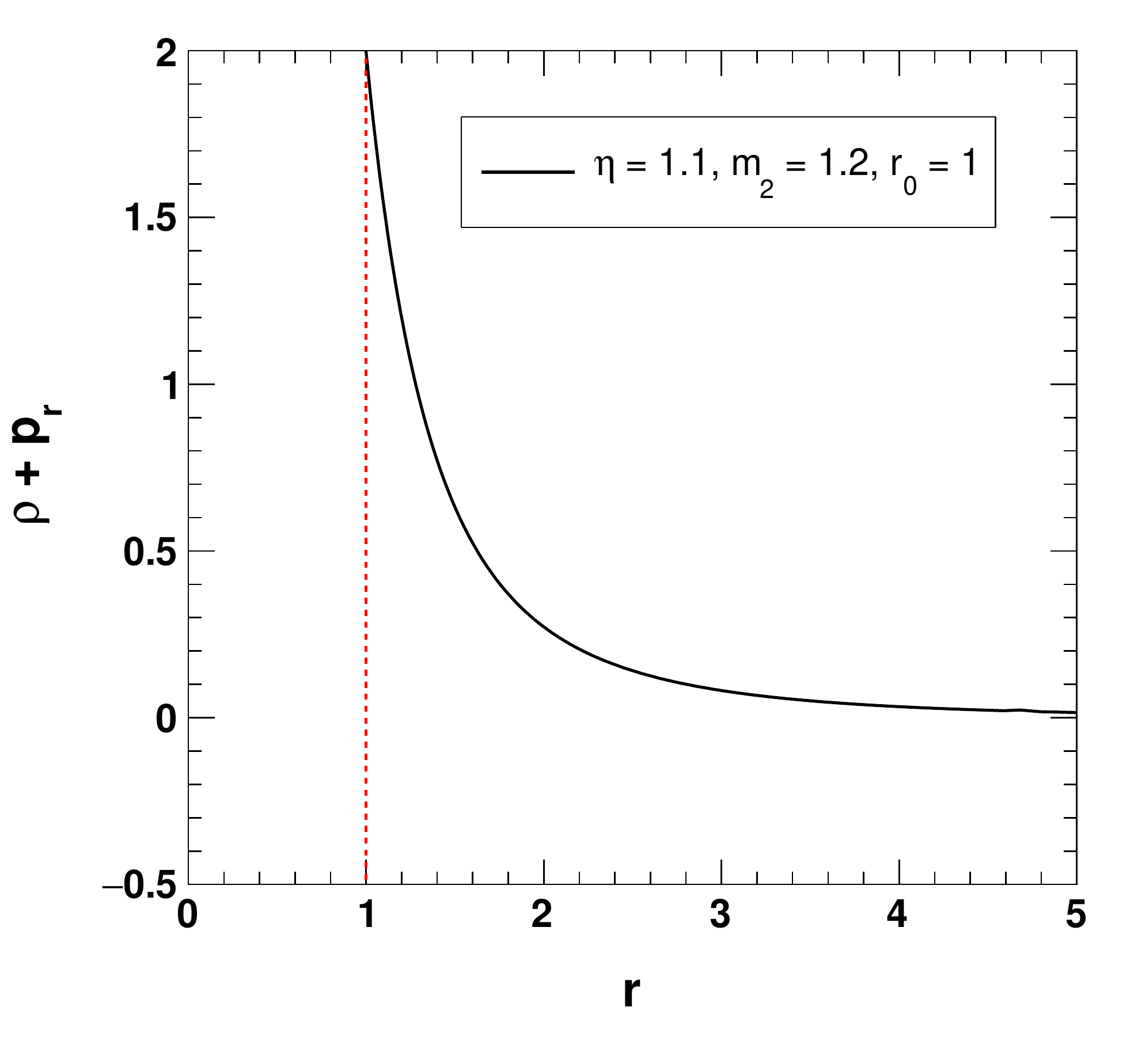}}
\vspace{-0.2cm}
\caption{Variation of $\rho+ p_r$ for shape function \eqref{toy_shape01} (on left) and \eqref{toy_shape02} (on right).}
\label{nec_01}
\end{figure}

\begin{figure}[htbp]
\centerline{
   \includegraphics[scale = 0.3]{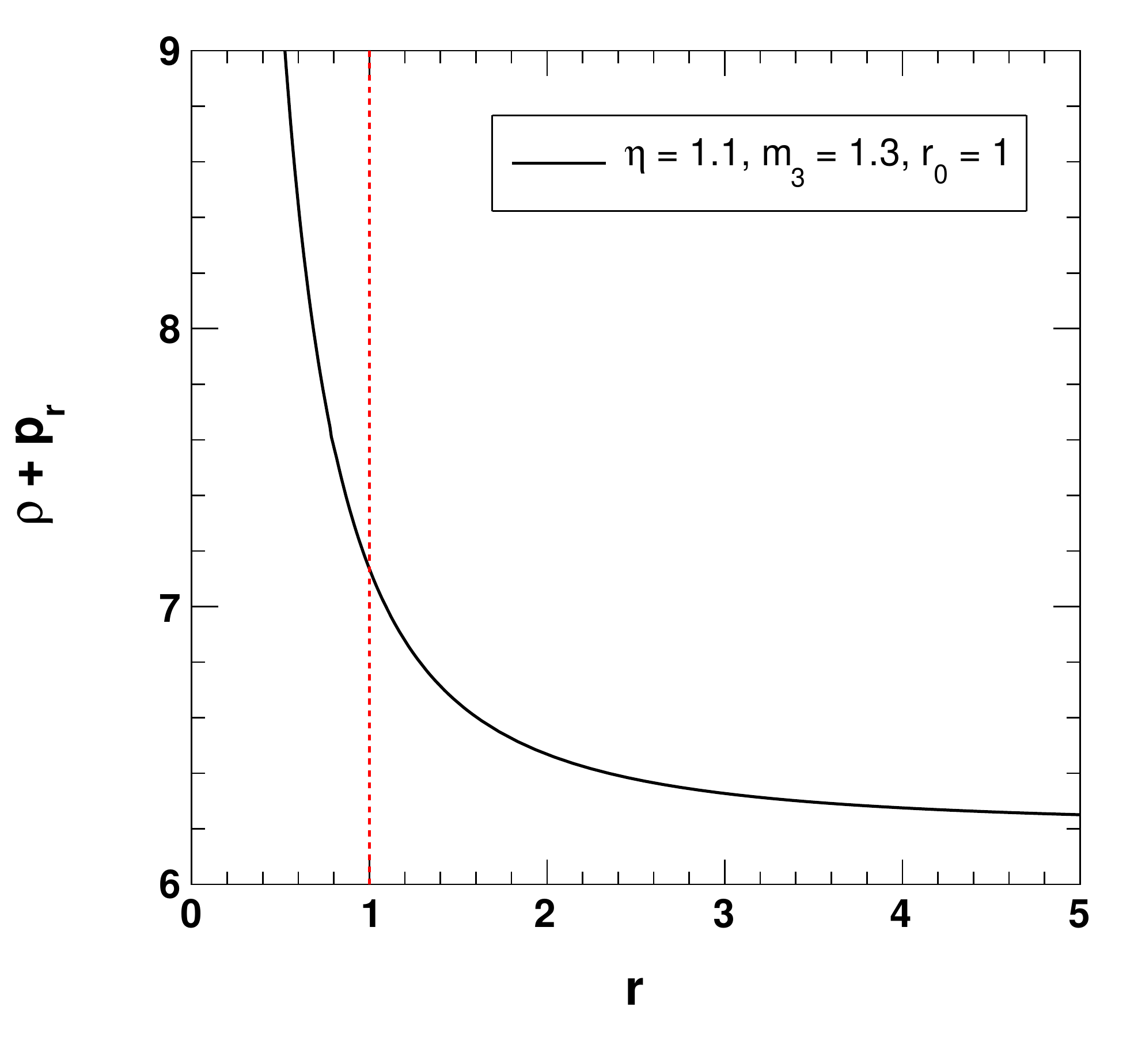}\hspace{1cm}
   \includegraphics[scale = 0.3]{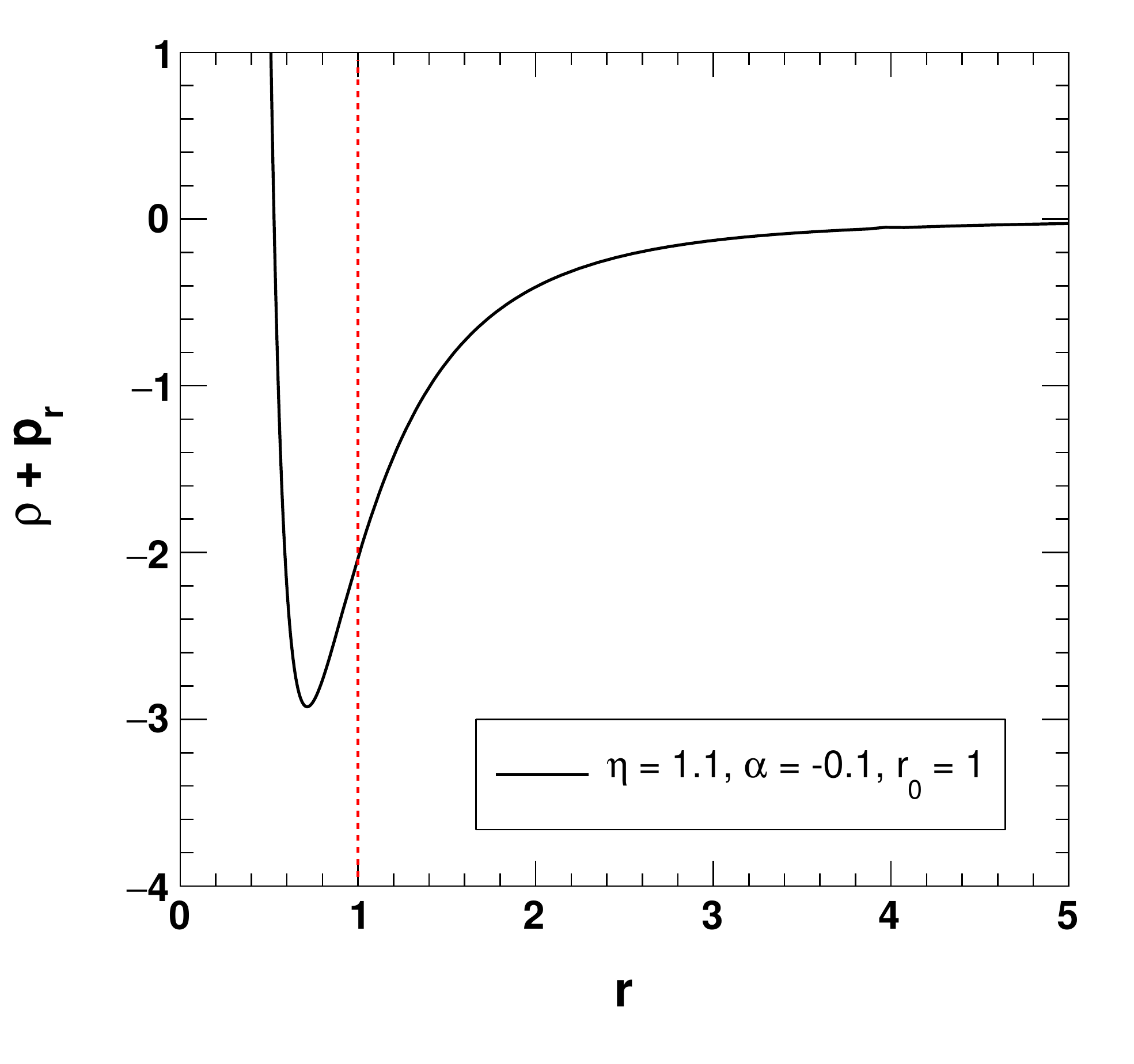}}
\vspace{-0.2cm}
\caption{Variation of $\rho+ p_r$ for shape function \eqref{toy_shape03} (on left) and \eqref{staro_shape} (on right).}
\label{nec_02}
\end{figure}

Now we shall study the NEC for our toy shape functions and the Starobinsky 
model. Since $\rho(r) + p_r(r)$ has a complex form, we shall see the condition 
only at the throat for simplicity. So at the throat of the wormhole, the first 
toy shape function defined in Eq.~\eqref{toy_shape01} gives,
\begin{equation}
\rho(r_0) + p_r(r_0) = -\frac{\eta ^2 m_1 \left(m_1^2 r_0^2-2\right)}{2 r_0 \left(m_1 r_0-1\right)^2}.
\end{equation}
From this expression, it is found that the wormhole solution violates NEC 
for $(m_1 r_0>\sqrt{2}) \; \cup \; (-\frac{\sqrt{2}}{r_0}<m_1<0)$ and beyond 
this range the solution respects the NEC. For the second shape function 
defined by Eq.~\eqref{toy_shape02}, we have at the throat of the wormhole,
\begin{equation}
\rho(r_0) + p_r(r_0) =\frac{\eta ^2 \Big[\log \left(2 m_2 r_0\right) \Big\lbrace 8 \log \left(2 m_2 r_0\right)-1\Big\rbrace-2\Big]}{4 \log \left(2 m_2 r_0\right) \Big\lbrace r_0-2 r_0 \log \left(2 m_2 r_0\right)\Big\rbrace ^2}.
\end{equation}
NEC is violated at the throat for the following range:
\begin{equation}
\log \left(2 m_2 r_0\right)<\frac{1}{16} \left(1-\sqrt{65}\right) \; \cup \; 0<\log \left(2 m_2 r_0\right)<\frac{1}{16} \left(\sqrt{65}+1\right)
\end{equation}
and beyond this range the NEC is respected. Similarly, for the third shape 
function defined in Eq.~\eqref{toy_shape03}, at the throat of the wormhole, 
\begin{equation}
\rho(r_0) + p_r(r_0) = \frac{\eta ^2 \left(m_3+r_0\right)}{m_3 r_0 \left(m_3-r_0\right)}.
\end{equation}
In this case the NEC is violated for $0<m_3<r_0\; \cup \; m_3+r_0<0.$
Finally for the Starobinsky model case, we have
\begin{equation}\label{starobinsky_nec}
\rho(r_0) + p_r(r_0) = \frac{4 \alpha  \eta ^2 \left(\sqrt{r_0^2-16 \alpha  \eta ^2}+2 \eta ^2 r_0+r_0\right)}{r_0^3 \left(-16 \alpha  \eta ^2-r_0 \sqrt{r_0^2-16 \alpha  \eta ^2}+r_0^2\right)}.
\end{equation}
In this situation the NEC is violated for $(\alpha<0) \;\cup\; (0<\alpha<\frac{r_0^2}{16 \eta ^2}).$ Moreover, one may note that for $\alpha>\frac{r_0^2}{16 \eta ^2}$, Eq.~\eqref{starobinsky_nec} results in imaginary values. Hence, for 
this case, NEC is violated for the complete feasible range in the parameter 
space.

We have also checked the NEC for the variable $r$ graphically in Fig.s 
\ref{nec_01} and \ref{nec_02} for the shape functions \eqref{toy_shape01}, 
\eqref{toy_shape02}, \eqref{toy_shape03} and \eqref{staro_shape} respectively 
for selected set of parameters. Since the expressions for $\rho+ p_r$ are 
complex, we have given their explicit forms along with the corresponding 
$d f(R)/dR$ in terms of $r$ in the appendix. On the left panel of 
Fig.~\ref{nec_01}, we have shown the variation of NEC for the first wormhole 
shape function with $\eta = 1.1$, 
$m_1 = 0.1$ and the throat radius of the wormhole $r_0 = 1$. The red dotted 
line represents the throat of the wormhole. One may note that for the chosen 
value of the parameters, NEC is respected at the throat of the wormhole. With 
an increase in the distance from the throat of the wormhole, $\rho + p_r$ 
increases gradually respecting NEC. On the right panel of Fig.~\ref{nec_01}, 
we have shown the variation of  $\rho + p_r$ for the second wormhole shape 
function defined by \eqref{toy_shape02} with the model parameters 
$\eta = 1.1$, $m_2 = 1.2$ and the throat radius of the wormhole $r_0 = 1$. In 
this case also, the selected values of the parameters respect the NEC at the 
throat as seen from the previous results. From the figure, one can see that 
in this case $\rho + p_r$ is positive at the throat of the wormhole and 
gradually decreases towards $0$ with an increase in $r$.

On the left panel of Fig.~\ref{nec_02}, we have shown variation of 
$\rho + p_r$ for the third shape function defined by \eqref{toy_shape03}. In 
this case we have chosen $\eta = 1.1$, $m_3 = 1.3$ and the throat radius of 
the wormhole $r_0 = 1$. These values of the parameters respect the NEC at the 
throat of the wormhole. We have seen that with an increase in the value of 
$r$, $\rho + p_r$ decreases gradually but the value of $\rho + p_r$ is much 
greater than $0$ for this particular shape function. Finally, on the right 
panel of Fig.~\ref{nec_02}, we have shown the variation of $\rho + p_r$ with 
respect to $r$ for the shape function obtained for the Starobinsky $f(R)$ 
gravity model surrounded by a cloud of strings. We have already seen that for 
the feasible $\alpha<0$ values and at the throat, this wormhole always 
violates the NEC.  Hence, 
with the selected values of the model parameters $\eta = 1.1$, $\alpha = -0.1$ 
and $r_0 = 1$, we have seen that $\rho + p_r$ violates NEC for a variable $r$ 
and with an increase in the value of $r$, $\rho + p_r$ approaches to $0$ 
gradually.

So, as a result we have seen that the wormhole solution obtained in $f(R)$ 
gravity Starobinsky model always violates NEC, while the three toy shape 
functions can violate NEC in a particular range of the model parameters only. 
This range has been calculated explicitly at the throat of the wormholes for
each of toy shape function. 

\section{Quasinormal modes}\label{section03}
Quasinormal modes of compact stellar objects, such as black holes, wormholes 
etc.~are the long lasting components of gravitational waves from such objects 
when the morphological symmetry of those objects are disturbed by some 
perturbative effects. Quasinormal mode frequencies are usually 
expressed in terms of complex numbers, where the real part represents the 
amplitude of the mode and the imaginary part represents the loss of energy of 
the objects. Thus by quasinormal modes compact objects try to regain their 
original state with the loss of energies. The details 
about the quasinormal modes can be found in Ref.s \cite{Kokkotas, Vishveshwara, Press, Chandrasekhar, Berti2015, Dreyer2004, Chen, Zhang, Zhang2, Ma, Zhang3, Graca}. To calculate the 
quasinormal mode frequencies of a compact object it is necessary to use an 
external field surrounding the object as a probe, which can give a measure of
the perturbation in the object. There are various possible probe fields, 
e.g.~scalar fields, vector fields, fermionic or Dirac fields \cite{Frolov2018, Jing2005, Chowdhury2018}  
etc.~surrounding the compact objects. Here, we shall study the quasinormal 
modes from the wormholes for the scalar (field) perturbations and Dirac (field) 
perturbations using WKB approximation method. We shall neglect any possible 
echoes from quantum corrections near the wormhole throat or from any 
matter far from it \cite{Konoplya2018, Konoplya2022}. 

The primary problem in the quasinormal modes calculations of wormholes is 
related with the potential, because in the absence of tidal forces and in the 
usual situation, the peak of the potential is at the throat of the wormhole. 
In normal coordinates it is difficult to visualise such peak of the potential 
properly. This issue can be simply avoided by converting the potential of 
the problem into tortoise coordinates. During the calculations, we assume 
that the test fields such as scalar field or Dirac field have negligible 
reaction on the spacetime.

\subsection{Scalar Perturbation}
Now we consider a massless scalar field $\zeta$ around the wormhole spacetime.
Assuming that the reaction of the scalar field on the spacetime negligible as
mentioned above, it is possible to describe the quasinormal modes of the 
wormholes by the Klein Gordon equation in curved spacetime as given by
\begin{equation}\label{scalar_KG}
\square \zeta = \dfrac{1}{\sqrt{-g}} \partial_\mu (\sqrt{-g} g^{\mu\nu} \partial_\nu \zeta) = 0.
\end{equation}
Using spherical harmonics, it is possible to decompose the scalar field in the 
following form:
\begin{equation}
\zeta(t,r,\theta, \phi) = \dfrac{1}{r} \sum_{l,m} \psi_l(t,r) Y_{lm}(\theta, \phi),
\end{equation}
where $\psi_l(t,r)$ is the radial time dependent wave function, and $l$ and 
$m$ are the indices of the spherical harmonics. Using this equation in 
\eqref{scalar_KG}, we get,
\begin{equation}
\partial^2_{r_*} \psi(r_*)_l + \omega^2 \psi(r_*)_l = V_s(r) \psi(r_*)_l,  
\end{equation}
where $r_*$ is the tortoise coordinate defined by
\begin{equation}\label{tortoise}
\dfrac{dr_*}{dr} = \sqrt{\frac{r \; e^{-2 \phi (r)}}{r-b(r)}}
\end{equation}
and the effective potential is given by
\begin{equation}\label{Vs}
V_s(r) = \frac{e^{2 \Phi (r)} \left[-\,r b'(r)+b(r) \left(1-2 r \Phi '(r)\right)+2 r \left\lbrace l (l+1)+r \Phi '(r)\right\rbrace \right]}{2 r^3}.
\end{equation}
Here $l$ is referred to as the multipole moment of the quasinormal modes of 
the wormhole.

\subsection{Dirac Perturbation}

For the perturbation of Dirac field with mass $m$, one can write the general 
equation as \cite{Brill1957, Cho2003}
\begin{equation}\label{dirac_field}
[\gamma^{a}e_{a}^{\mu}(\partial_{\mu}+\Gamma_{\mu})+m]\psi=0,
\end{equation}
where $\Gamma_{\mu}=\frac{1}{8}[\gamma^{a},\gamma^{b}]e_{a}^{\nu}e_{b\nu;\mu}$
is the spin connection, $\gamma^{a}$ are the Dirac metrices and
$e_{b\nu;\mu}=\partial_{\mu}e_{b\nu}-\Gamma_{\mu\nu}^{\alpha}e_{b\alpha}$.
Using our ansatz \eqref{metric}, $e_{\nu}^{a}$ can be taken to be
\begin{equation}
e_{\nu}^{a}={\rm diag}(e^{\Phi(r)},(1-b(r)/r)^{-1/2},r,r\sin \theta).
\end{equation}
Using this and now considering massless Dirac field i.e. $m=0$, 
Eq.~\eqref{dirac_field} can be further written in the following form:
\begin{equation}
\partial^2_{r_*} \psi(r_*)_l + \omega^2 \psi(r_*)_l = V_{d \pm}(r) \psi(r_*)_l, \end{equation}
where $r_*$ is the tortoise coordinate and two isospectral potentials,
\begin{equation}
V_{d \pm}(r) = \frac{k}{r} \left(\frac{k e^{2\Phi(r)}}{r}\mp\frac{e^{2\Phi(r)}\sqrt{(1-b(r)/r)}}{r}\pm e^{\Phi(r)}\sqrt{(1-b(r)/r)}\dfrac{d \;e^{\Phi(r)}}{dr}\right).
\end{equation}
Here $k=1,2,3 \hdots $ are called the multipole numbers with $k=\ell+1/2$. The 
potentials can be transformed from one to another form by using the Darboux 
transformation as shown below:
\begin{equation}\label{psi}
\psi_{+}=A \left(W+\dfrac{d}{dr_*}\right) \psi_{-}, \quad W=\sqrt{e^{\Phi(r)}\sqrt{(1-b(r)/r)}},
\end{equation}
where $A$ is a constant. As $+$ and $-$ wave equations are isospectral, we 
shall consider only one of the two effective potentials for the Dirac case.

\subsection{WKB method for Quasinormal modes}
The first order WKB method for calculating the quasinormal modes was first 
suggested by Schutz and Will in \cite{Schutz}. Later, the method was developed 
to higher orders \cite{Will_wkb, Konoplya_wkb, Maty_wkb}. In this work, we 
shall use higher order WKB methods, more specifically the 3rd order and 5th 
order WKB approximation methods to calculate the quasinormal modes from the 
wormholes defined in the previous sections.

\begin{table}[ht]
\caption{Fundamental quasinormal modes of the wormhole defined by the
shape function \eqref{toy_shape01} for the scalar perturbation with 
$m_1=0.3$ and $r_0=1$.}
\label{tab01}
\begin{center}
\begin{small}{
\begin{tabular}{ c c c c c}
    \hline
    \hline
    $l$  & WKB $3$rd order & WKB $5$th order & $\Delta_3$ & $\Delta_5$  \\
    \hline
    \hline
   \multirow{1}{*}{\;\;\;$l=1$\;\;\;} & $1.4317 - 0.1853 i$ & $1.4322 - 0.1851 i$ & $0.000863134$ & $0.000111803$ \\
    \multirow{1}{*}{$l=2$} & $2.4434 - 0.1862 i$ & $2.4435 - 0.1862 i$ & $0.000280943$ & $0.0000132872$\\
     \multirow{1}{*}{$l=3$} & $3.4426 - 0.1864 i$ & $3.4427 - 0.1864 i$ & $0.000139003$ & $3.40181\times10^{-6}$ \\
     \multirow{1}{*}{$l=4$} & $4.4378 - 0.1865  i$ & $4.4378 - 0.1865  i$ & $0.0000830125$ & $1.23455\times10^{-6}$\\
    \hline
\end{tabular}
}\end{small}
\end{center}
\end{table}

\begin{table}[ht]
\caption{Fundamental quasinormal modes of the wormhole defined by the
shape function \eqref{toy_shape01} for the Dirac perturbation with 
$m_1=0.3$ and $r_0=1$.}
\label{tab02}
\begin{center}
\begin{small}{
\begin{tabular}{ c c c c c}
    \hline
    \hline
    $k$  & WKB $3$rd order & WKB $5$th order & $\Delta_3$ & $\Delta_5$  \\
    \hline
    \hline
   \multirow{1}{*}{\;\;\;$k=8$\;\;\;} & $7.9443 - 0.1882 i$ & $7.9455 - 0.1697 i$ & $0.000801561$ & $0.12791$ \\
    \multirow{1}{*}{$k=9$} & $8.9348 - 0.1880 i$ & $8.9356 - 0.1777 i$ & $0.000502494$ & $0.0628711$\\
     \multirow{1}{*}{$k=10$} & $9.9252 - 0.1879  i$ & $9.9257 - 0.1817 i$ & $0.00035$ & $0.0334861$ \\
     \multirow{1}{*}{$k=11$} & $10.9155 - 0.1877 i$ & $10.9159 - 0.1839  i$ & $0.00025$ & $0.0189136$\\
    \hline
\end{tabular}
}\end{small}
\end{center}
\end{table}

\begin{table}[ht]
\caption{Fundamental quasinormal modes of the wormhole defined by the
shape function \eqref{toy_shape02} for the scalar perturbation with 
$m_2=1.3$ and $r_0=1$.}
\label{tab03}
\begin{center}
\begin{small}{
\begin{tabular}{ c c c c c}
    \hline
    \hline
    $l$  & WKB $3$rd order & WKB $5$th order & $\Delta_3$ & $\Delta_5$  \\
    \hline
    \hline
   \multirow{1}{*}{\;\;\;$l=1$\;\;\;} & $1.45101 - 0.251715 i$ & $1.45255 - 0.25038 i$ & $0.00181116$ & $0.000638004$ \\
    \multirow{1}{*}{$l=2$} & $2.45502 - 0.247206 i$ & $2.45537 - 0.247046 i$ & $0.000629713$ & $0.0000668254$\\
     \multirow{1}{*}{$l=3$} & $3.45096 - 0.246039 i$ & $3.45109 - 0.245998 i$ & $0.000318384$ & $0.0000163396$ \\
     \multirow{1}{*}{$l=4$} & $4.44427 - 0.245568 i$ & $4.44434 - 0.245553 i$ & $0.000191903$ & $5.81055\times10^{-6}$\\
    \hline
\end{tabular}
}\end{small}
\end{center}
\end{table}

\begin{table}[ht]
\caption{Fundamental quasinormal modes of the wormhole defined by the
shape function \eqref{toy_shape02} for the Dirac perturbation with 
$m_2=1.3$ and $r_0=1$.}
\label{tab04}
\begin{center}
\begin{small}{
\begin{tabular}{ c c c c c}
    \hline
    \hline
    $k$  & WKB $3$rd order & WKB $5$th order & $\Delta_3$ & $\Delta_5$  \\
    \hline
    \hline
   \multirow{1}{*}{\;\;\;$k=10$\;\;\;} & $9.9313 - 0.2489 i$ & $9.9421 - 0.0908 i$ & $0.00676849$ & $1.29907$ \\
    \multirow{1}{*}{$k=11$} & $10.9219 - 0.2483 i$ & $10.9293 - 0.1497  i$ & $0.00461736$ & $0.683984$\\
     \multirow{1}{*}{$k=12$} & $11.9124 - 0.2479 i$ & $11.9178 - 0.1839 i$ & $0.00326382$ & $0.395367$ \\
     \multirow{1}{*}{$k=13$} & $12.9028 - 0.2475 i$ & $12.9068 - 0.2045 i$ & $0.00241299$ & $0.241984$\\
    \hline
\end{tabular}
}\end{small}
\end{center}
\end{table}

\begin{table}[ht]
\caption{Fundamental quasinormal modes of the wormhole defined by the
shape function \eqref{toy_shape03} for the scalar perturbation with 
$m_3=1.3$ and $r_0=1$.}
\label{tab05}
\begin{center}
\begin{small}{
\begin{tabular}{ c c c c c}
    \hline
    \hline
    $l$  & WKB $3$rd order & WKB $5$th order & $\Delta_3$ & $\Delta_5$  \\
    \hline
    \hline
   \multirow{1}{*}{\;\;\;$l=1$\;\;\;} & $1.47458 - 0.308873  i$ & $1.47748 - 0.305617 i$ & $0.00307541$ & $0.00168329$ \\
    \multirow{1}{*}{$l=2$} & $2.46893 - 0.301043 i$ & $2.46966 - 0.300633 i$ & $0.00110078$ & $0.000177451$\\
     \multirow{1}{*}{$l=3$} & $3.46087 - 0.298982 i$ & $3.46114 - 0.298877 i$ & $0.000561802$ & $0.0000431066$ \\
     \multirow{1}{*}{$l=4$} & $4.45198 - 0.298148 i$ & $4.45211 - 0.298109 i$ & $0.000339966$ & $0.0000152743$\\
    \hline
\end{tabular}
}\end{small}
\end{center}
\end{table}

\begin{table}[ht]
\caption{Fundamental quasinormal modes of the wormhole defined by the
shape function \eqref{toy_shape03} for the Dirac perturbation with 
$m_3=1.3$ and $r_0=1$.}
\label{tab06}
\begin{center}
\begin{small}{
\begin{tabular}{ c c c c c}
    \hline
    \hline
    $k$  & WKB $3$rd order & WKB $5$th order & $\Delta_3$ & $\Delta_5$  \\
    \hline
    \hline
   \multirow{1}{*}{\;\;\;$k=13$\;\;\;} & $12.9085 - 0.3013 i$ & $12.9185 - 0.1644 i$ & $0.00627415$ & $0.988848$ \\
    \multirow{1}{*}{$k=14$} & $13.899 - 0.3008 i$ & $13.9066 - 0.2060 i$ & $0.00467172$ & $0.615035$\\
     \multirow{1}{*}{$k=15$} & $14.8895 - 0.3004 i$ & $14.8953 - 0.2331 i$ & $0.00357246$ & $0.400761$ \\
     \multirow{1}{*}{$k=16$} & $15.8799 - 0.3001 i$ & $15.8845 - 0.2512 i$ & $0.00277218$ & $0.270291$\\
    \hline
\end{tabular}
}\end{small}
\end{center}
\end{table}

\begin{table}[ht]
\caption{Fundamental quasinormal modes of the wormhole in Starobinsky model 
defined by the shape function \eqref{staro_shape} for the scalar perturbation 
with $\alpha=-0.3$, $\eta=0.5$ and $r_0=1$.}
\label{tab07}
\begin{center}
\begin{small}{
\begin{tabular}{ c c c c c}
    \hline
    \hline
    $l$  & WKB $3$rd order & WKB $5$th order & $\Delta_3$ & $\Delta_5$  \\
    \hline
    \hline
   \multirow{1}{*}{\;\;\;$l=1$\;\;\;} & $1.50652 - 0.401418 i$ & $1.51379 - 0.389575 i$ & $0.00693889$ & $0.00702422$ \\
    \multirow{1}{*}{$l=2$} & $2.49009 - 0.381394 i$ & $2.492 - 0.380002 i$ & $0.00230204$ & $0.000643374$\\
     \multirow{1}{*}{$l=3$} & $3.47658 - 0.375938 i$ & $3.47729 - 0.375589 i$ & $0.00114282$ & $0.000147775$ \\
     \multirow{1}{*}{$l=4$} & $4.46441 - 0.373693 i$ & $4.46474 - 0.373568 i$ & $0.000682844$ & $0.0000511278$\\
    \hline
\end{tabular}
}\end{small}
\end{center}
\end{table}

\begin{table}[ht]
\caption{Fundamental quasinormal modes of the wormhole in Starobinsky model 
defined by the shape function \eqref{staro_shape} for the Dirac perturbation 
with $\alpha=-0.3$, $\eta=0.5$ and $r_0=1$.}
\label{tab08}
\begin{center}
\begin{small}{
\begin{tabular}{ c c c c c}
    \hline
    \hline
    $k$  & WKB $3$rd order & WKB $5$th order & $\Delta_3$ & $\Delta_5$  \\
    \hline
    \hline
   \multirow{1}{*}{\;\;\;$k=16$\;\;\;} & $15.8877 - 0.376659  i$ & $15.9049 - 0.143713 i$ & $0.0105442$ & $1.77187$ \\
    \multirow{1}{*}{$k=17$} & $16.8783 - 0.376026 i$ & $16.8918 - 0.203532 i$ & $0.00829676$ & $1.18099$\\
     \multirow{1}{*}{$k=18$} & $17.8688 - 0.375502  i$ & $17.8796 - 0.245583 i$ & $0.00661801$ & $0.819652$ \\
     \multirow{1}{*}{$k=19$} & $18.8592 - 0.375063 i$ & $18.8681 - 0.275713 i$ & $0.0053437$ & $0.585226$\\
    \hline
\end{tabular}
}\end{small}
\end{center}
\end{table}

\begin{figure}[htbp]
\centerline{
   \includegraphics[scale = 0.3]{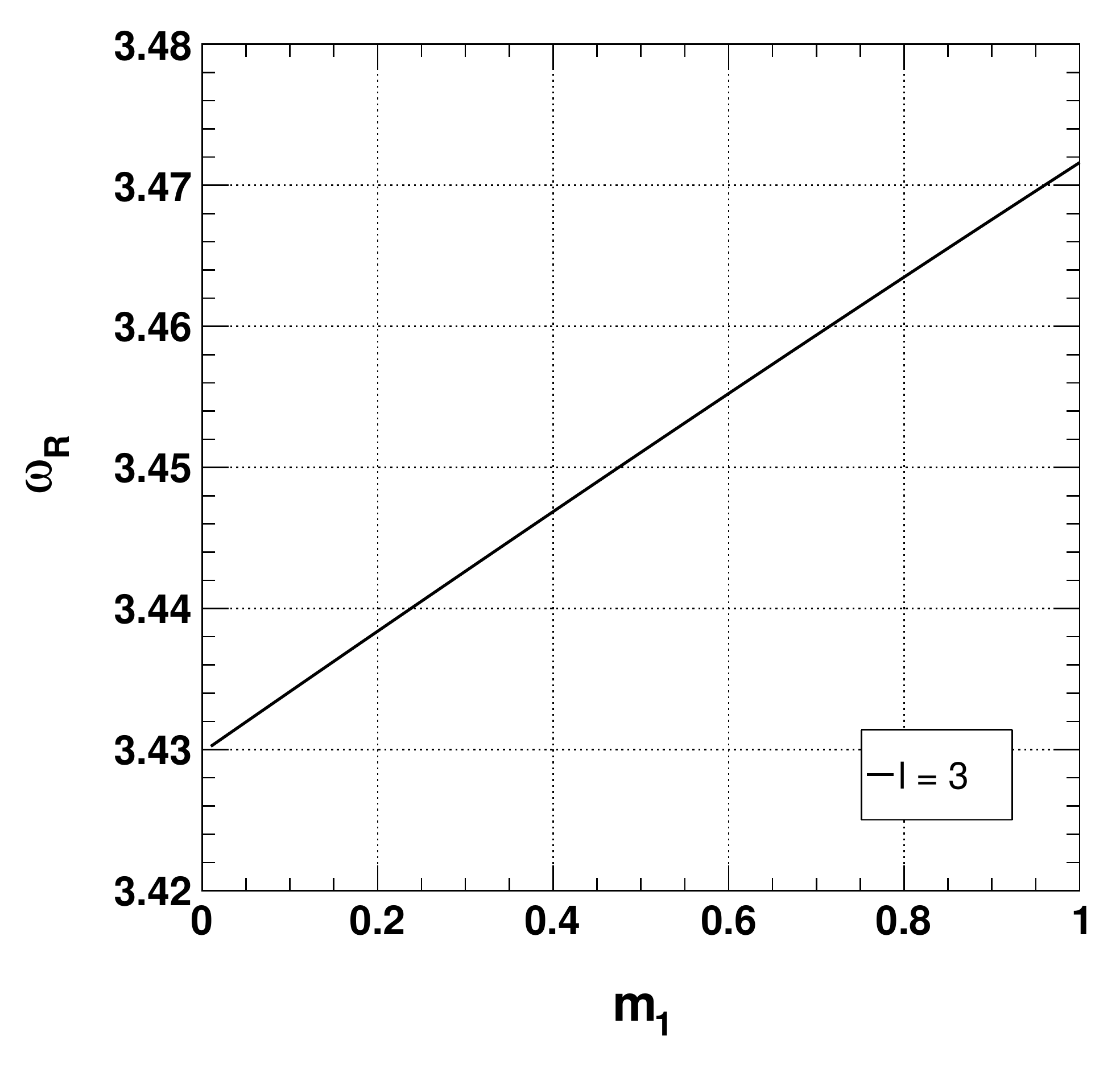}\hspace{1cm}
   \includegraphics[scale = 0.3]{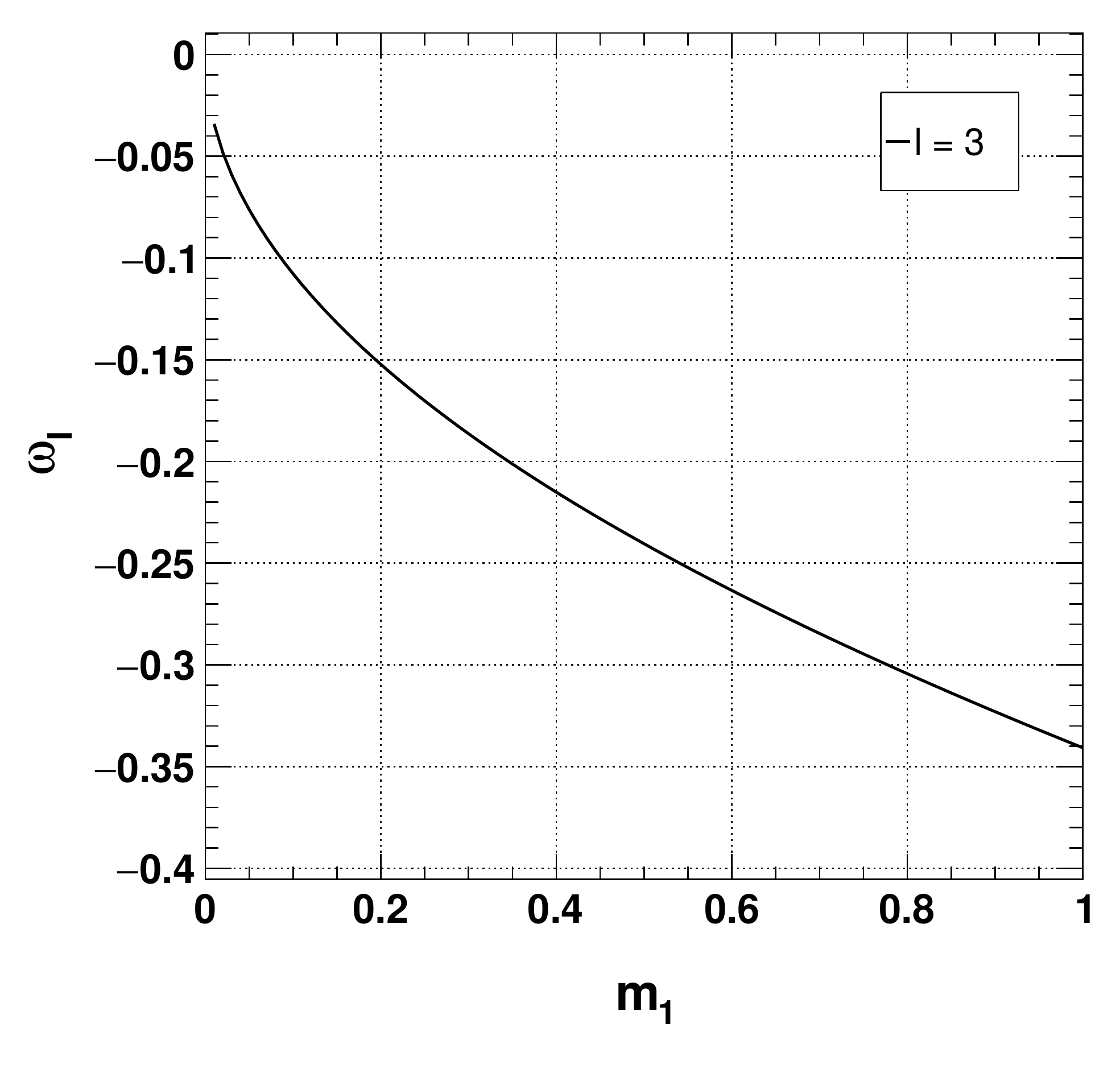}}
\vspace{-0.2cm}
\caption{Variation of the fundamental scalar quasinormal modes with model 
parameter $m_1$ for the wormhole defined by the shape function 
\eqref{toy_shape01}, calculated with 3rd order WKB approximation method.}
\label{toy01_qnms_scalar}
\end{figure}

\begin{figure}[htbp]
\centerline{
   \includegraphics[scale = 0.3]{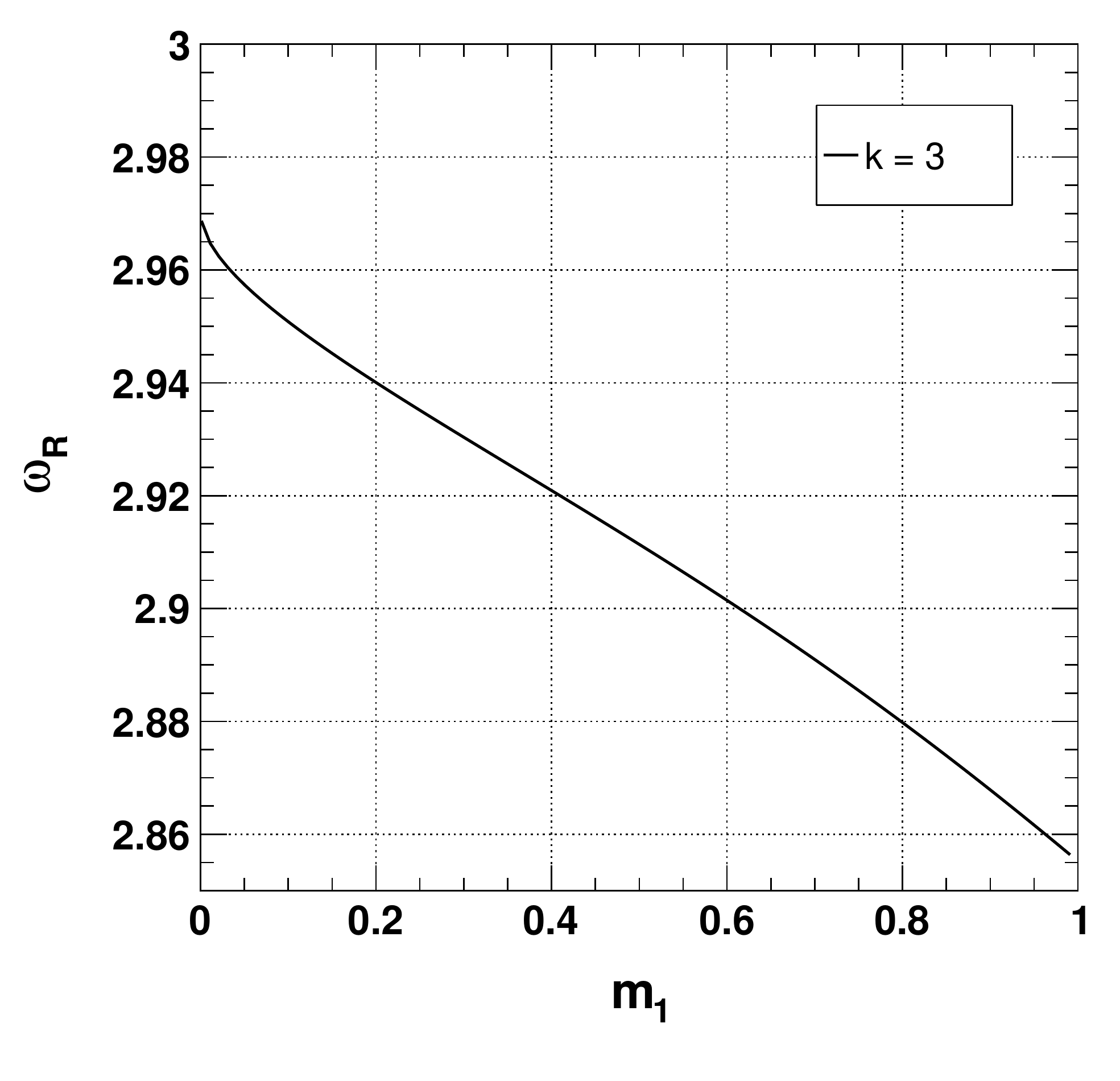}\hspace{1cm}
   \includegraphics[scale = 0.3]{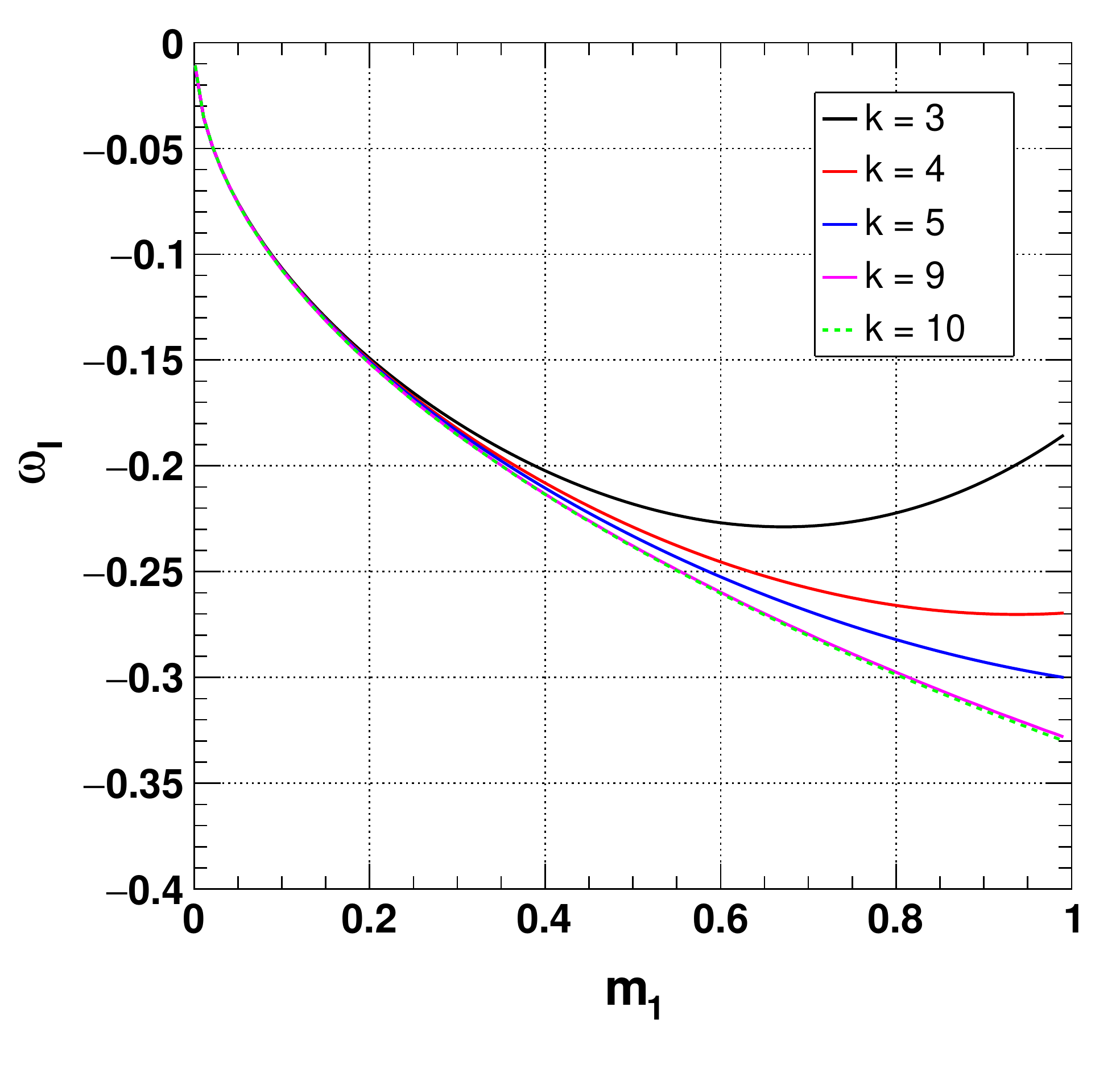}}
\vspace{-0.2cm}
\caption{Variation of the fundamental Dirac quasinormal modes with model
parameter $m_1$ for the wormhole defined by the shape function \eqref{toy_shape01}, calculated with 3rd order WKB approximation method.}
\label{toy01_qnms_dirac}
\end{figure}

\begin{figure}[htbp]
\centerline{
   \includegraphics[scale = 0.3]{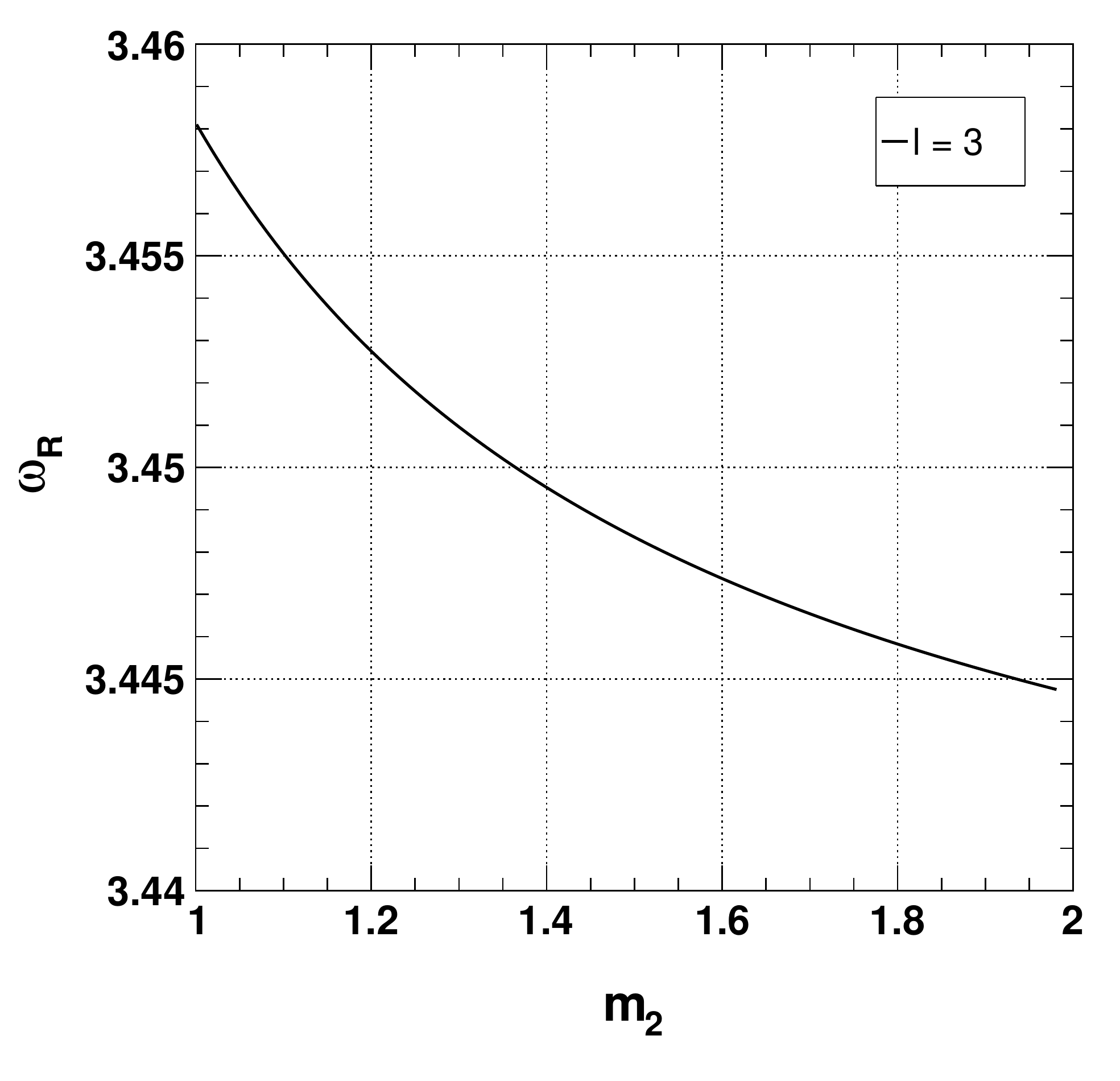}\hspace{1cm}
   \includegraphics[scale = 0.3]{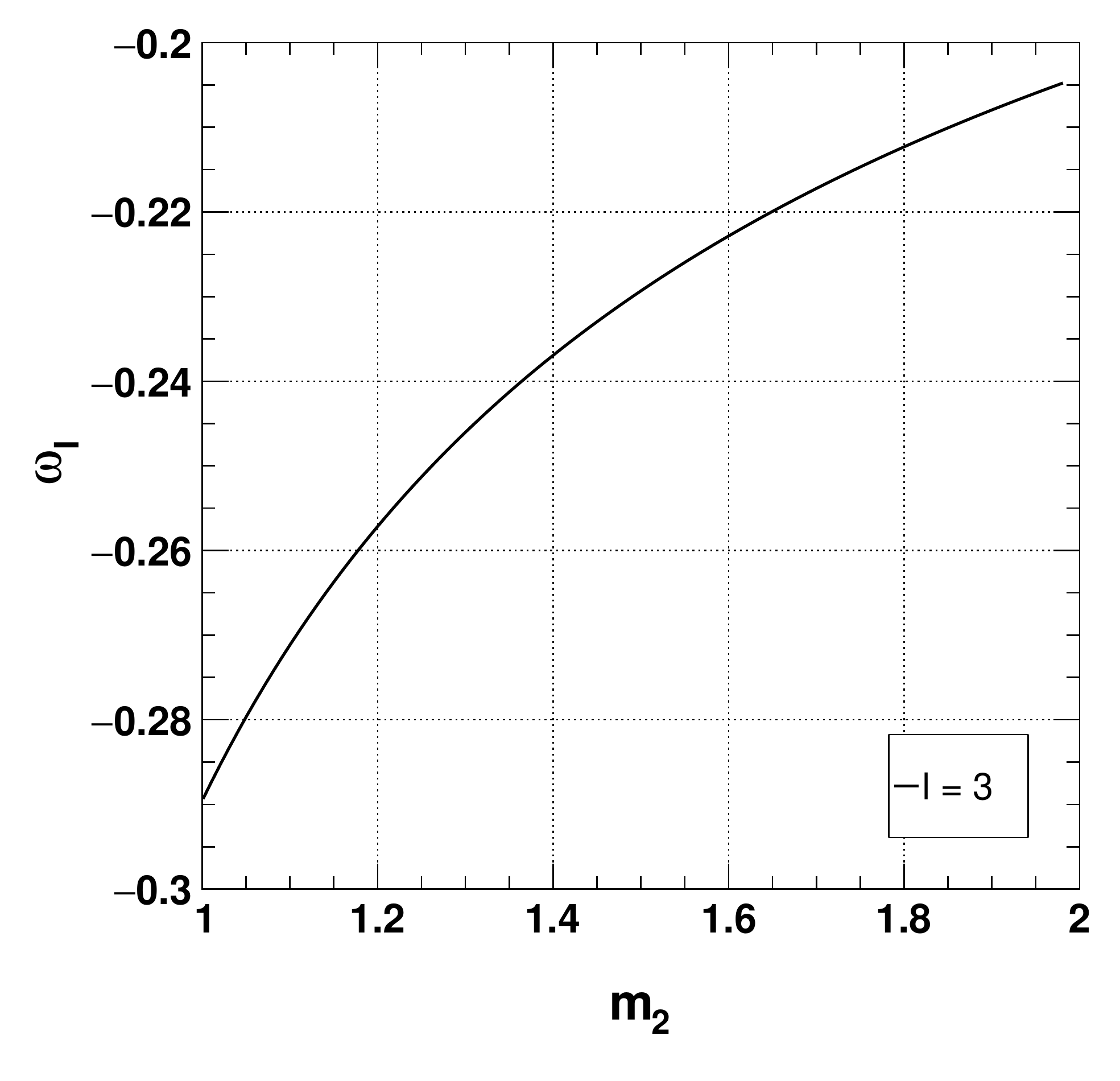}}
\vspace{-0.2cm}
\caption{Variation of the fundamental scalar quasinormal modes with model 
parameter $m_2$ for the wormhole defined by the shape function 
\eqref{toy_shape02}, calculated with 3rd order WKB approximation method.}
\label{toy02_qnms_scalar}
\end{figure}

\begin{figure}[htbp]
\centerline{
   \includegraphics[scale = 0.3]{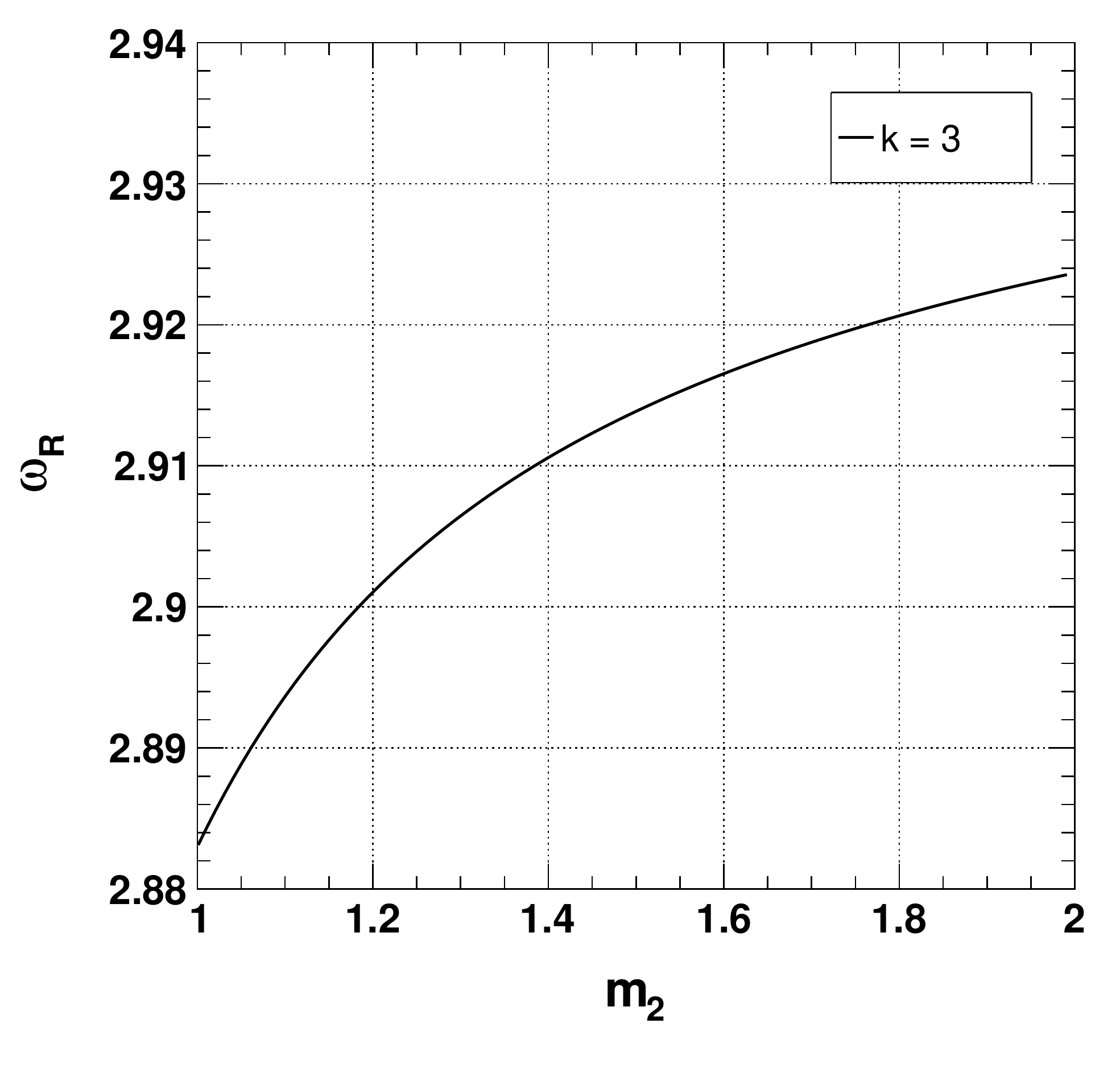}\hspace{1cm}
   \includegraphics[scale = 0.3]{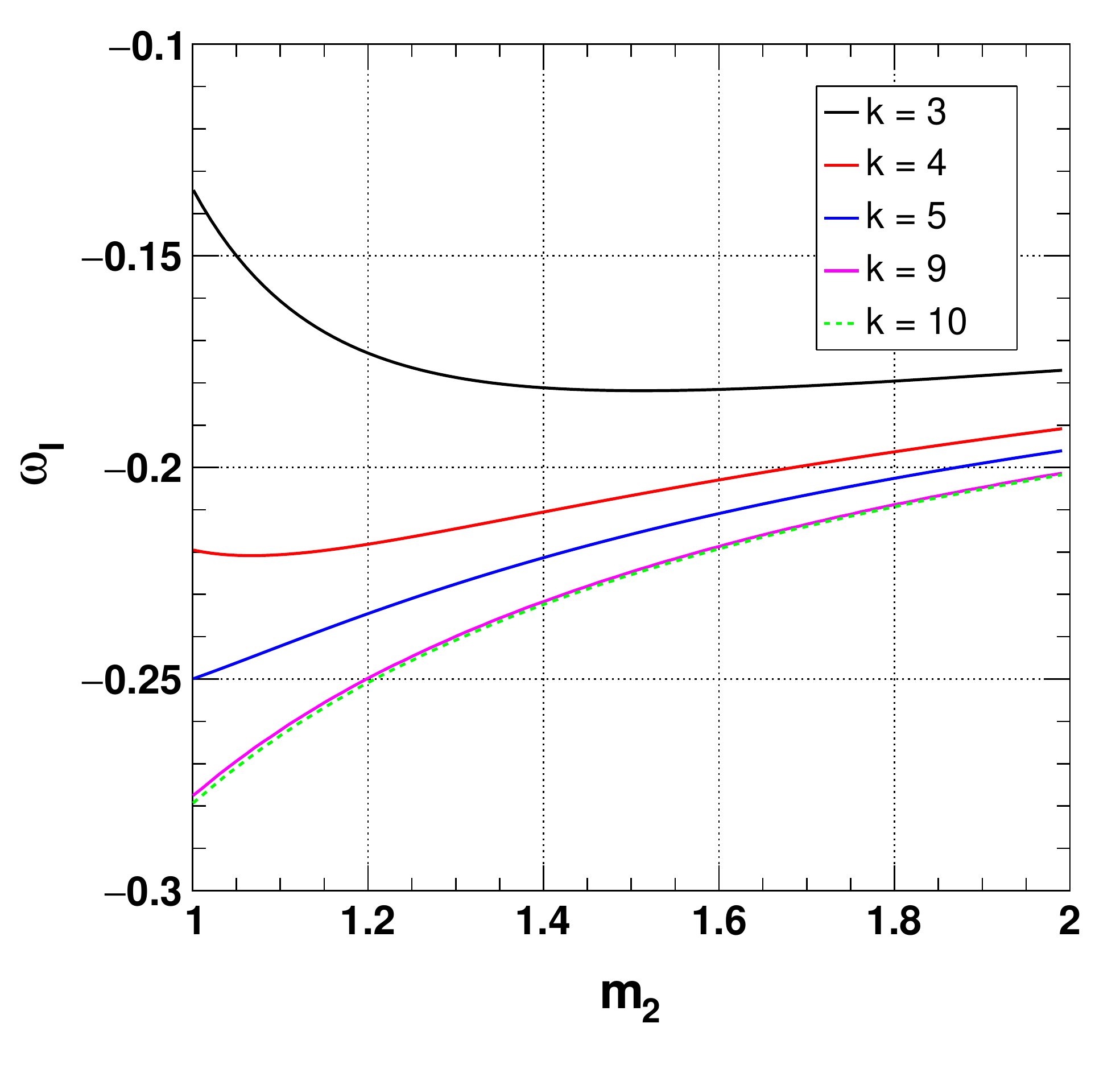}}
\vspace{-0.2cm}
\caption{Variation of the fundamental Dirac quasinormal modes with model
parameter $m_2$ for the wormhole defined by the shape function
\eqref{toy_shape02}, calculated with 3rd order WKB approximation method.}
\label{toy02_qnms_dirac}
\end{figure}

\begin{figure}[htbp]
\centerline{
   \includegraphics[scale = 0.3]{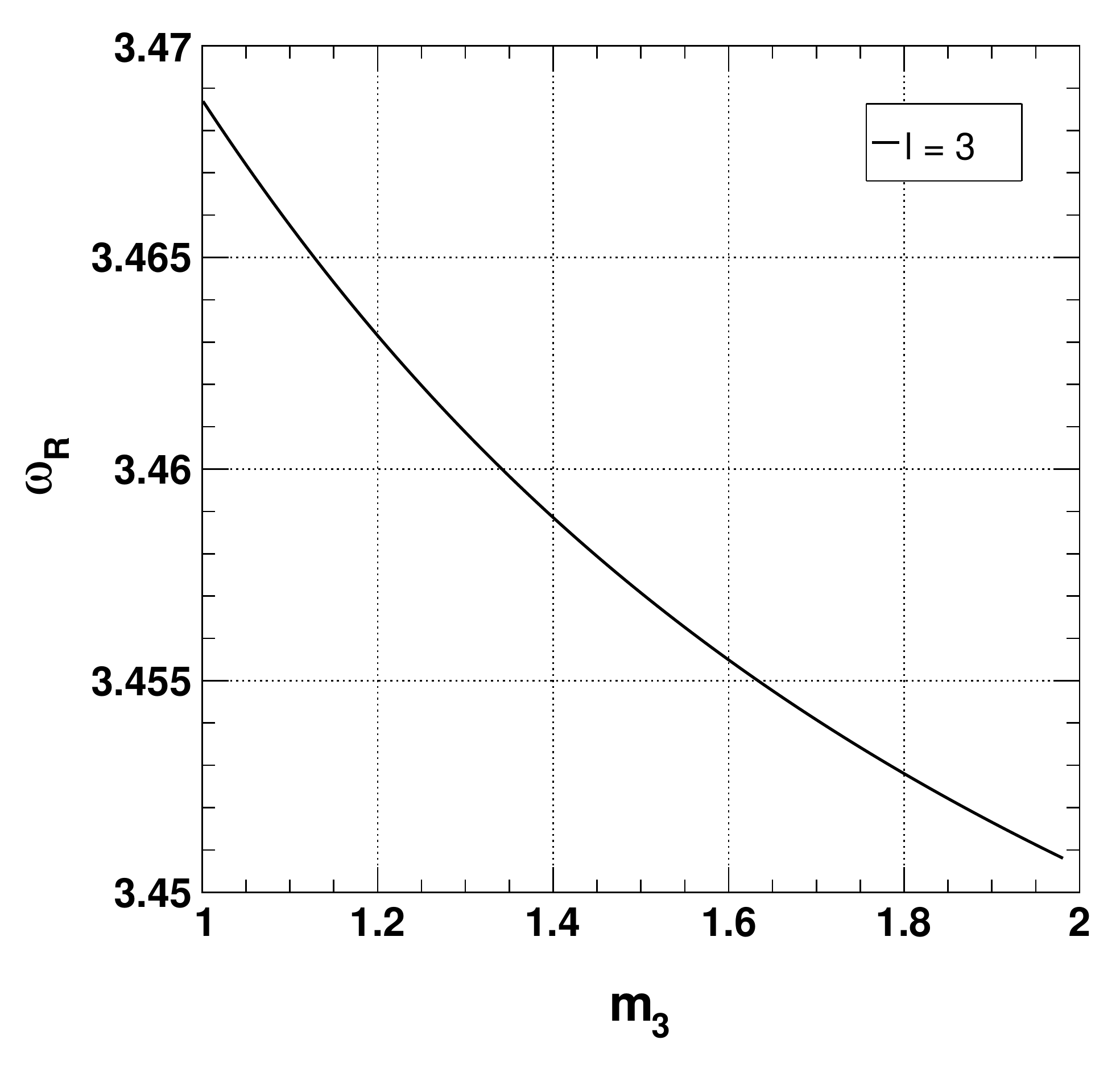}\hspace{1cm}
   \includegraphics[scale = 0.3]{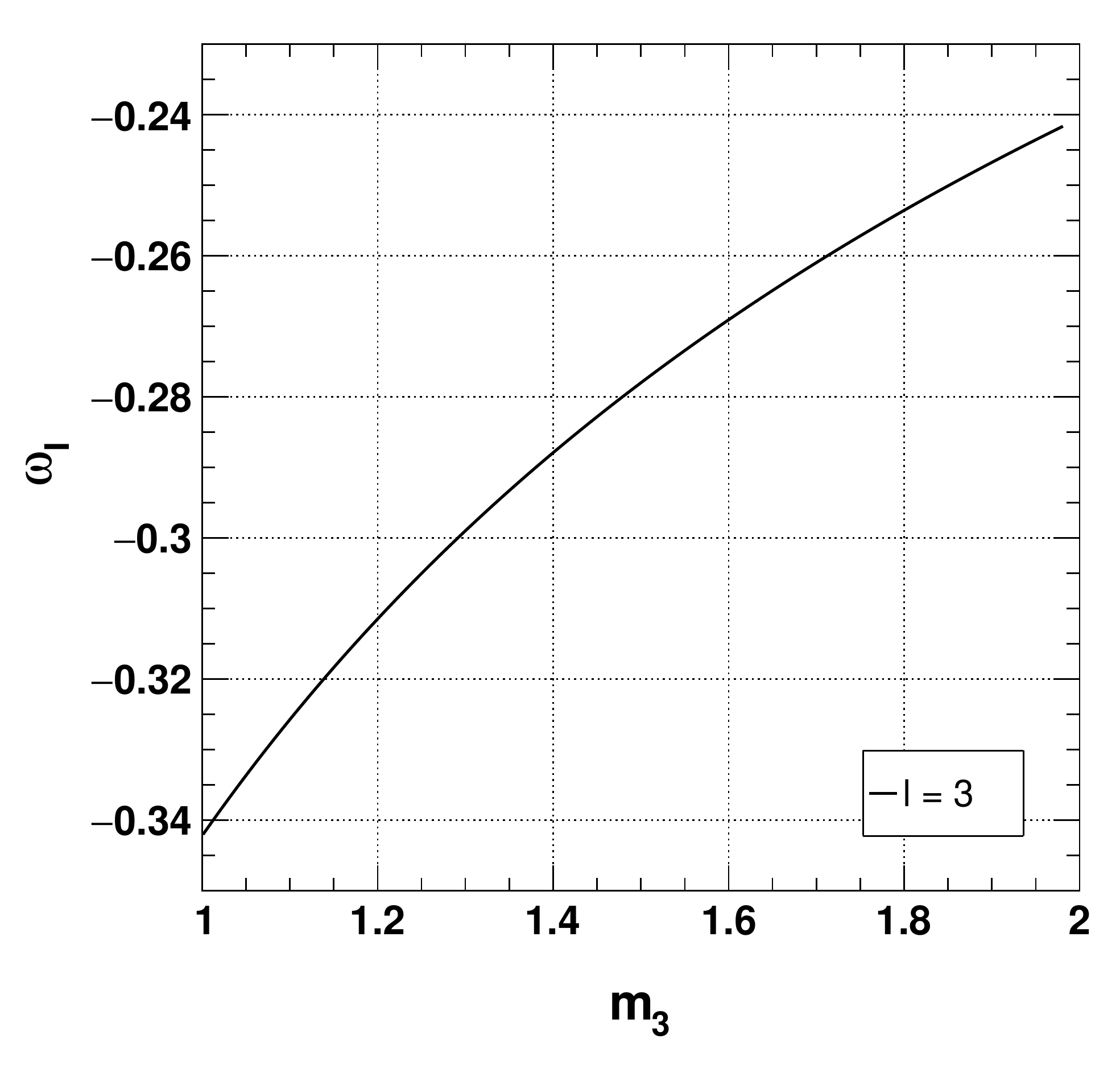}}
\vspace{-0.2cm}
\caption{Variation of the fundamental scalar quasinormal modes with model 
parameter $m_3$ for the wormhole defined by the shape function 
\eqref{toy_shape03}, calculated with 3rd order WKB approximation method.}
\label{toy03_qnms_scalar}
\end{figure}

\begin{figure}[htbp]
\centerline{
   \includegraphics[scale = 0.3]{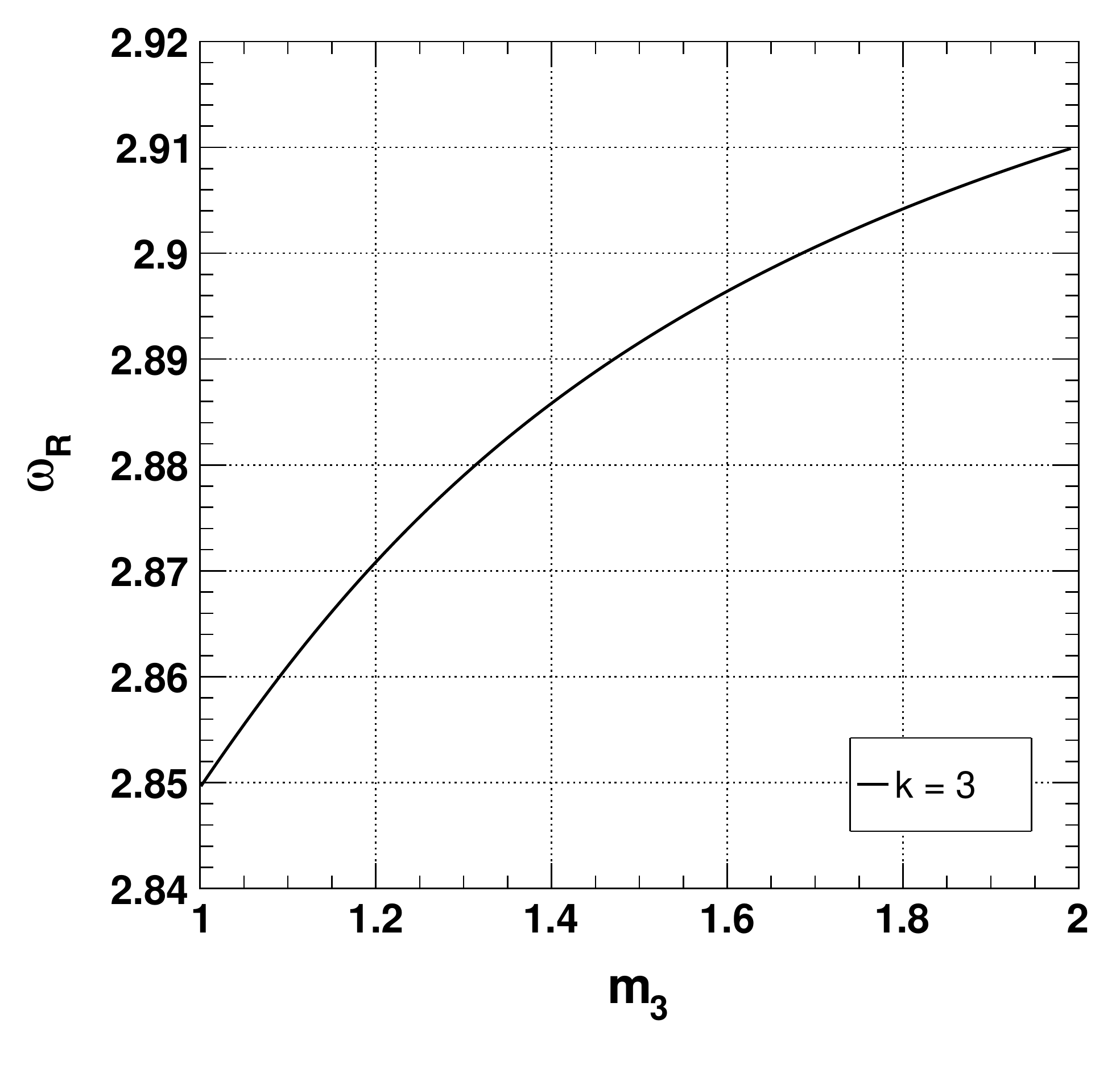}\hspace{1cm}
   \includegraphics[scale = 0.3]{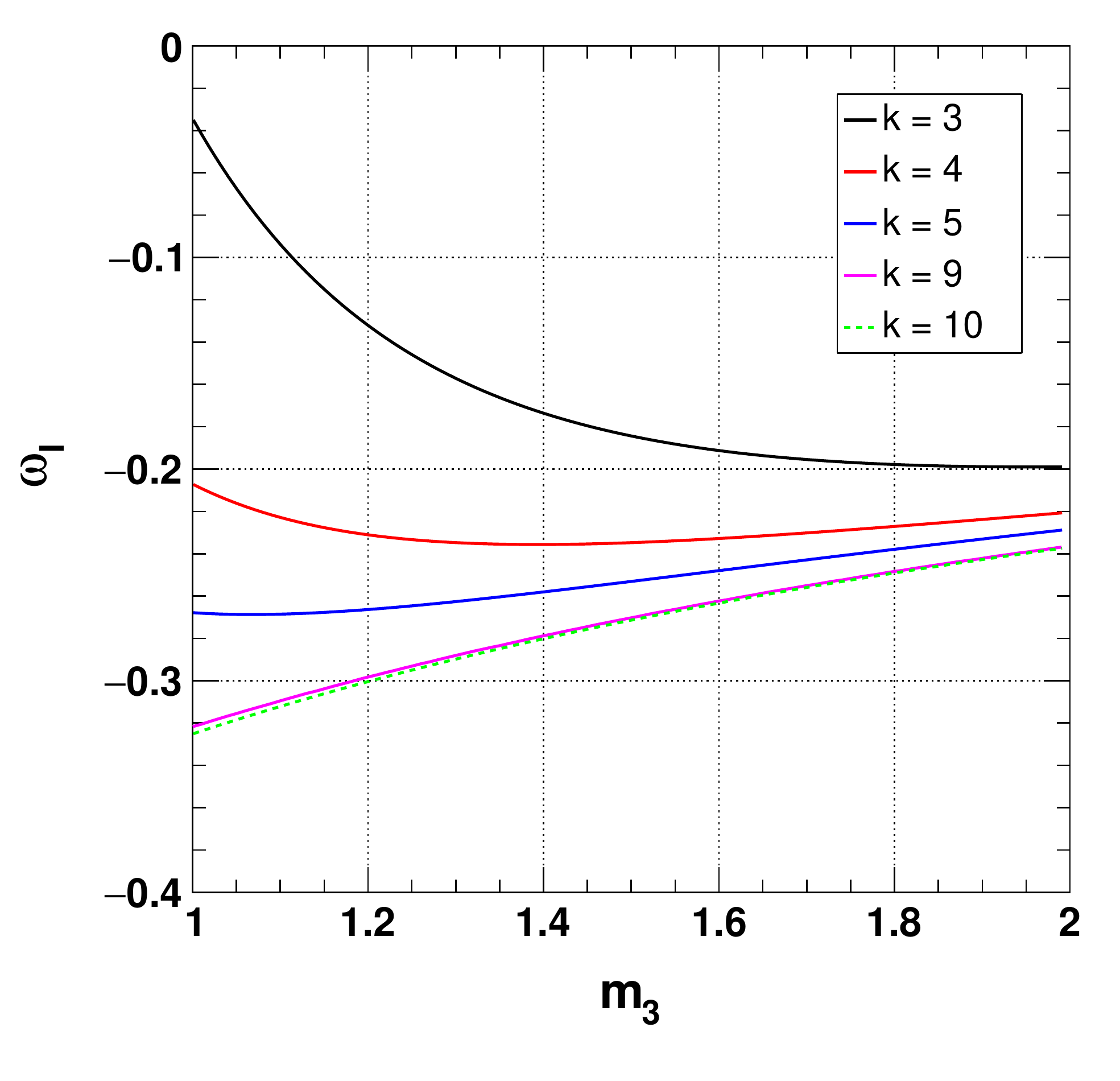}}
\vspace{-0.2cm}
\caption{Variation of the fundamental Dirac quasinormal modes with model 
parameter $m_3$ for the wormhole defined by the shape function 
\eqref{toy_shape03}, calculated with 3rd order WKB approximation method.}
\label{toy03_qnms_dirac}
\end{figure}

\begin{figure}[htbp]
\centerline{
   \includegraphics[scale = 0.3]{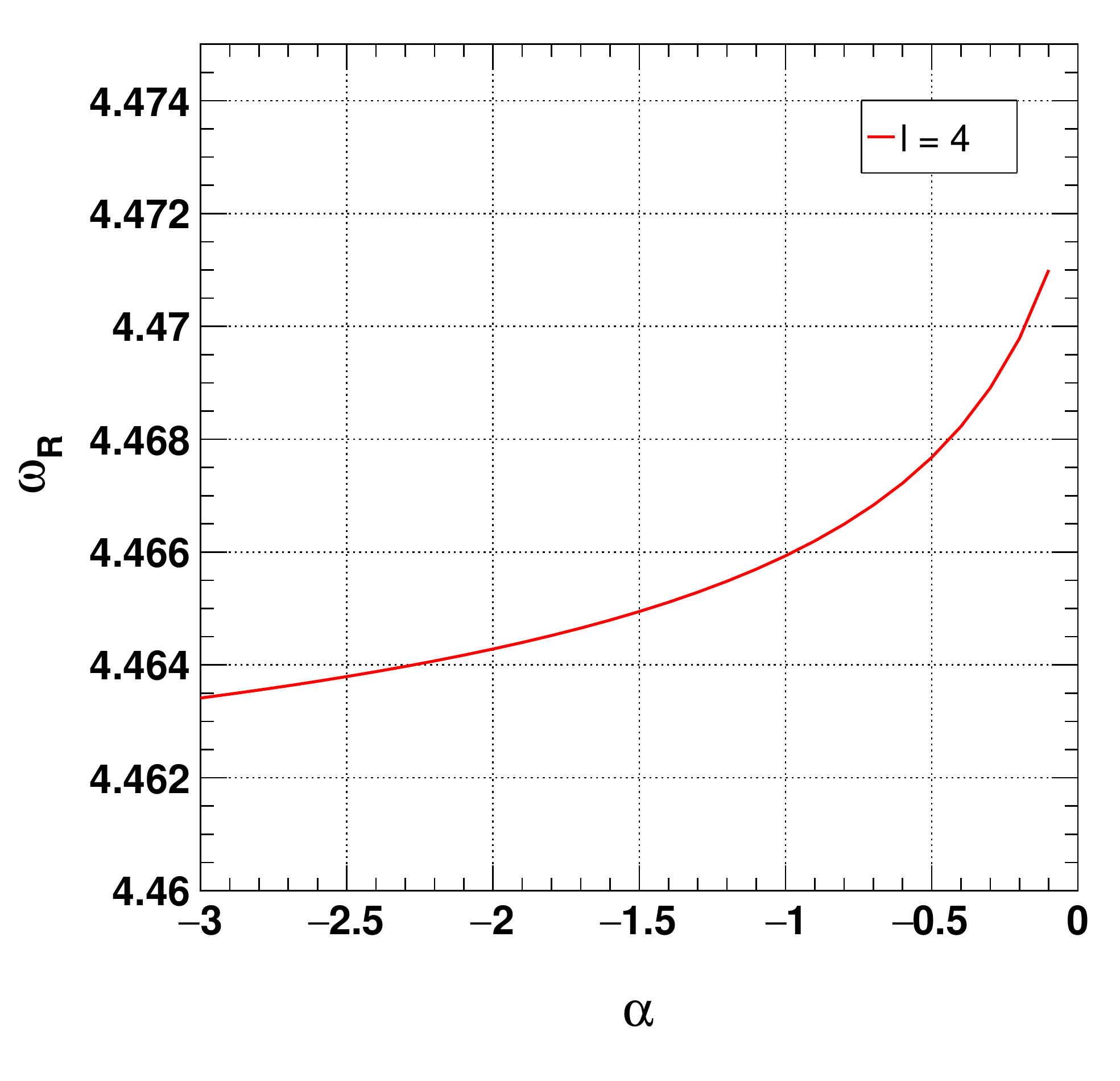}\hspace{1cm}
   \includegraphics[scale = 0.3]{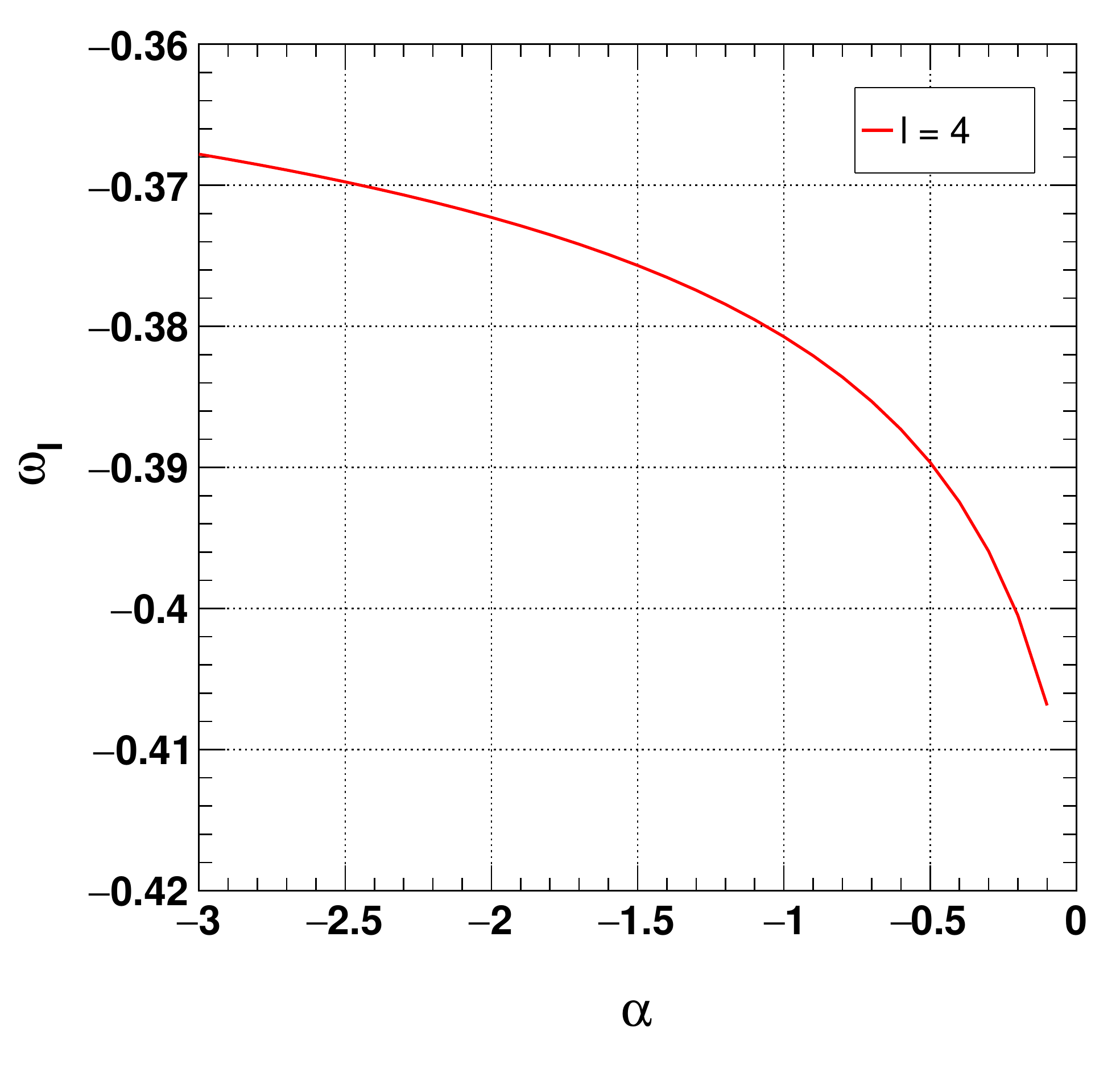}}
\vspace{-0.2cm}
\caption{Variation of the fundamental scalar quasinormal modes with model 
parameter $\alpha$ for the wormhole defined by the shape function 
\eqref{staro_shape}, calculated with 3rd order WKB approximation method using 
$\eta = 0.7$ and $r_0 = 1$.}
\label{staro_qnms_scalar_a}
\end{figure}

\begin{figure}[htbp]
\centerline{
   \includegraphics[scale = 0.3]{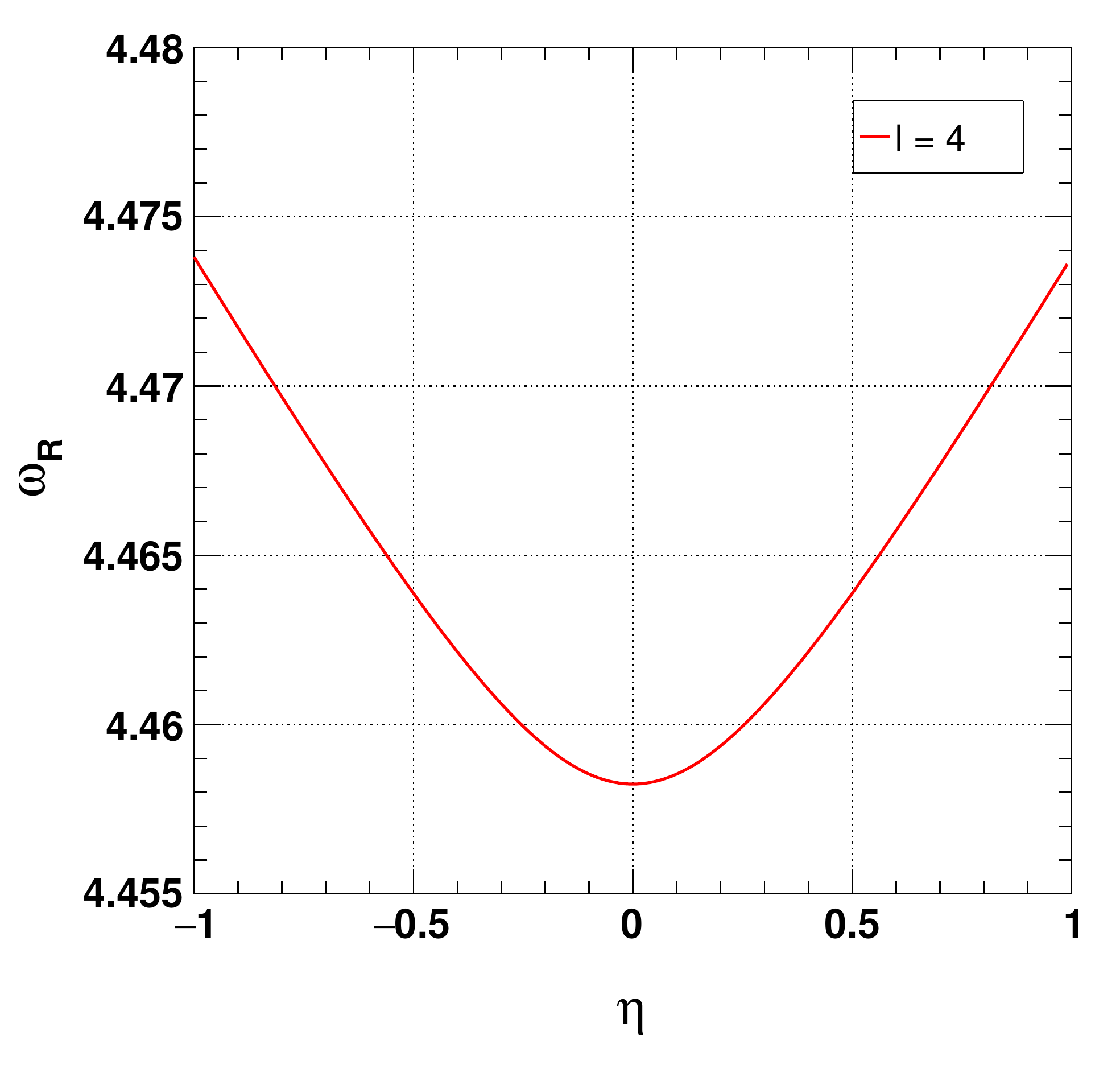}\hspace{1cm}
   \includegraphics[scale = 0.3]{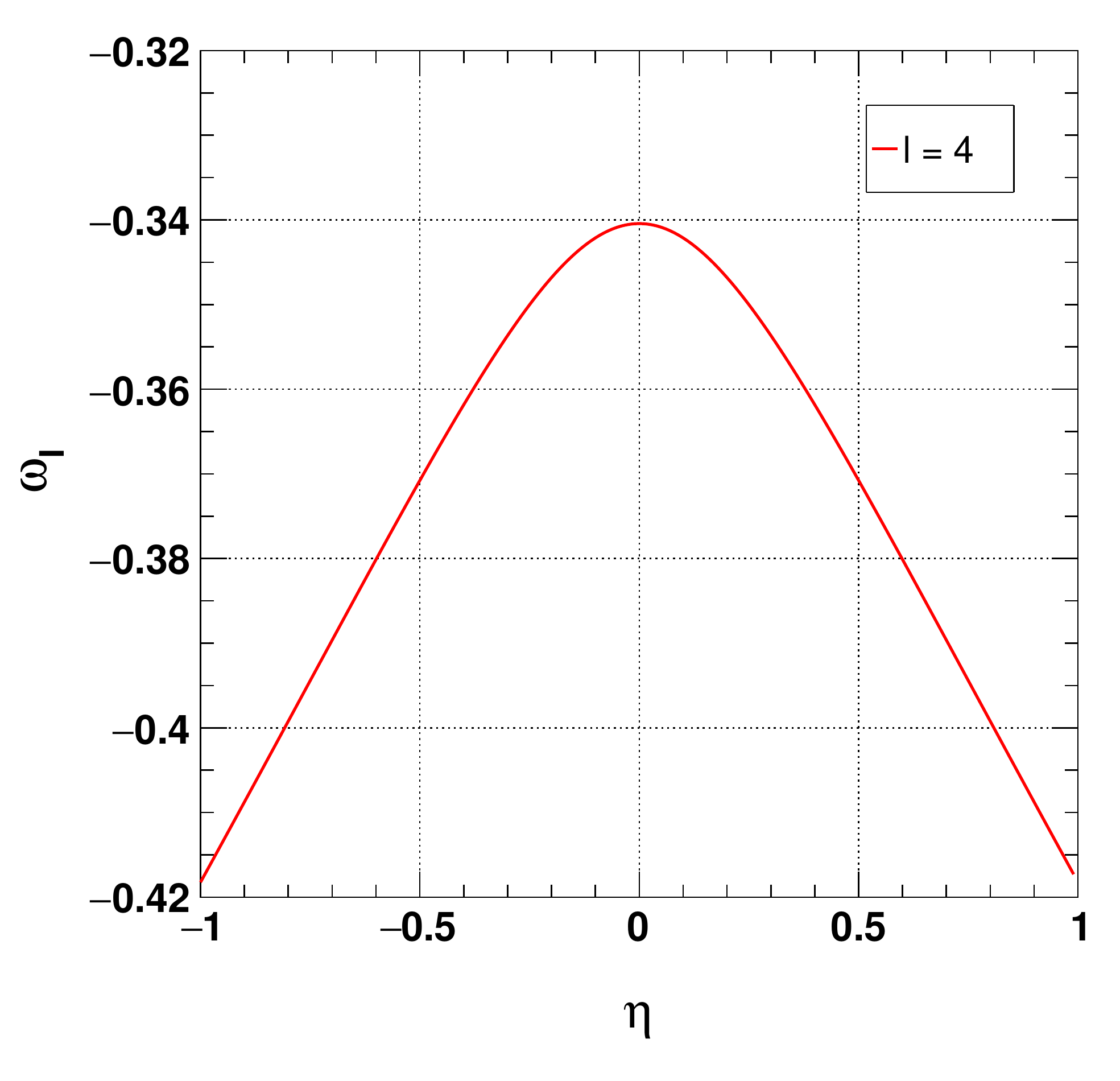}}
\vspace{-0.2cm}
\caption{Variation of the fundamental scalar quasinormal modes with model 
parameter $\eta$ for the wormhole defined by the shape function 
\eqref{staro_shape}, calculated with 3rd order WKB approximation method 
using $\alpha = -0.5$ and $r_0 = 1$.}
\label{staro_qnms_scalar_eta}
\end{figure}

\begin{figure}[htbp]
\centerline{
   \includegraphics[scale = 0.3]{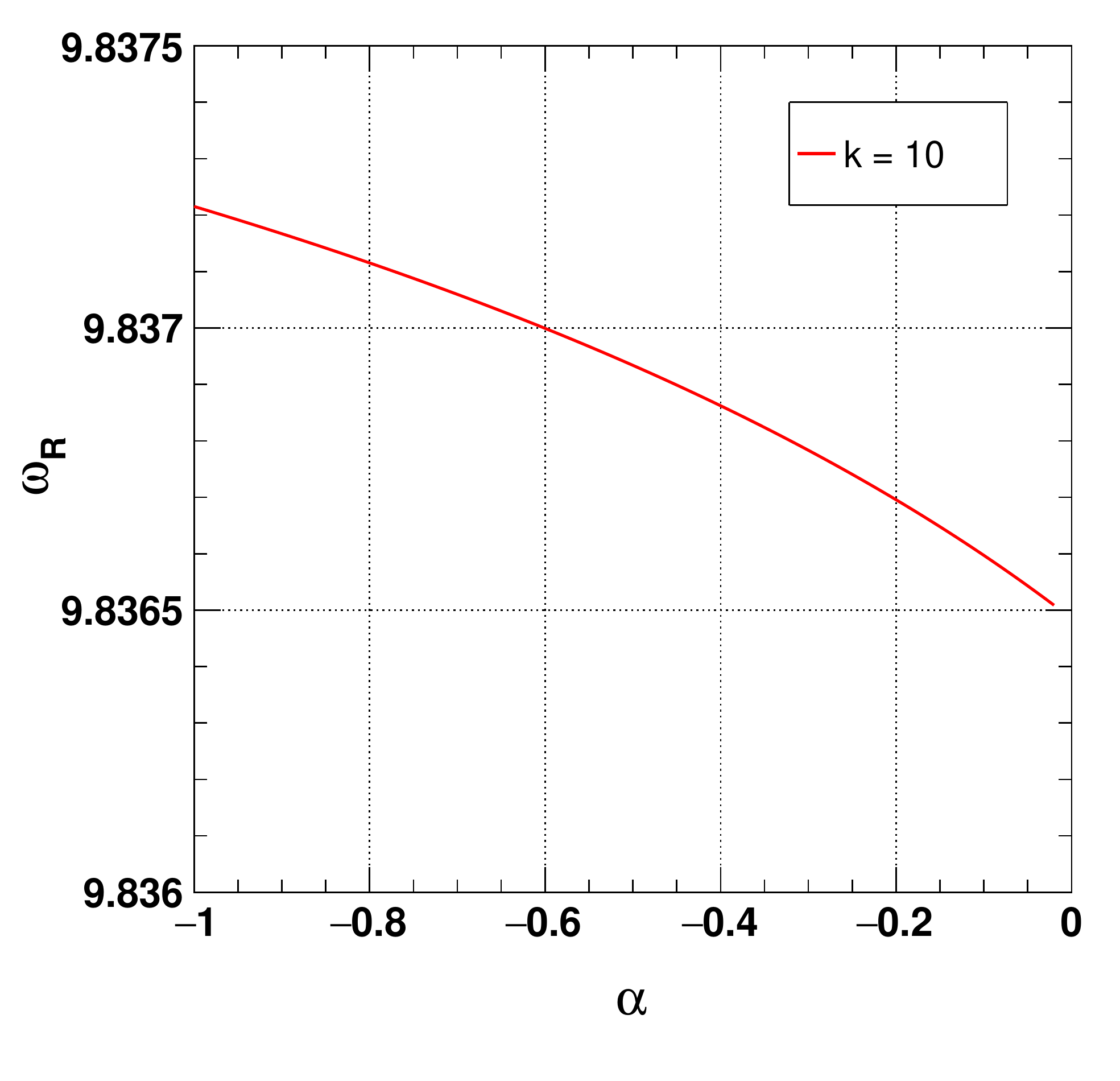}\hspace{1cm}
   \includegraphics[scale = 0.3]{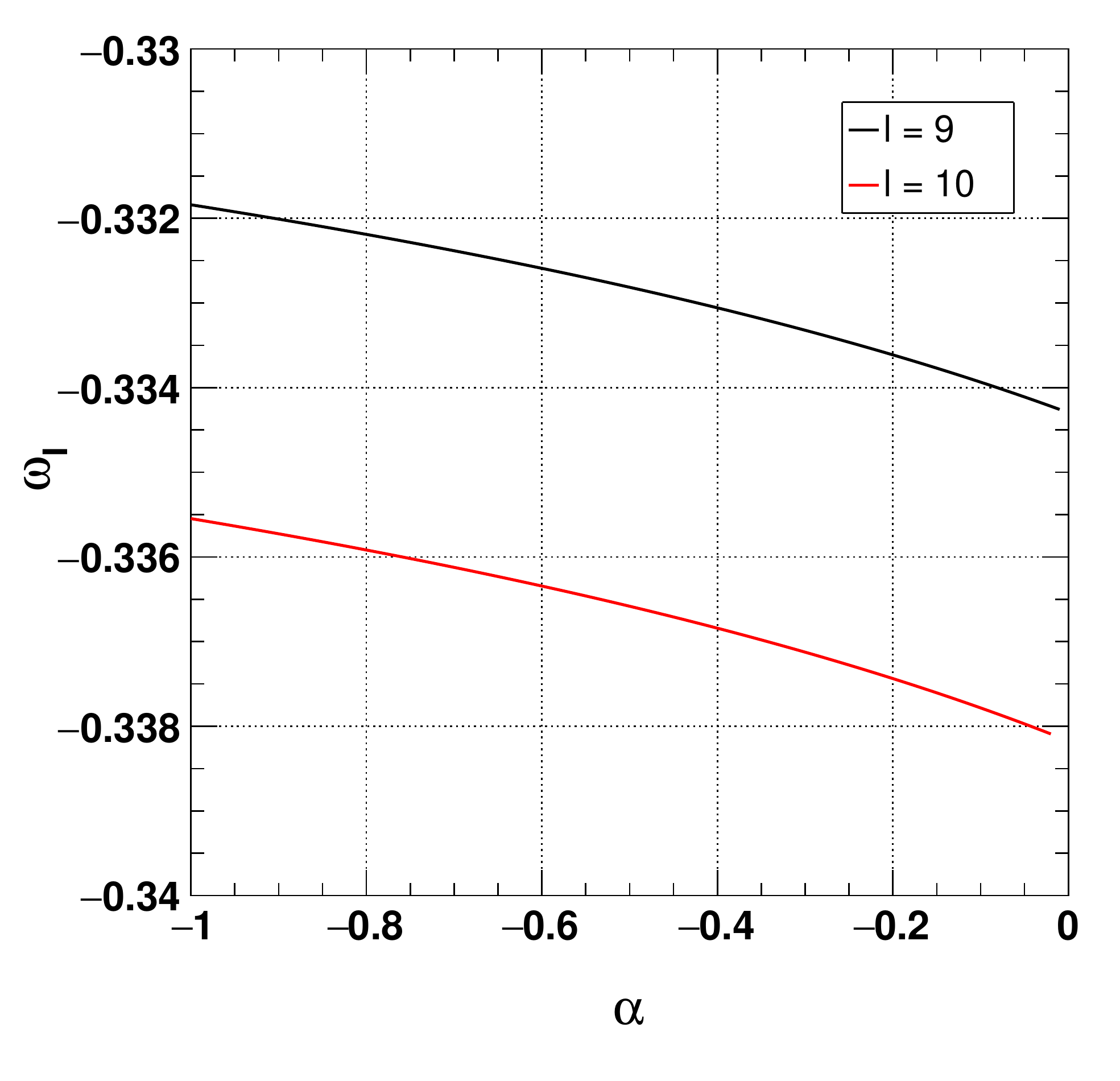}}
\vspace{-0.2cm}
\caption{Variation of the fundamental Dirac quasinormal modes with model 
parameter $\alpha$ for the wormhole defined by the shape function 
\eqref{staro_shape}, calculated with 3rd order WKB approximation method 
using $\eta = 0.3$ and $r_0 = 1$.}
\label{staro_qnms_dirac_a}
\end{figure}

\begin{figure}[h!]
\centerline{
   \includegraphics[scale = 0.3]{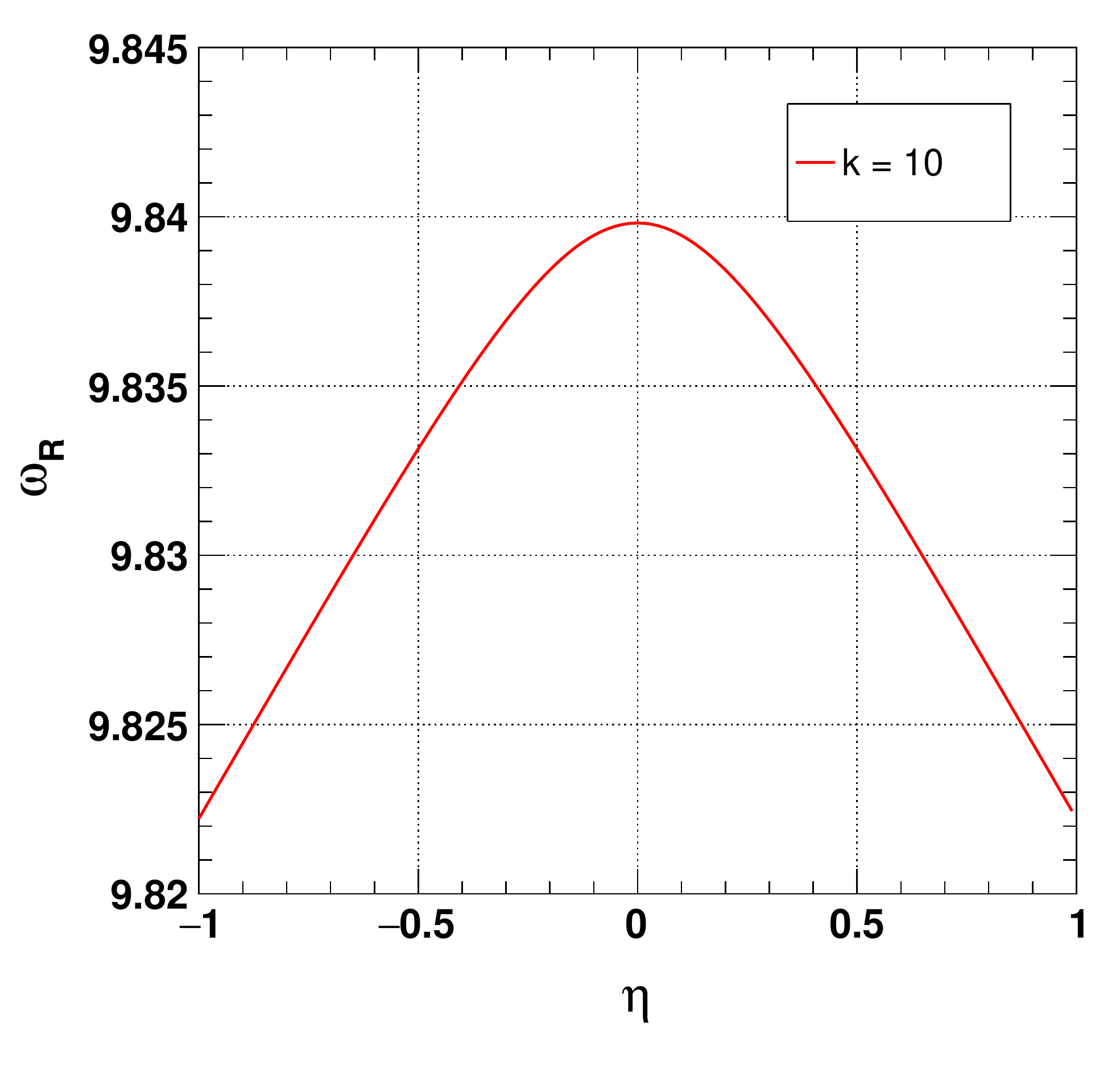}\hspace{1cm}
   \includegraphics[scale = 0.3]{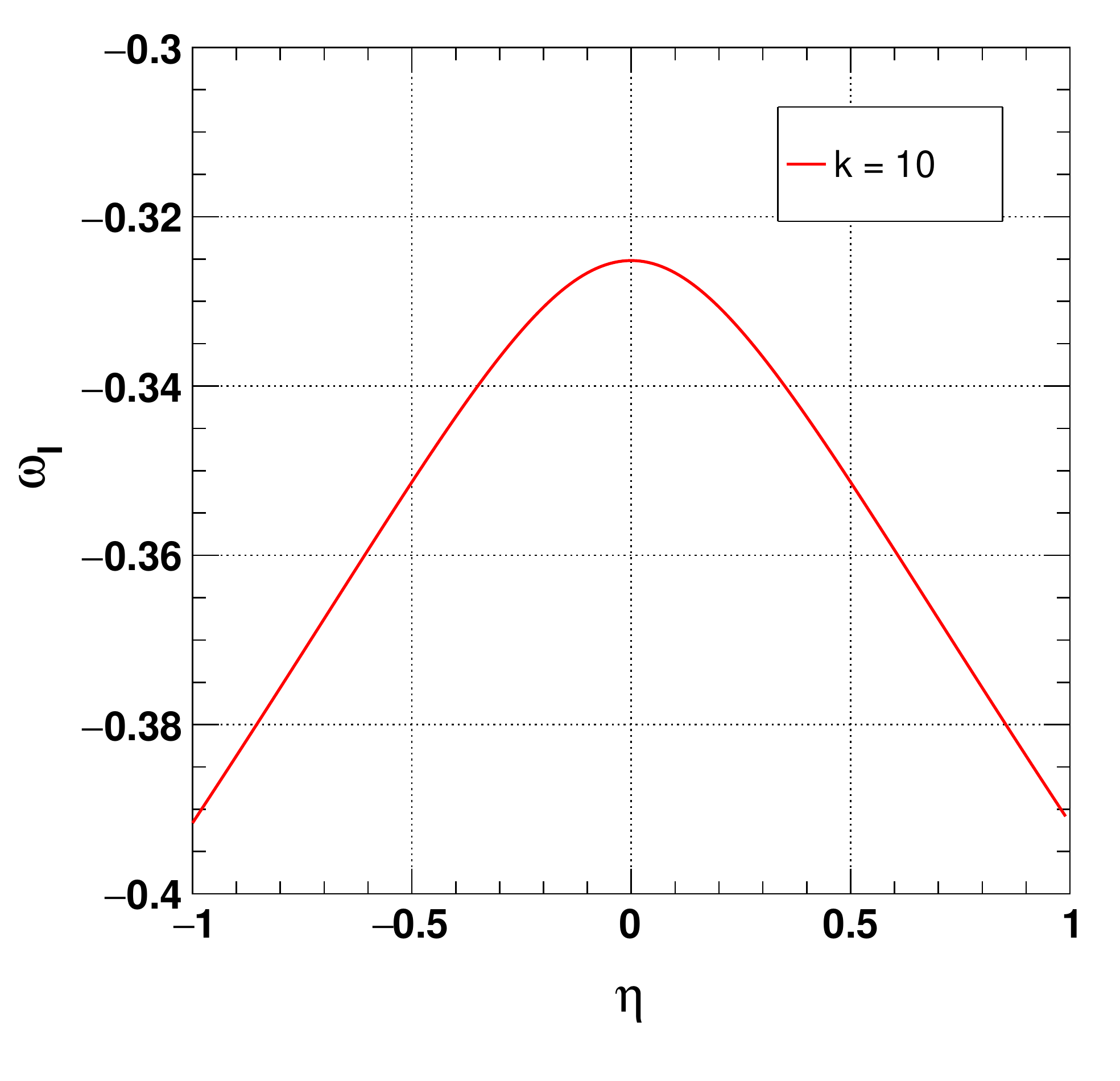}}
\vspace{-0.2cm}
\caption{Variation of the fundamental Dirac quasinormal modes with model 
parameter $\eta$ for the wormhole defined by the shape function 
\eqref{staro_shape}, calculated with 3rd order WKB approximation method 
using $\alpha = -0.5 $ and $r_0 = 1$.}
\label{staro_qnms_dirac_eta}
\end{figure}

At first, we calculated the quasinormal modes for scalar perturbation for 
the wormhole defined by the shape function \eqref{toy_shape01} using $m_1=0.3$ 
and throat $r_0=1$. The value of the parameter $m_1$ is chosen in such a way 
that it satisfies the viability conditions for a wormhole mentioned earlier. 
The results for different multipole numbers are shown in Table \ref{tab01}. In
this table and following ones $\Delta_3$ is defined as
\begin{equation}
\Delta_3 = \dfrac{\vline \; \omega_4 - \omega_2 \; \vline}{2},
\end{equation}
and similarly,
\begin{equation}
\Delta_5 = \dfrac{\vline \; \omega_6 - \omega_4 \; \vline}{2},
\end{equation}
where $\omega_2, \omega_4$ and $\omega_6$ represent the quasinormal modes 
obtained from 2nd order, 4th order and 6th order WKB approximation methods 
respectively. $\Delta_3$ and $\Delta_5$ give a measurement of the errors 
associated with the quasinormal modes obtained from 3rd order and 5th order WKB approximation methods. It is clearly visible from the Table \ref{tab01} that 
with an increase in $l$, magnitudes of both the real and imaginary frequencies 
increase gradually. So, for higher multipole numbers the decay rate of the 
quasinormal frequency increases slowly. Another point to note is that the 
corresponding errors also decrease as the value of $l$ increases. In this 
case, 5th order WKB method seems to be more accurate with less errors. For the 
same wormhole, we have calculated the Dirac quasinormal modes in Table 
\ref{tab02}. In this scenario, for lower $k$ values, the WKB approximation 
gives less accurate results due to the behaviour of the corresponding 
potentials. Hence, we have listed the quasinormal modes for some higher 
values of $k$. In this case, considering the 3rd order WKB results which have 
less errors, with an increase in the value of $k$, we observe a decrease 
in the decay rate of the quasinormal frequencies while the real part increases 
gradually following the previous trend of scalar perturbation. One may note 
that in case of Dirac perturbation, 5th order WKB method is less accurate in 
comparison to the 3rd order WKB method. The errors, as expected, decrease with 
increase in the value of $k$. Similarly, we have calculated the quasinormal 
modes for the second wormhole \eqref{toy_shape02} and listed them in Table 
\ref{tab03} for scalar perturbation. In this case, with an increase in the 
multipole number $l$, the decay rate decreases following an opposite trend 
in comparison to the first case. For this wormhole, the Dirac quasinormal 
modes are listed in Table \ref{tab04}. For the third wormhole, defined by 
Eq.~\eqref{toy_shape03}, the scalar and Dirac quasinormal modes are listed 
in Tables \ref{tab05} and \ref{tab06} respectively. For this case also, the 
scalar quasinormal modes decrease with an increase in the value of $l$. 

To get a clear picture, we have plotted the real and imaginary quasinormal 
modes of the wormhole \eqref{toy_shape01} with respect to the model parameter 
$m_1$ in Fig.~\ref{toy01_qnms_scalar} for scalar perturbations and in 
Fig.~\ref{toy01_qnms_dirac} for Dirac perturbations using 3rd order WKB 
approximation method. The reason for choosing 3rd order WKB method in the 
plots can be justified from the previous tables, where we have seen that in 
general the 5th order WKB method results in higher magnitudes of errors 
basically in Dirac perturbations. In Fig.~\ref{toy01_qnms_scalar}, it is 
seen that the 
real quasinormal mode increases linearly with respect to the model parameter 
$m_1$. However, at the same time, the decay rate also increases following a 
non-linear pattern. The decay rate is smaller when the parameter $m_1$ 
negligibly small. The Dirac quasinormal modes are shown in 
Fig.~\ref{toy01_qnms_dirac}. The real quasinormal modes decrease with an 
increase in the model parameter $m_1$. However, in case of decay rates, we 
observe a multipole dependency towards $m_1=1$. Initially, for all multipole 
modes, decay rate increases with an increase in $m_1$. But, for smaller 
multipole modes, a reverse pattern is observed near $m_1=1$. The decay rates 
for $l=3$ start increasing gradually near $m_1=1$. In the case of $l=4$, the 
slope of the decay rate curve becomes very small near $m_1=1$. However, for 
higher multipole modes, such a pattern vanishes and the decay rates become 
almost identical to each other.

We have shown the variation of the scalar quasinormal modes for the second 
shape function defined by Eq.~\eqref{toy_shape02} in 
Fig.~\ref{toy02_qnms_scalar}. Here, we have chosen the range of the parameter 
$m_2$ in such a way that the shape function satisfies the viability 
conditions. It is seen that an increase in the parameter from $m_2=1$ to $2$ 
results in decrease of the quasinormal frequencies and at the same time the 
decay rate also decreases gradually. In case of Dirac quasinormal modes 
(see Fig.~\ref{toy02_qnms_dirac}), however, an increase in the parameter $m_2$ 
results in an increase of the quasinormal frequencies. But in decay rate, we 
observe an anomalous pattern which depends on the multipole number $l$. For 
$l=3$, the decay rate initially increases with an increase in $m_2$ and beyond 
$m_2=1.4$, the decay rate starts to decrease very slowly with an increase in 
$m_2$. For $l=4$, decay rate increases very slowly near $m_2=1.1$ and beyond 
this, decay rate decreases. In the case of $l=5$ and above, the decay rate only 
decreases gradually from $m_2=1$ to $2$. It may be noted that for higher $l$ 
values, the decay rates are almost identical to each other and they start 
decreasing with an increase in $m_2$ beyond $1$. For the third shape function 
\eqref{toy_shape03} also, the scalar quasinormal modes follow a similar 
pattern to that of the second shape function as seen from the 
Fig.~\ref{toy03_qnms_scalar}. In case of the Dirac quasinormal modes as seen 
from Fig.~\ref{toy03_qnms_dirac}, the quasinormal frequencies increase with 
increase in the parameter $m_3$ and the decay rate follows a similar trend as 
that of the Dirac perturbation case for the shape function 
\eqref{toy_shape02}. Hence, the analysis shows that the shape functions 
\eqref{toy_shape02} and \eqref{toy_shape03} show a similar behaviour in terms 
of the quasinormal modes of the wormholes in both scalar and Dirac quasinormal 
modes. 

We now move to the new shape function that has been obtained from the 
Starobinsky $f(R)$ gravity model surrounded by a cloud of strings. We have 
calculated the quasinormal modes for different values of multipole moments in 
Tables \ref{tab07} and \ref{tab08} for the scalar and Dirac perturbations 
respectively. In this case also, we observe that the errors associated with 
the scalar perturbation are comparatively smaller than those found in the 
Dirac perturbation. In the scalar perturbation, 5th order WKB has smaller 
errors. The real quasinormal mode increases and the decay rate decreases with 
an increase in the value of multipole moment, $l$. In the case of Dirac 
perturbation also, with an increase in $k$, real quasinormal mode increases. 
But according to the results from 3rd WKB, the decay rate decreases and for 
5th WKB, decay rate increases with an increase in $k$. However, one may note 
that the errors associated with 5th order WKB are higher and this is basically 
due to the form of the potential. For smaller values of $k$, the errors are 
large. Since, the error associated with 3rd order WKB is comparatively smaller,
 we can use the trend provided by this method for the implications from 
quasinormal modes. Moreover, we use 3rd order WKB in case of this new shape 
function also, in order to analyse the quasinormal modes graphically.
For this shape function defined by \eqref{staro_shape}, we have observed the 
variation of real and imaginary quasinormal modes with respect to the 
Starobinsky model parameter $\alpha$ in Fig.~\ref{staro_qnms_scalar_a}. The 
real quasinormal modes increase non-linearly with an increase in the parameter 
$\alpha$ towards $0$. One may note that it is not possible to obtain a 
wormhole solution and hence the corresponding quasinormal modes for positive 
values of the parameter $\alpha$. The decay rate also increases non-linearly 
with an increase in the parameter $\alpha$. In 
Fig.~\ref{staro_qnms_scalar_eta}, we have shown the dependency of the scalar 
quasinormal frequencies and the decay rates with respect to the cloud of strings parameter present in the wormhole shape function. It is obvious from the figure 
that he quasinormal frequencies and the decay rates depend on the magnitude of 
the parameter $\eta$ only. With a decrease in the magnitude of the parameter 
$\eta$ towards $0$, both the quasinormal frequencies and the decay rates 
decrease to a minimum value.

The variation of Dirac quasinormal modes with respect to the parameter $\alpha$
for the wormhole obtained in the Starobinsky model is shown in 
Fig.~\ref{staro_qnms_dirac_a}. In this case, the real quasinormal modes 
decrease with an increase in $\alpha$. The decay rate, on the other hand, 
increases with an increase in $\alpha$. So, in both types of perturbations, 
i.e.\ in scalar and Dirac perturbations, an increase in the Starobinsky 
parameter, the decay rate increases.

Finally, we have plotted the real and imaginary Dirac quasinormal modes with 
respect to $\eta$ for the wormhole in the Starobinsky model in 
Fig.~\ref{staro_qnms_dirac_eta}. It is seen that an increase in the magnitude 
of $\eta$, decreases the quasinormal frequencies gradually. On the other hand, 
the decay rate increases with an increase in the magnitude of $\eta$. So, in 
general for both the type of perturbations, i.e.\ for the scalar and Dirac 
perturbations $\eta$ has a similar type of impact over the decay rate of the 
quasinormal frequencies, while the impact of $\eta$ over the real quasinormal 
modes is opposite.

\section{Time Domain Analysis} \label{section04}
We study the evolution of the scalar and Dirac perturbations, especially the
time domain profiles for these perturbations in this section for the 
considered wormhole solutions. To study the time domain profiles for the 
respective perturbation schemes we implement the time domain integration method 
introduced in Ref.~\cite{gundlach}. We define the associated wavefunction 
as $\psi(r_*, t) = \psi(i \Delta r_*, j \Delta t) \equiv \psi_{i,j} $ and the 
potential as $V(r(r_*)) = V(r_*,t) \equiv V_{i,j}$ to write 
Eq.~\eqref{scalar_KG} in the form:
\begin{equation}
\dfrac{\psi_{i+1,j} - 2\psi_{i,j} + \psi_{i-1,j}}{\Delta r_*^2} - \dfrac{\psi_{i,j+1} - 2\psi_{i,j} + \psi_{i,j-1}}{\Delta t^2} - V_i\psi_{i,j} = 0.
\end{equation}

\begin{figure}[h!]
\centerline{
   \includegraphics[scale = 0.3]{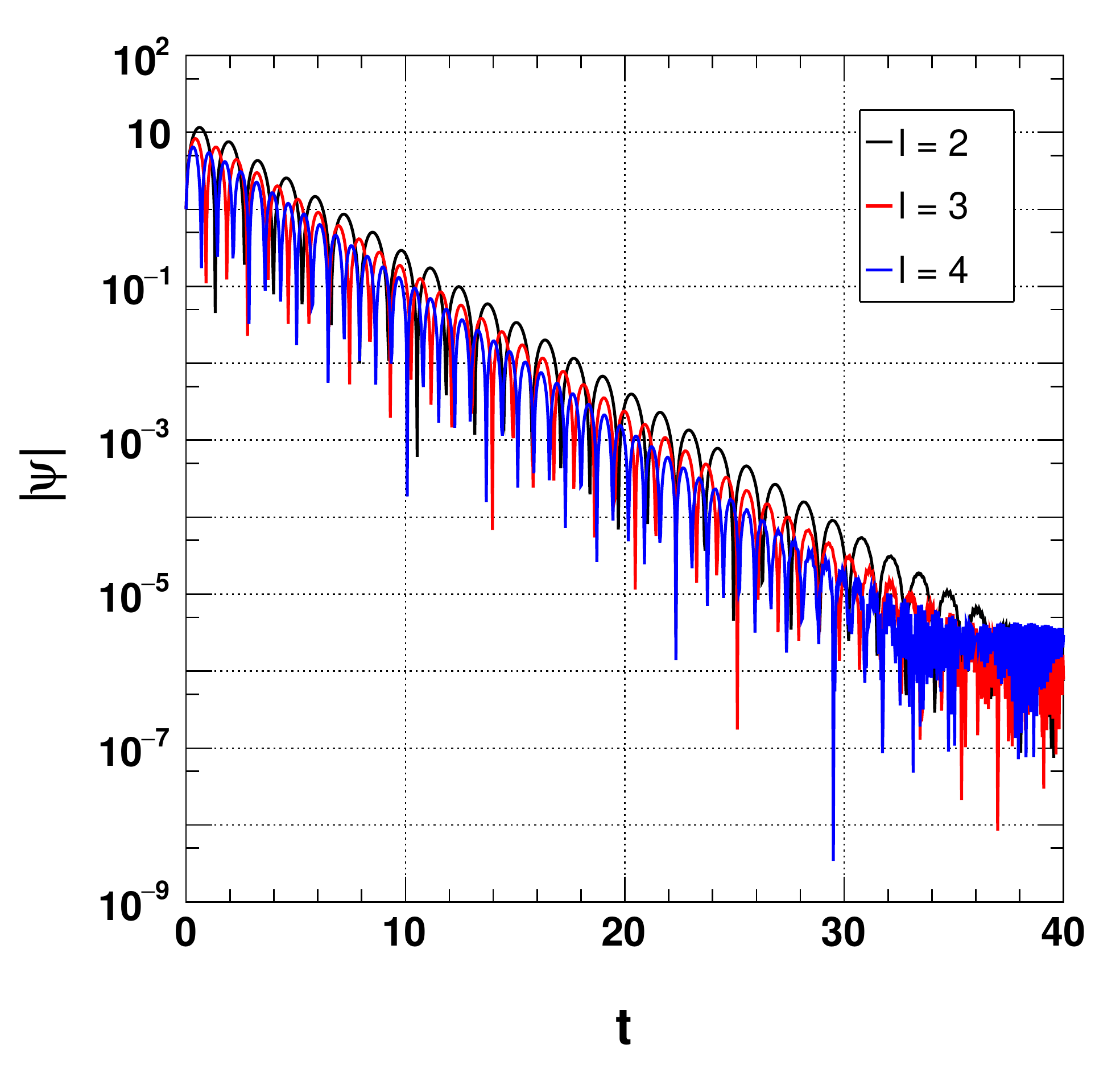}\hspace{1cm}
   \includegraphics[scale = 0.3]{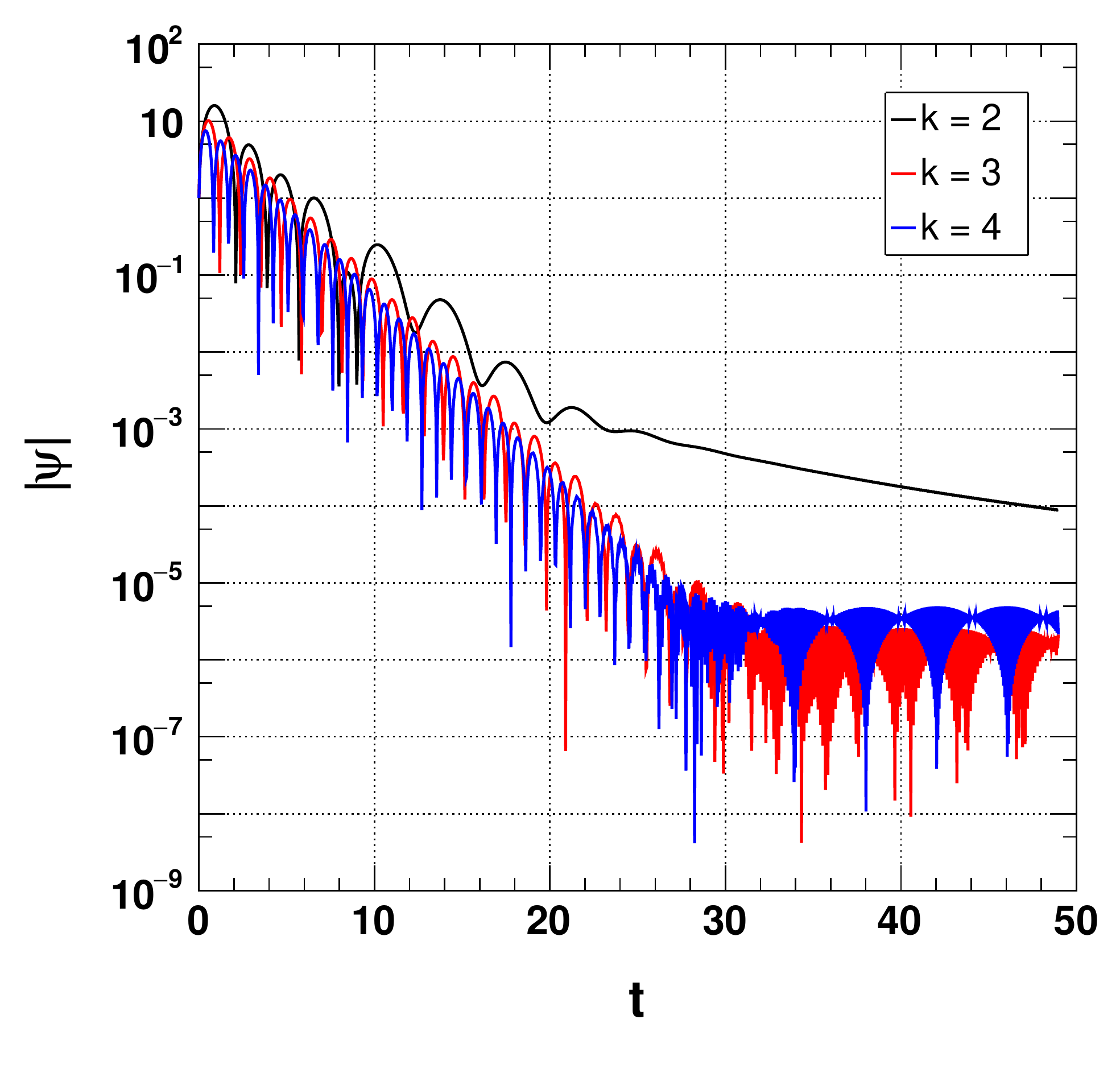}}
\vspace{-0.2cm}
\caption{Time domain profile for the wormhole defined by the shape function 
\eqref{staro_shape} for the scalar perturbation with $\alpha = -0.5$, 
$\eta = 0.1$ and $r_0=1$ (on the left panel) and for the Dirac perturbation 
with $\alpha = -0.3, \eta = 0.5$ and $r_0 = 1$ (on the right panel).}
\label{time_domain01}
\end{figure}

\begin{figure}[h!]
\centerline{
   \includegraphics[scale = 0.3]{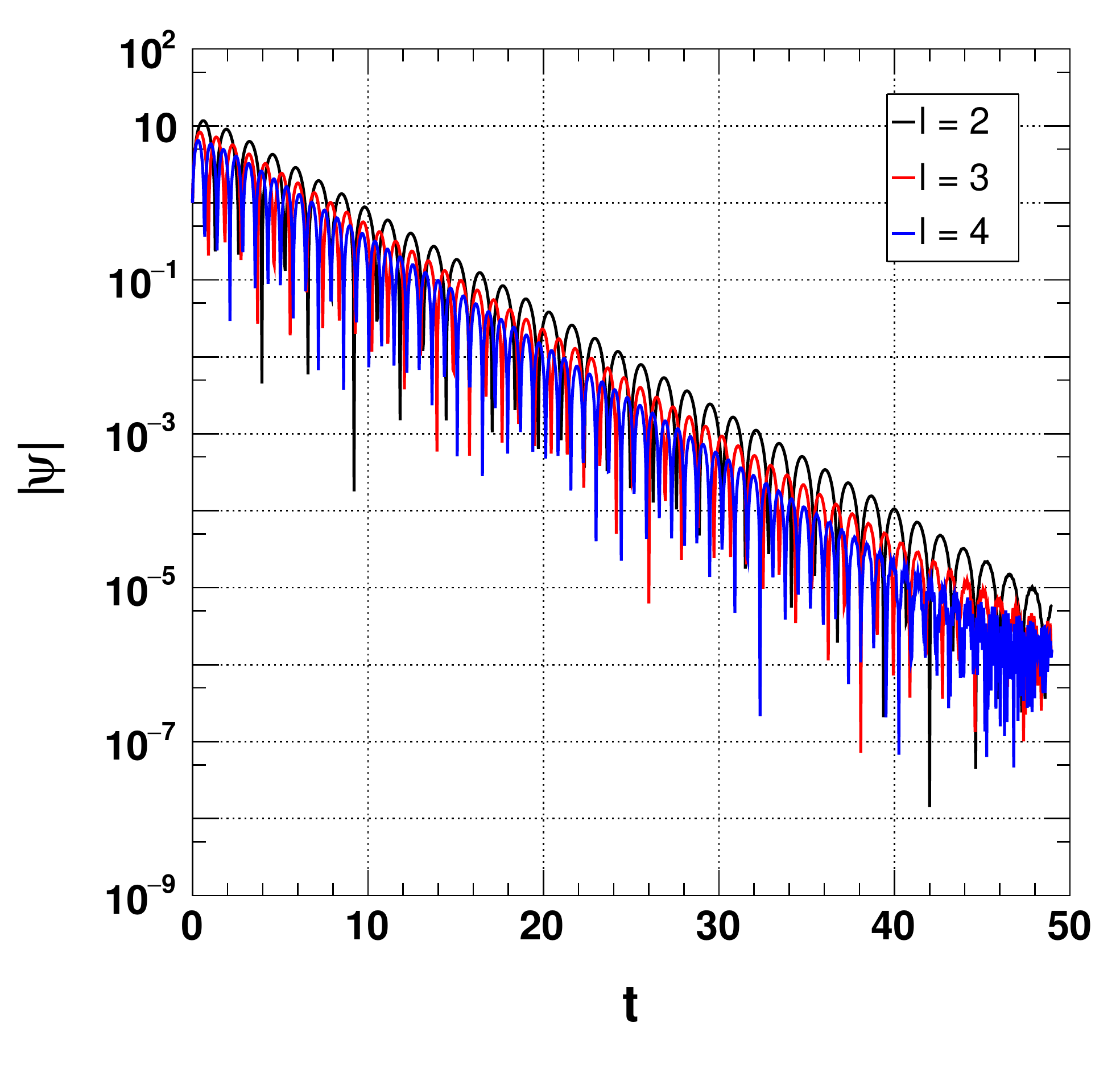}\hspace{1cm}
   \includegraphics[scale = 0.3]{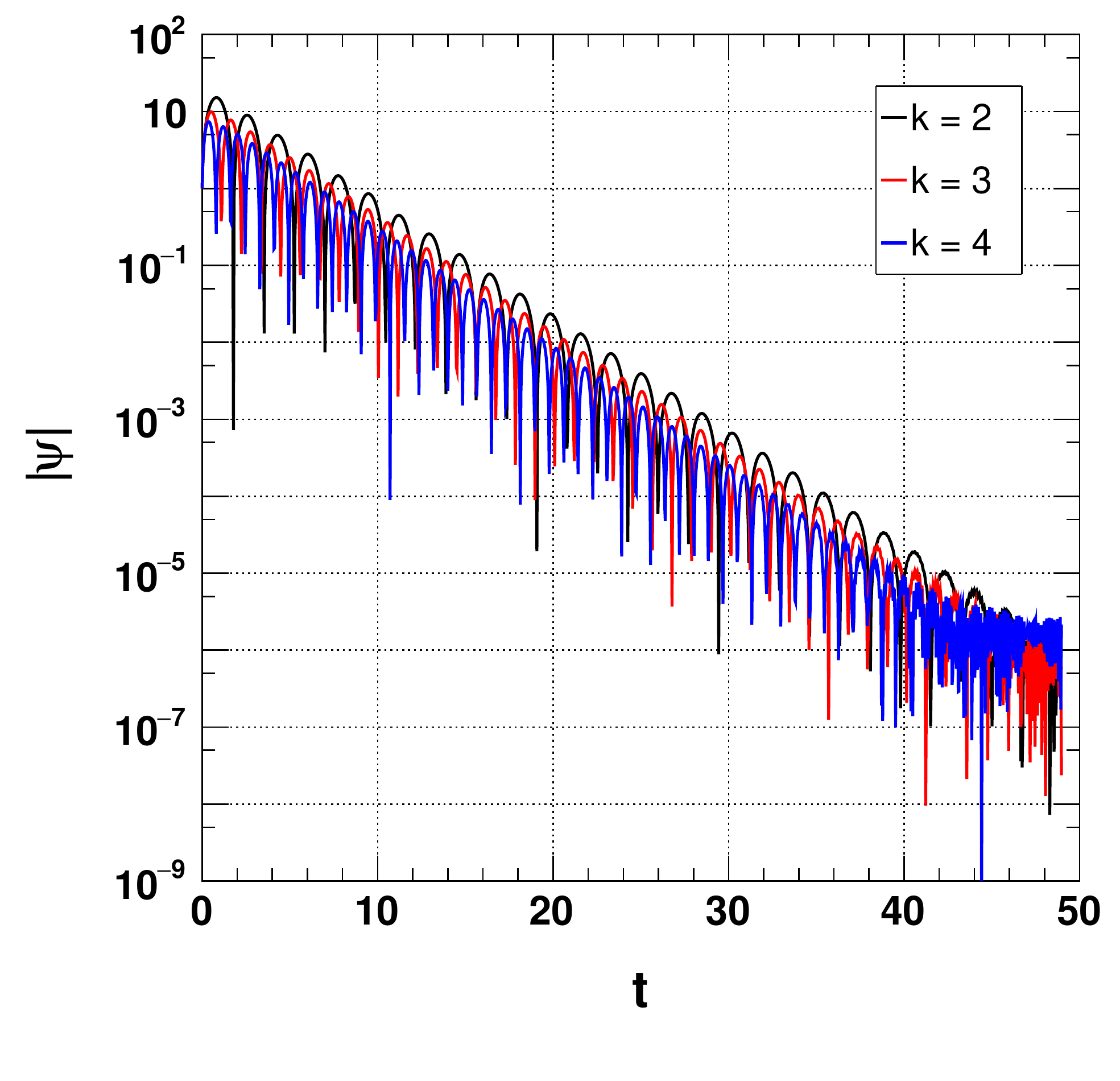}}
\vspace{-0.2cm}
\caption{Time domain profile for the wormhole defined by the shape function 
\eqref{toy_shape01} for the scalar perturbation (on the left panel) and the Dirac perturbation (on the right panel) with 
$m_1 = 0.5$ and $r_0=1$.}
\label{time_domain02}
\end{figure}

\begin{figure}[h!]
\centerline{
   \includegraphics[scale = 0.3]{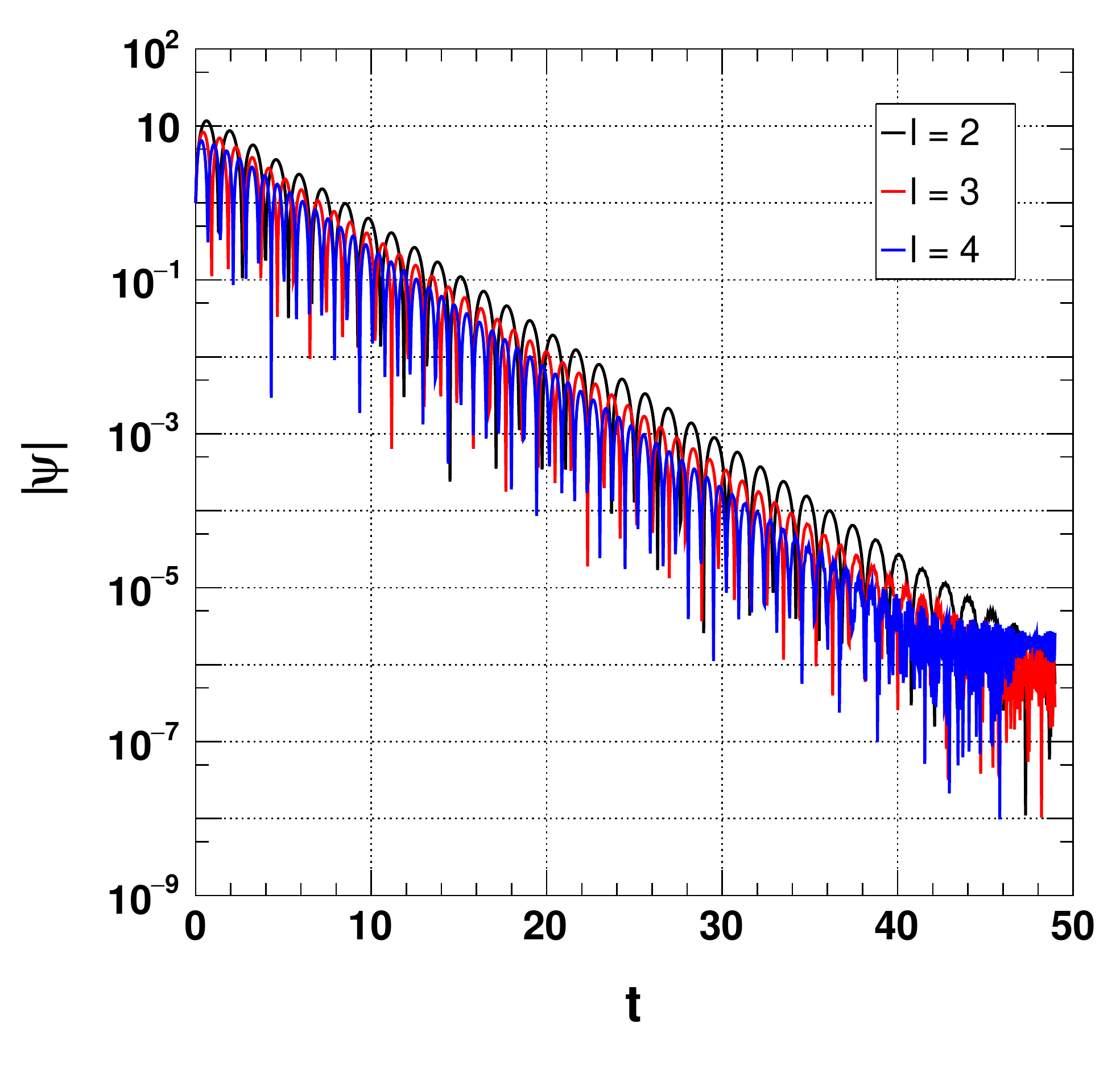}\hspace{1cm}
   \includegraphics[scale = 0.3]{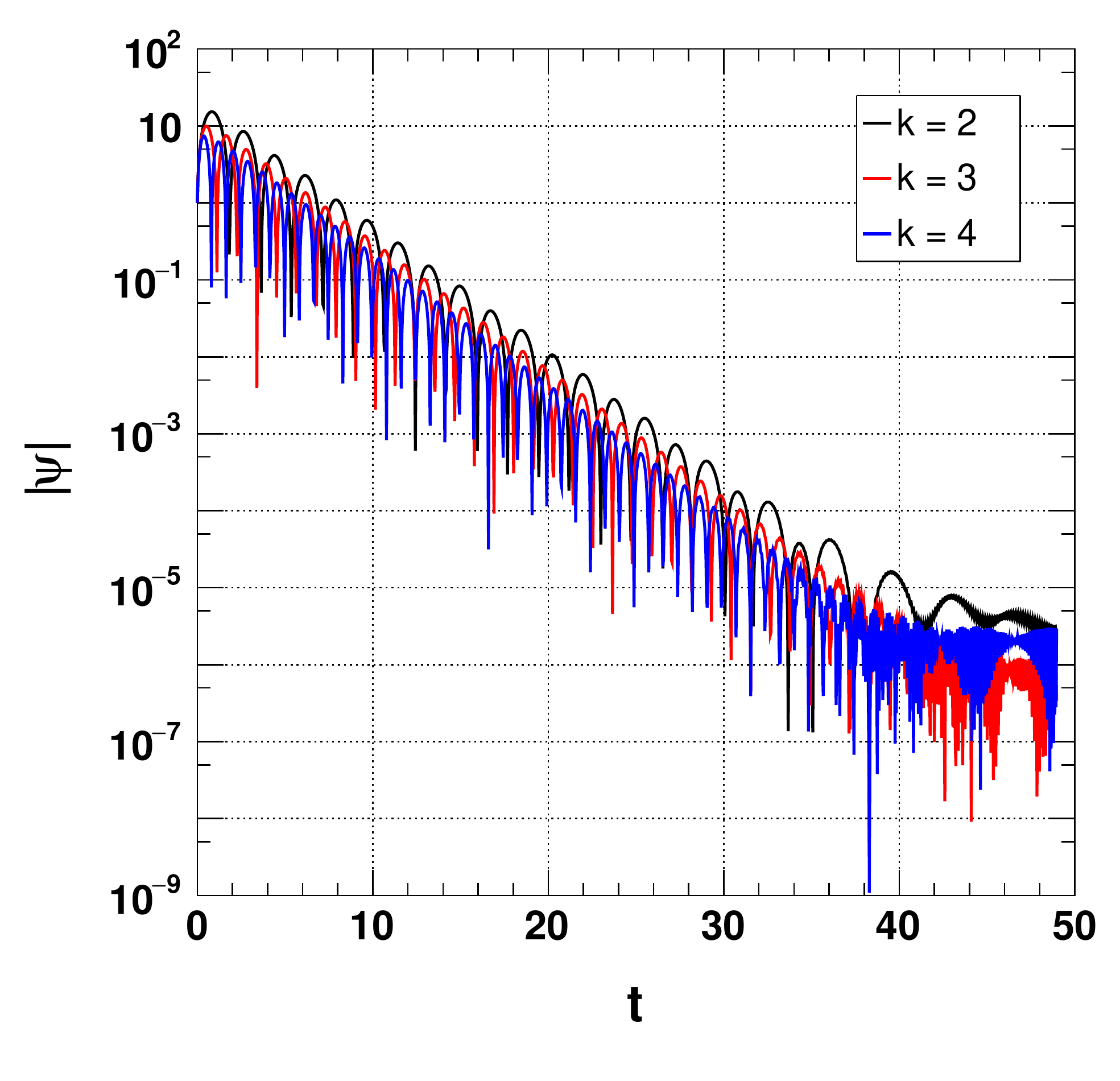}}
\vspace{-0.2cm}
\caption{Time domain profile for the wormhole defined by the shape function 
\eqref{toy_shape02} for the scalar perturbation (on the left panel) and the Dirac perturbation (on the right panel) with 
$m_2 = 1.1$ and $r_0=1$.}
\label{time_domain03}
\end{figure}

\begin{figure}[h!]
\centerline{
   \includegraphics[scale = 0.3]{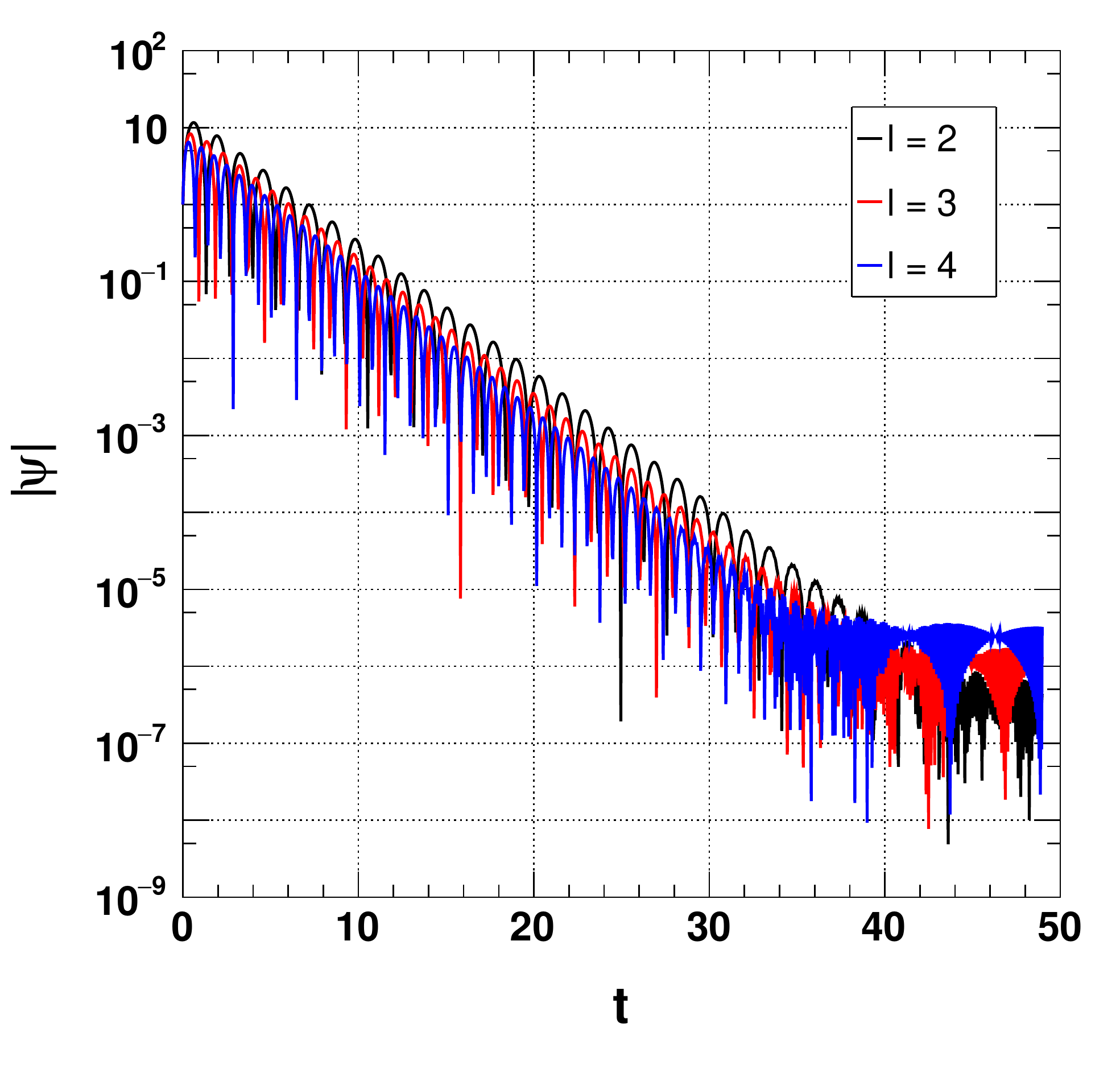}\hspace{1cm}
   \includegraphics[scale = 0.3]{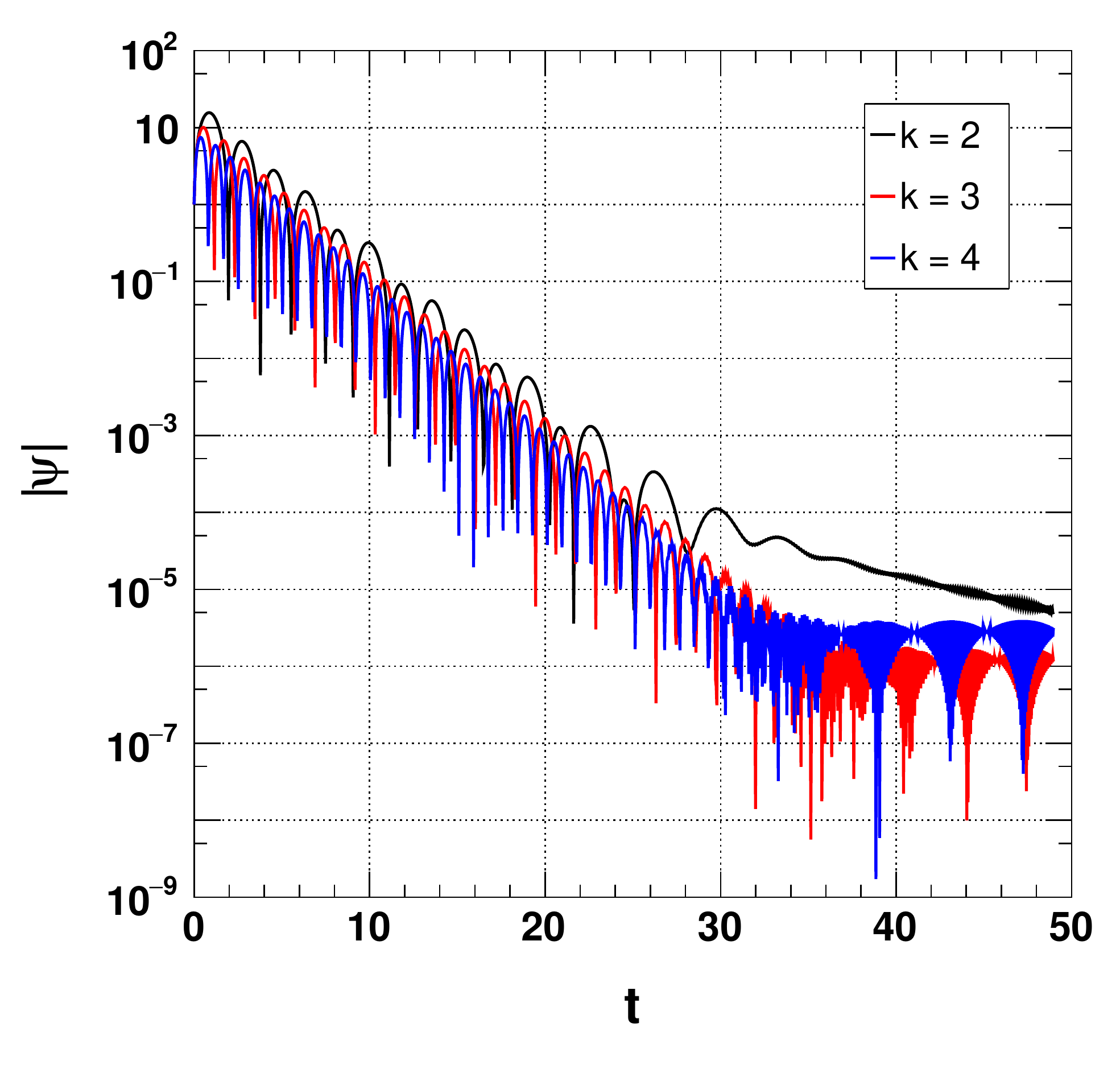}}
\vspace{-0.2cm}
\caption{Time domain profile for the wormhole defined by the shape function 
\eqref{toy_shape03} for the scalar perturbation (on the left panel) and the Dirac perturbation (on the right panel) with 
$m_3 = 1.1$ and $r_0=1$.}
\label{time_domain04}
\end{figure}

Identifying the initial conditions as $\psi(r_*,t) = \exp \left[ -\dfrac{(r_*-k_1)^2}{2\sigma^2} \right]$ and $\psi(r_*,t)\vert_{t<0} = 0$ (here $k_1$ and 
$\sigma$ are median and width of the initial wave-packet), it is possible to 
express the time evolution of the scalar field as

\begin{equation}
\psi_{i,j+1} = -\psi_{i, j-1} + \left( \dfrac{\Delta t}{\Delta r_*} \right)^2 (\psi_{i+1, j + \psi_{i-1, j}}) + \left( 2-2\left( \dfrac{\Delta t}{\Delta r_*} \right)^2 - V_i \Delta t^2 \right) \psi_{i,j}.
\end{equation}
In order to comply with the Von Neumann stability condition we have chosen 
$\frac{\Delta t}{\Delta r_*} < 1$ during the numerical procedure. We use the 
same method for the Dirac perturbation scheme also to obtain the time domain 
profiles or the time evolution of the perturbations for a selected set of 
parameters. The time profiles for scalar and Dirac perturbations are shown in 
Fig.s~\ref{time_domain01}, \ref{time_domain02}, \ref{time_domain03} and \ref{time_domain04} for shape function obtained from Starobinsky model and toy shape functions respectively. The profiles show that with an increase in the 
number of the multipole modes, the oscillation frequency increases. Another 
important implication is that the corresponding oscillation frequency for the 
Dirac perturbation case is smaller than that obtained for the scalar 
perturbation. The damping rate or decay rate also increases slowly with an 
increase in the multipole modes for both types of perturbations. One may also 
note that the times domain profiles for the Starobinsky model case are 
distinguishable from the time domain profiles of the toy wormhole models. It 
is because of the impact of the cloud of strings parameter. In the Starobinsky 
model case, the structure of the wormhole highly depends on the surrounding 
cloud of strings and any variation in the cloud of strings parameter will 
impact the quasinormal modes and time domain profiles obtained from the 
perturbation of the wormhole geometry. But in case of the toy models, we have 
well defined the shape functions initially and hence the cloud of strings 
parameter appears as a contribution towards the geometry modification i.e. the 
corresponding $f(R)$ gravity model has the cloud of strings parameter as a 
model parameter in it. This basically provides an $f(R)$ gravity model which 
can compensate the effect of cloud of strings for the chosen shape function of 
the wormhole. In Fig.~\ref{time_domain02}, for both type of perturbations we 
have chosen $m_1 = 0.5$ and throat radius of the wormhole $r_0 = 1$. Similar 
to the previous case, we observe that with increase in the multipole moment, 
quasinormal frequencies and decay rate increases. However, the variation of 
decay rate is comparatively small. In Fig.s~\ref{time_domain03} and \ref{time_domain04}, we have used $m_2 = 1.1$ and $m_3 = 1.1$ respectively. We observe 
that for both cases, the time domain profiles are not identical and the 
variation is more prominent in case of the Dirac perturbation. These results 
support the previous results obtained from the WKB analysis. So, at last we 
can summarise that the structure of the wormholes are connected with the 
quasinormal modes and hence also on the time domain profiles. Next generation 
GWs detectors like LISA may play a prominent role \cite{Gogoi_bumblebee} in 
distinguishing between quasinormal modes from black holes and wormholes and 
hopefully may be able to constrain the wormhole shape functions to a 
satisfactory order.
 
However, one can also consider constraining the gravity models using 
cosmological or other experimental data to see how tightly the quasinormal 
spectrum is bounded. Since in the wormhole configuration we have considered a 
cloud of strings as a possible relic, a proper constraint from cosmological 
data may give useful insights here. We keep this as a future prospect of this 
study. In a recent study, from GW150914 with $90\%$ credibility, the following 
bound on the fractional deviations of quasinormal modes for Kerr black holes 
are obtained \cite{Ghosh2021}:
\begin{equation}
 \delta f_{220} = 0.05^{+0.11}_{-0.07}, \;\;\; \delta \tau_{220} = 0.07^{+0.26}_{-0.23}, 
\end{equation}
where $f$ stands for real frequencies and $\tau$ stands for damping time.
Assuming this bound to be valid for the case of wormholes also, one can see 
that it puts a weak limit on the model parameters. For example, in the
Starobinsky model, we can see that $\delta \alpha \approx 71.43$ and 
$\delta \eta \approx 2.78$ for the Dirac perturbation using real quasinormal 
mode only. It may be noted that the values of these two parameters considered 
in our study lie within these weak limits of their values. However, to get a 
proper constraint, we need to wait for LISA.

\section{Conclusion}\label{section05}
In this study, we have used three toy models of wormhole shape functions along 
with a wormhole solution obtained from the Starobinsky model surrounded by a 
cloud of strings in the $f(R)$ gravity metric formalism. At first we studied 
the viability conditions for this new wormhole solution along with the 
three ad-hoc shape functions. We have also plotted the embedded diagrams of 
the wormholes for a better visualisation of the model parameter dependencies 
on the structure of the wormhole. We see that the first toy model 
\eqref{toy_shape01} has a unique behaviour in terms of the model parameter 
dependency, which is opposite to the cases observed for the other two toy 
shape functions. On the other hand, the new shape function obtained for the 
Starobinsky model has two model parameters apart from the throat radius $r_0$. 
They are the Starobinsky parameter $\alpha$ and the cloud of strings parameter 
$\eta$. The viability functions as well as the structure of the wormhole 
depends highly on the cloud of strings parameter $\eta$. However, the 
Starobinsky parameter $\alpha$ has a smaller dependency on the same. 
Thereafter, we 
considered two types of perturbations in the wormhole spacetime, viz.~the 
scalar and Dirac perturbations. We have calculated the corresponding 
potentials in the tortoise coordinates, as in the normal coordinates, it is 
not possible to predict the peak of the potential due to the property of the 
wormhole spacetime. We have seen that the quasinormal modes for the scalar 
perturbation are more accurate than those obtained in the Dirac perturbation 
in terms of the error parameter defined earlier. This is due to the different 
behaviour of the Dirac perturbation for which the quasinormal modes with 
smaller multipole modes become unstable in the WKB approximation method and 
for smaller values of $k$, sometimes positive imaginary quasinormal modes are 
obtained denoting unstable quasinormal modes. Hence, in our study, we have 
considered higher values of $k$ in case of Dirac perturbation so that we can 
avoid unstable modes and obtain quasinormal modes with higher accuracy. One 
may note that in WKB approximation method, one may not always get higher 
accuracy for higher order corrections. We have observed that for the scalar 
perturbation, 5th order WKB gives higher accuracy but in the case of Dirac 
perturbation, 3rd order WKB gives better accuracy than 5th order WKB. Since, 
in general for both the cases, errors associated with the 3rd order WKB is in 
acceptable range, we have considered 3rd order WKB to analyse the quasinormal 
modes with respect to the model parameters in the graphs.

Our study shows that the cloud of strings parameter can have a significant 
impact on the quasinormal modes from a wormhole in $f(R)$ gravity Starobinsky 
model. The impacts are opposite for the real quasinormal modes for the scalar 
and Dirac perturbations. So, in such a situation, quasinormal modes may be a 
probe in the near future to distinguish between scalar and Dirac perturbations 
in different wormhole geometries. However, in case of the toy shape functions 
defined in \eqref{toy_shape01}, \eqref{toy_shape02} and \eqref{toy_shape03}, 
the quasinormal modes do not depend on the cloud of strings parameter. It is 
because, the shape functions are defined at first and the definitions are kept 
rigid throughout the whole calculation. So, the impact of the relic cloud of 
strings appears in the explicit form of the $f(R)$ gravity model. Therefore, 
it is impossible to see the impact of cloud of strings parameter on the 
quasinormal modes in case of the toy shape functions. So, in such a situation 
astrophysical constraints on the corresponding $f(R)$ gravity model may 
provide some useful insights including viability of such configurations. The 
other model parameters $m_1$, $m_2$ and $m_3$ have significant impacts on the 
quasinormal modes from the corresponding wormholes. The impact on the 
quasinormal modes of $m_1$ of the first toy shape function is however, 
opposite in comparison to the second and third toy shape functions. On 
the other hand, one may note that the wormhole solution found for the 
Starobinsky model depends highly on the cloud of strings parameter and this 
solution is valid only in presence of this relic. So, as expected the wormhole 
shape function does not have a GR limit, or the solution is completely 
different from a GR scenario. Moreover, the solution is different significantly
from a standard Schwarzschild solution and hence it is expected to be easily 
differentiable from a GR black hole with the help of quasinormal modes. 
Another important point to note is that the Starobinsky wormhole differs 
significantly from the black hole solutions obtained in Ref.~\cite{Graca2018},
where black holes in $f(R)$ gravity have been taken into account along with a 
cloud of strings. Therefore, 
it might be possible in the near future to distinguish between wormhole 
shape functions and a black hole using the GW observations from the next generation detectors 
like LISA \cite{Gogoi_bumblebee}. This study will also help to test the viability of the Starobinsky model and to constrain it in the near future using the observational data from quasinormal modes. The Starobinsky model provides a unique wormhole shape function in presence of a cloud of strings and a study of several other properties of this wormhole such as weak deflection angle, geodesic equations etc. will shed some more light on such configurations.

A basic question has been raised several times in literature if it is possible 
to distinguish quasinormal modes from a black hole and a wormhole spacetime
\cite{Cardoso2016, Cardoso2016_2, Chirenti2016, Konoplya2016, Nandi2017}. 
Although it is seen that the quasinormal spectra for wormholes varies from 
those for black holes, for some wormholes the quasinormal modes can be very
close to those found in black holes and in such a case quasinormal modes may 
not be very helpful in distinguishing between black holes and wormholes 
\cite{Konoplya2016, Nandi2017, Volkel2018}. But it might be possible to 
distinguish between wormholes and black holes effectively using other ways 
\cite{Hong2021}. In such a situation, experimental results from quasinormal 
modes may be used to study the variations and as a supporting evidence.

\section*{Acknowledgements} DJG is thankful to Prof.~Peter Kuhfittig, 
Emeritus Professor, Department of Mathematics, Milwaukee School of Engineering, United States for some useful discussions. UDG is thankful to the 
Inter-University Centre for Astronomy and Astrophysics (IUCAA), Pune, India 
for the Visiting Associateship of the institute.

\appendix*
\section{Expressions of $\rho + p_r$}
\subsection{Shape function 01}
For the first shape function defined by \eqref{toy_shape01}, we have 
\begin{equation}
\mathcal{F}(r) = -\frac{\eta ^2 e^{-m_1 \left(r_0-r\right)}}{1-m_1 r}.
\end{equation}
Using this expression, Eq.s \eqref{fieldeq01} and \eqref{generic2} can provide,
\begin{equation}
\rho + p_r = \frac{\eta ^2 m_1 e^{m_1 \left(r-r_0\right)} \left(e^{m_1 \left(r_0-r\right)} \left(m_1^3 r^3-7 m_1^2 r^2+12 m_1 r-2\right)-2 m_1 r \left(m_1^2 r^2-4 m_1 r+5\right)\right)}{2 r \left(m_1 r-1\right){}^3},
\end{equation}
which has been used for checking the violation of NEC.

\subsection{Shape function 02}
For the second shape function defined by \eqref{toy_shape02}, using the field equations, we have obtained,
\begin{equation}
\mathcal{F}(r) = -\,\frac{\eta ^2 \left(r+r_0\right) \log ^2\left(m_2 \left(r+r_0\right)\right)}{\log \left(2 m_2 r_0\right) \left(r \log \left(m_2 \left(r+r_0\right)\right)+r_0 \log \left(m_2 \left(r+r_0\right)\right)-r\right)}.
\end{equation}
This expression along with the field Eq.s \eqref{fieldeq01} and \eqref{generic2}, we have
\begin{align}
\rho + p_r =& \frac{\eta ^2}{2 r \left(r+r_0\right) \log \left(2 m_2 r_0\right) \log \left(m_2 \left(r+r_0\right)\right) \left(\left(r+r_0\right) \log \left(m_2 \left(r+r_0\right)\right)-r\right){}^3}
\Big[-2 r^3 \log \left(2 m_2 r_0\right)
\\ \notag &+2 \left(r+r_0\right) \left(\left(2 r^2+5 r_0 r+r_0^2\right) \log \left(2 m_2 r_0\right)+r \left(3 r+2 r_0\right)\right) \log ^3\left(m_2 \left(r+r_0\right)\right)
\\ \notag &+r^2 \left(3 \left(r+2 r_0\right) \log \left(2 m_2 r_0\right)+4 r\right) \log \left(m_2 \left(r+r_0\right)\right)-2 r \left(r+r_0\right) \left(r+3 r_0\right) \log ^4\left(m_2 \left(r+r_0\right)\right)
\\ \notag &-r \left(3 \left(r+r_0\right) \left(3 r+2 r_0\right) \log \left(2 m_2 r_0\right)+4 r \left(r+2 r_0\right)\right) \log ^2\left(m_2 \left(r+r_0\right)\right) \Big].
\end{align}

\subsection{Shape function 03}
Similarly, for the shape function \eqref{toy_shape03}, we have 
\begin{equation}
\mathcal{F}(r) = -\,\frac{\eta ^2 \left(m_3+r-r_0\right){}^2}{m_3 \left(m_3-r_0\right)},
\end{equation}
which along with the field Eq.s \eqref{fieldeq01} and \eqref{generic2} results,
\begin{equation}
\rho + p_r = \frac{\eta ^2 \left(m_3 \left(2 r-r_0\right)+m_3^2+2 r \left(r-r_0\right)\right)}{m_3 r \left(m_3-r_0\right) \left(m_3+r-r_0\right)}.
\end{equation}

\subsection{Shape function 04 (Starobinsky model shape function)}
Finally, for the Starobinsky model,
\begin{equation}
\mathcal{F}(r) = -\,\frac{8 \alpha  \eta ^2}{r \left(\sqrt{r^2-16 \alpha  \eta ^2}-r\right)}.
\end{equation}
Using this along with the field Eq.s \eqref{fieldeq01} and \eqref{generic2}, we can obtain
\begin{align}
\rho + p_r =& \frac{\eta ^2}{6 r^4 \left(r^2-16 \alpha  \eta ^2\right)^{3/2} \left(\sqrt{r^2-16 \alpha  \eta ^2}-r\right)} 
 \Big[ 12288 \alpha ^3 \eta ^6 +128 \alpha ^2 \eta ^2 \Big\lbrace  r \Big(10 \eta ^2 \left(\sqrt{r^2-16 \alpha  \eta ^2}-\sqrt{r_0^2-16 \alpha  \eta ^2}\right)
\\ \notag &-12 \sqrt{r^2-16 \alpha  \eta ^2}-15 r_0  \Big) +3 \sqrt{r^2-16 \alpha  \eta ^2} \left(2 \eta ^2 \sqrt{r_0^2-16 \alpha  \eta ^2}+3 r_0\right)+4 \left(3-7 \eta ^2\right) r^2  \Big\rbrace 
\\ \notag &+r^2 \left(-12 r^4+5 r_0^2 \sqrt{r^2-16 \alpha  \eta ^2} \left(\sqrt{r_0^2-16 \alpha  \eta ^2}-r_0\right)+12 r^3 \sqrt{r^2-16 \alpha  \eta ^2}+7 r_0^2 r \left(r_0-\sqrt{r_0^2-16 \alpha  \eta ^2}\right)\right)
\\ \notag &-8 \alpha  \Big\lbrace  6 \left(3-7 \eta ^2\right) r^4+5 r^2 \sqrt{r^2-16 \alpha  \eta ^2} \left(2 \eta ^2 \sqrt{r_0^2-16 \alpha  \eta ^2}+3 r_0\right)+6 \eta ^2 r_0^2 \sqrt{r^2-16 \alpha  \eta ^2} \left(\sqrt{r_0^2-16 \alpha  \eta ^2}-r_0\right)
\\ \notag &+r^3 \left(2 \eta ^2 \left(15 \sqrt{r^2-16 \alpha  \eta ^2}-7 \sqrt{r_0^2-16 \alpha  \eta ^2}\right)-18 \sqrt{r^2-16 \alpha  \eta ^2}-21 r_0\right)+10 \eta ^2 r_0^2 r \left(r_0-\sqrt{r_0^2-16 \alpha  \eta ^2}\right) \Big\rbrace \Big].
\end{align}



\bibliographystyle{apsrev}
\end{document}